\documentclass[a4paper,11pt]{article}
\usepackage{jheppub}

\usepackage{graphicx}
\usepackage{graphics}
\usepackage{dcolumn}
\usepackage{bm}
\usepackage{epstopdf}
\usepackage{mathrsfs}
\usepackage{amssymb}
\usepackage{amsmath}
\usepackage{natbib}

\newcommand{\beq}{\begin{eqnarray}}
\newcommand{\eeq}{\end{eqnarray}}
\newcommand{\be}{\begin{equation}}
\newcommand{\ee}{\end{equation}}

\title{Constraints on the gluon PDF from top quark pair production at hadron colliders}

\author[a]{Michal Czakon,}
\author[b]{Michelangelo L. Mangano,}
\author[b]{Alexander Mitov}
\author[b]{and Juan Rojo}

\affiliation[a]{Institut fur Theoretische Teilchenphysik und Kosmologie, RWTH Aachen University, D-52056 Aachen, Germany}
\affiliation[b]{Physics Department, Theory Unit, CERN, CH-1211 Gen\`eve, Switzerland}

\emailAdd{mczakon@physik.rwth-aachen.de}
\emailAdd{michelangelo.mangano@cern.ch}
\emailAdd{alexander.mitov@cern.ch}
\emailAdd{juan.rojo@cern.ch}

\preprint{CERN-PH-TH-2013-036, TTK-13-09}

\abstract{Using the recently derived NNLO cross sections~\cite{Czakon:2013goa},
 we provide 
NNLO+NNLL theoretical predictions for top quark
  pair production based on all the available NNLO PDF sets, and
  compare them with the most precise LHC and Tevatron data. 
In this
  comparison we study in detail the PDF uncertainty and the scale,
  $m_t$ and $\alpha_s$ dependence of the theoretical predictions for
  each PDF set. 
Next, we observe that top quark pair
  production provides a powerful direct constraint on the gluon PDF at
  large $x$, and include Tevatron and LHC top pair data consistently into a global NNLO
  PDF fit. 
We then explore the phenomenological consequences of the
  reduced gluon PDF uncertainties, by showing how they can improve
  predictions for Beyond the Standard Model processes at the LHC. 
Finally,
  we update to full NNLO+NNLL the theoretical predictions for the ratio of
  top quark cross sections between different LHC center of mass
  energies, as well as the cross sections for 
hypothetical heavy fourth-generation quark
  production at the LHC.}

\begin{document} 
\maketitle
\flushbottom

\section{Introduction} 

The total $t\bar{t}$ production cross section is an important observable at hadron
colliders, which has been recently computed in full next-to-next-to leading order (NNLO) QCD \cite{Baernreuther:2012ws,Czakon:2012zr,Czakon:2012pz,Czakon:2013goa}. Thanks to a much improved control over higher order terms, including soft gluon emissions through next-to-next-to leading log (NNLL) \cite{Beneke:2009rj,Czakon:2009zw,Cacciari:2011hy}, scale uncertainties are now controlled down to the 2.2 (3) percent level at the Tevatron (LHC), enabling a number of precision phenomenology applications to SM and BSM physics.

In this work we present an in-depth study of the theoretical uncertainties
that affect the total cross section computed 
with NNLO+NNLL precision.
These uncertainties are the parton distributions
of the proton, the value of the strong coupling
$\alpha_s(M_Z)$, the value of the top quark mass
$m_t$ and the scale uncertainties from missing higher perturbative orders. 
We then compare the theoretical predictions with the most precise available data
from the Tevatron and the LHC at 7 and 8 TeV.  We also provide
predictions for LHC at 14 TeV, as well as for the ratio
of cross sections between 7, 8 and 14 TeV and for
the production of  heavy top-like fermions. Previous
phenomenological studies of the total $t\bar{t}$ cross sections,
based on different approximations to the full NNLO 
calculation were presented in
Refs.~\cite{Cacciari:2008zb,Cacciari:2011hy,Moch:2012mk,Aliev:2010zk,Beneke:2012wb,Ahrens:2011px,Beneke:2011mq,Kidonakis:2011ca,Watt:2011kp,Watt:2012np,Forte:2013wc}.

As a first phenomenological application of the full NNLO calculation 
we study the impact of top quark cross section data in the
parton distribution analysis.
Indeed, 
top quark production is directly sensitive to the large-$x$ gluon PDF,
which at present is affected by substantial uncertainties.
In turn, large-$x$ gluons play an important role in theoretical predictions
of many BSM scenarios like gluino pair production~\cite{Kramer:2012bx}, 
high-mass Kaluza-Klein graviton production~\cite{Agashe:2007zd,Randall:1999vf,Giudice:2000av},
resonances in the $t\bar{t}$ invariant mass spectrum~\cite{Frederix:2007gi,Barger:2006hm}, quark compositeness in inclusive jet and dijet production~\cite{Chiappetta:1990jd,Chatrchyan:2013muj,Chatrchyan:2012bf,ATLAS:2012pu} and many others.
The availability of the full NNLO calculation makes top quark pair production the only
hadron collider process that is both sensitive to the gluon
and can be consistently included in a NNLO PDF fit
without any approximations. 
Hadronic constraints
on the gluon PDF are provided also by inclusive jet
and dijet production~\cite{D0:2008hua,Aaltonen:2008eq,Chatrchyan:2012bja,Aad:2011fc}
and isolated photon production~\cite{d'Enterria:2012yj,Carminati:2012mm}, 
though these two processes are only known to NLO
and affected by substantial scale uncertainties.\footnote{Recent progress
on the NNLO cross section for jet production was presented
in Ref.~\cite{Ridder:2013mf}, so in the near future
it should also be possible to consistently include this process
in NNLO PDF fits.}

The focus of this paper is, on the one hand, to provide
an up-to-date summary of the theoretical
uncertainties on the total $t\bar{t}$ cross section,
and on the other hand, to show how top quark
data can be used to constrain the large-$x$ gluon PDF.
Indeed, unlike the Tevatron, top quark pair
production at the LHC is dominated by $gg$ scattering, thus providing
a complementary probe of the gluon PDF.
 As shown in 
Table~\ref{tab:subproc}, at the LHC the 
relative contribution of the $gg$ subprocess 
is between 85\% and
90\% depending on the beam energy, with
$qq$
being about 10-15\%, almost the opposite of the Tevatron.

\begin{table}[h]
\centering
\begin{tabular}{c|c|c|c|c}
\hline
 & TeVatron  & LHC 7 TeV  & LHC 8 TeV & LHC 14 TeV\\ [1ex]
\hline
\hline
$gg$  & 15.4\%  & 84.8\% &  86.2\%   &   90.2\% \\ [1ex]
$qg+\bar{q}g$  & -1.7\%  &   -1.6\%  &  -1.1\%   & 0.5\% \\ [1ex]
$qq$ & 86.3\%  & 16.8\% &  14.9\%  &   9.3\%  \\ [1ex]
\hline
\end{tabular}
\caption{\small The relative contribution of the
various partonic sub-channels to the NNLO+NNLL
cross section for different colliders and collider energies,
computed with the MSTW2008NNLO PDFs.
We loosely label with $qq$ the sum of all processes
  without gluons in the initial state. 
 \label{tab:subproc}}
\end{table}

\begin{figure}[h]
\centering
\includegraphics[width=0.49\textwidth]{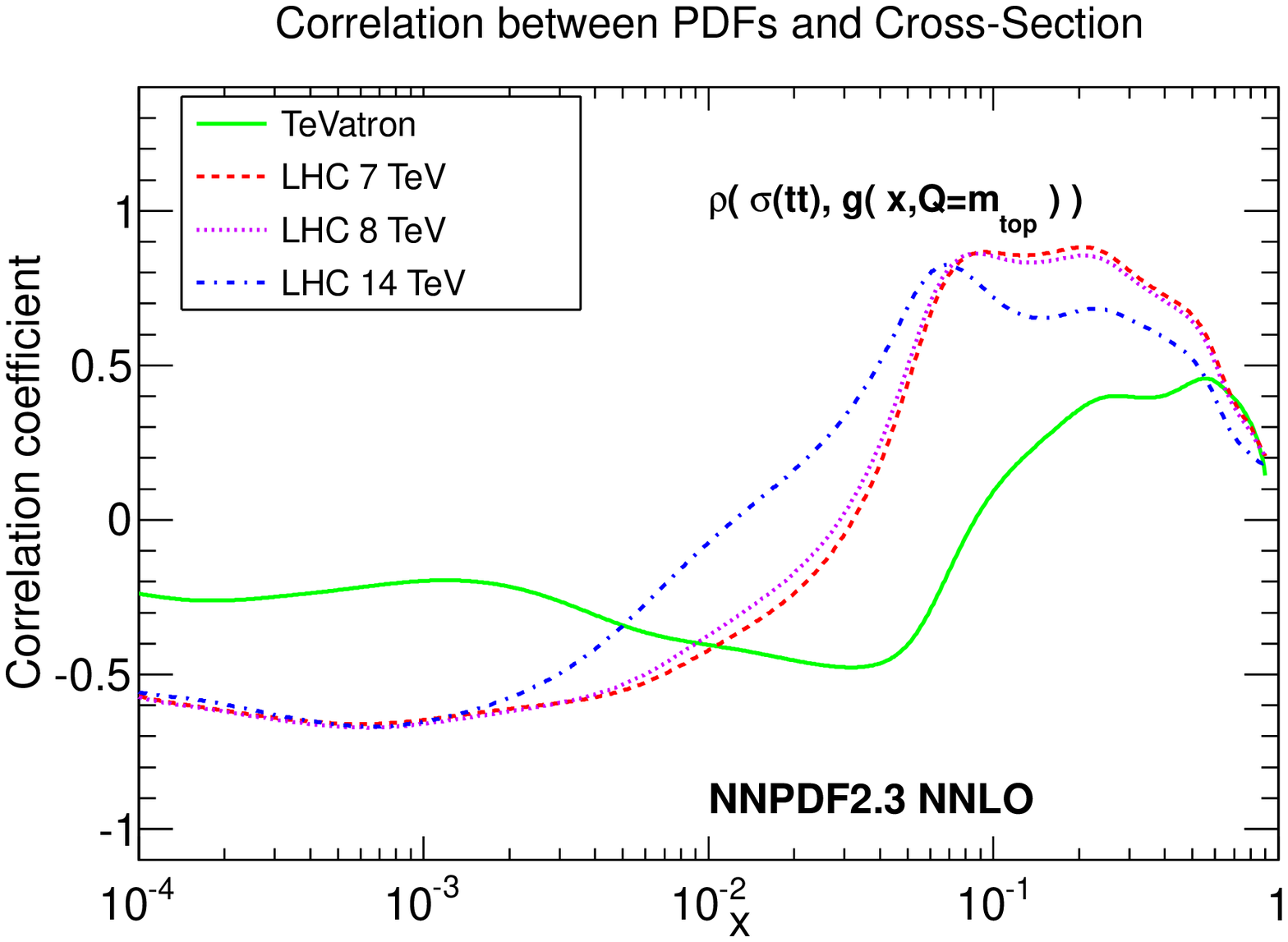}
\includegraphics[width=0.49\textwidth]{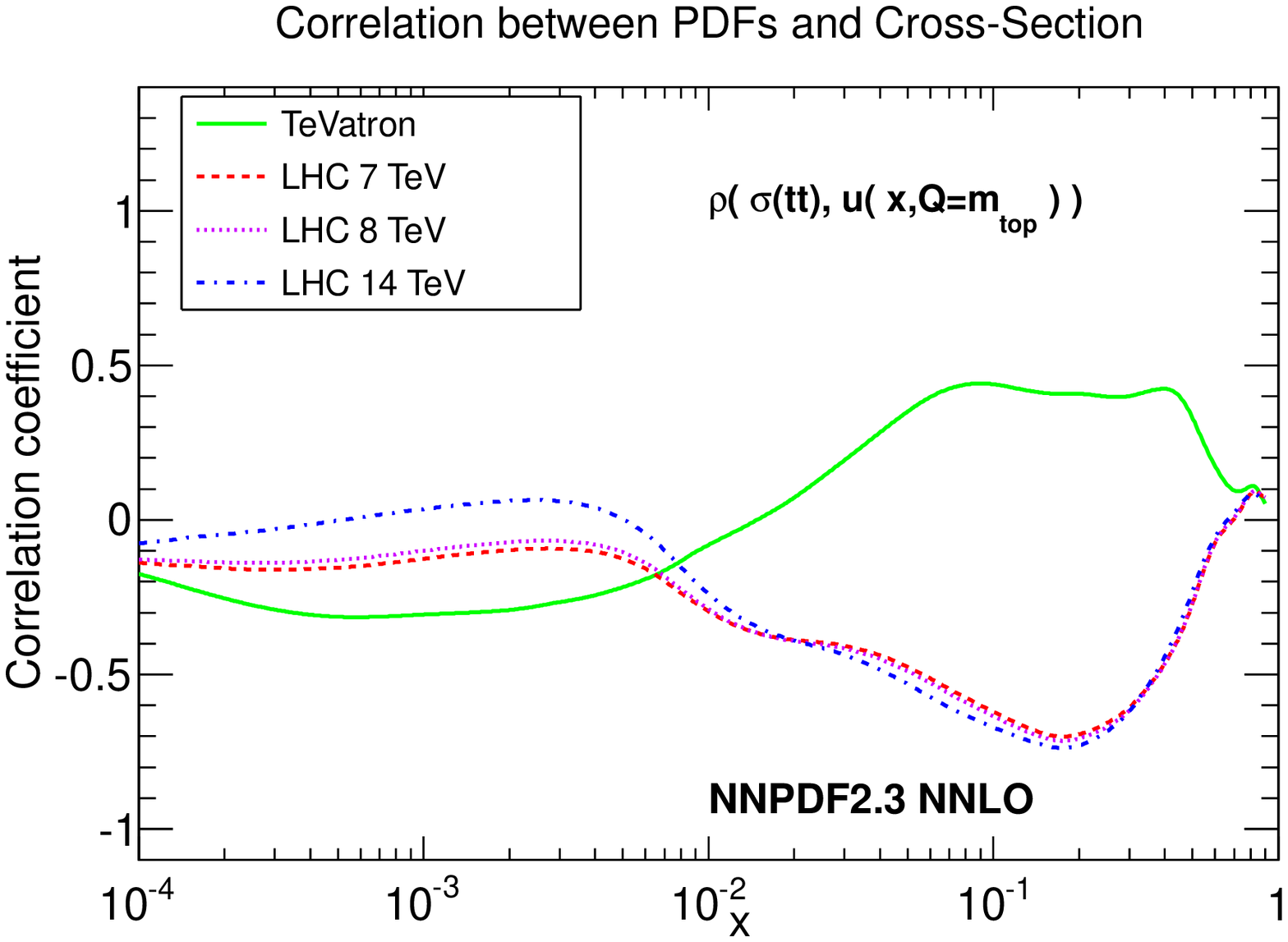}
\caption{\small The correlation between the gluon PDF (left plot)
and the up quark PDF (right plot) from NNPDF2.3 with the total
NNLO+NNLL top quark production cross sections at the
Tevatron and the LHC for different center of mass energies. 
The correlations are computed
for $Q=m_t=173.3$ GeV. }
\label{fig:pdf-xsec-corr}
\end{figure}

To illustrate the range of Bjorken-$x$'s to which the top cross
section is sensitive, the correlation~\cite{Ball:2010de} between the
top quark production cross section and the gluon and the up quark PDFs
is shown in Fig.~\ref{fig:pdf-xsec-corr} for the various cases that we
will discuss in the paper: Tevatron Run II, LHC 7, 8 and 14 TeV.
 A
correlation whose absolute magnitude is close to 1 indicates that
variations of PDFs with a particular value of $x$ will in turn
translate into cross-section variations.  
It is clear from
Fig.~\ref{fig:pdf-xsec-corr} that for the LHC the top quark cross
section directly probes the gluon in the range of $x$ between $x=0.1$
and $x=0.5$, where gluon PDF uncertainties are relatively large.

The outline of this paper is as follows. In Sect.~\ref{sec:settings}
we discuss the settings of the calculation and the treatment
of the various theoretical uncertainties. 
In Sect.~\ref{sec:results}
we provide up-to-date predictions for the $t\bar{t}$ cross section
at the Tevatron and LHC and compare with the most recent
experimental data. 
In Sect.~\ref{sec:rw} we quantify the
impact of the available top data on the gluon PDF, show
how it reduces the gluon PDF's large-$x$ uncertainties,
and study the phenomenological consequences of this
improvement. 
In Sect.~\ref{sec:ratios}
we provide up-to-date predictions for the ratio of top
quark cross sections between 7, 8 and 14 TeV and
in Sect.~\ref{sec:highmass} we provide
predictions for a heavy top-like fermion $T$.

\section{Settings} 
\label{sec:settings}

In the following we present the settings of the
computation. We use the program {\tt Top++}, {\tt v2.0}~\cite{Czakon:2011xx}
to evaluate the total top quark production cross section
at NNLO+NNLL accuracy.
We use as input the most up-to-date NNLO PDF sets from
each PDF collaboration: 
ABM11~\cite{Alekhin:2012ig}, CT10~\cite{Nadolsky:2012ia}, 
HERAPDF1.5~\cite{CooperSarkar:2011aa}, JR09~\cite{JimenezDelgado:2008hf} 
MSTW08~\cite{Martin:2009iq} and 
NNPDF2.3~\cite{Ball:2012cx}.
 The value of $\alpha_s(M_Z)$
is set to the preferred value of each group, respectively
0.1134, 0.118, 0.1176, 0.120, 0.1171 and 0.118. For NNPDF2.3 we use the
set with a maximum of $N_f=5$ active flavors.
When providing the predictions for each PDF set, we will use
the default $\alpha_s$ in each case, while
 later in Sect.~\ref{sec:asdep} we 
will explore the dependence of the results on the strong
 coupling as $\alpha_s$ is varied. 
A detailed benchmark comparison
of these five NNLO PDF sets was recently presented in~\cite{Ball:2012wy},
where the similarities and differences between each of the five
sets are discussed.

We consider the following sources of theoretical
uncertainties in the top quark pair production
cross section:

\begin{itemize}

\item Higher perturbative orders. 

The central scales of
the NNLO+NNLL computation are set to $\mu_R=\mu_F=m_t$. 
As customary, we explore the effect of missing higher perturbative orders
by varying the scales independently by a factor of two upwards and downwards,
with the constraint that the ratio of the two scales can never
be larger than two.
 The envelope of the resulting cross section
defines the scale uncertainty of the computation.
As shown in Ref.~\cite{Czakon:2013goa},  scale
variations at the LHC with the full NNLO+NNLL
result are substantially smaller that with the NLO
or previous approximated NNLO computations.

\item Parton Distributions. 

We use the corresponding prescription
from each group to provide the 68\% confidence level PDF uncertainties.
For the Hessian sets, we use
the asymmetric expression for PDF uncertainties~\cite{Nadolsky:2001yg}.
The CT10 errors are rescaled by a factor 1.642 since they
are provided as 90\% CL.  
In the case of HERAPDF1.5, we consider
only the experimental uncertainties, but not the model
and parametrization uncertainties.\footnote{If the HERAPDF1.5
uncertainties for the $t\bar{t}$ cross section include
the model and parametrization uncertainties the total
PDF uncertainty increases to about 20\%~\cite{Ball:2012wy}.}
Parton distributions are accessed through the LHAPDF
interface, version 5.8.9~\cite{lhapdf}.

\item Strong coupling constant.

 We assume the 68\% CL
uncertainty on $\alpha_s(M_Z)$ to be $\Delta \alpha_s=0.0007$,
as indicated by the 2012 PDG average~\cite{Beringer:1900zz}. For each PDF set,
we add in quadrature PDF and $\alpha_s$ uncertainties,
except for ABM11 and JR09, where $\alpha_s$ errors are already
part of the total PDF uncertainty, and thus we avoid double counting.
For Hessian PDF sets the addition in quadrature of PDF
and $\alpha_s$ uncertainties is known to be exact~\cite{Lai:2010nw},
while for Monte Carlo sets it 
is a good approximation~\cite{Demartin:2010er}.
The slope of the cross section with $\alpha_s$ for each PDF set
is extracted from a linear fit in a common range of
$\alpha_s(M_Z)$ between 0.116 and 0.120, close to the
PDG average and where all PDF groups provide predictions.

\item Top quark mass.

We take as central value $m_t=173.3$ GeV, with an uncertainty of
$\Delta m_t=\pm 1 $ GeV. 
 This is consistent both with the latest
Tevatron average,\footnote{See
\cite{Aaltonen:2012ra}, and the update prepared for the 2013 Rencontres de Moriond
QCD, presented by G.~Petrillo.}  $m_t=173.20\pm 0.87$ GeV, as
well as with the 2012 PDG average \cite{Beringer:1900zz} of the top
quark mass $m_t=173.5\pm 0.6 \pm 0.8$ GeV. 
These masses are to be
understood as pole masses. To determine the top-mass related
uncertainty $\delta_{\rm m_t}$ of the top cross section, we compute
the central value of the cross-section for top masses in a 1 GeV
range around the central value $m_t=173.3$.

\end{itemize}

Following the recommendations of the Higgs Cross Section
Working Group~\cite{Dittmaier:2011ti} regarding theoretical uncertainties, 
we adopt the most conservative method to combine
them: we add all parametric uncertainties
(PDFs, $\alpha_s$ and $m_t$) in quadrature and then
the total parametric error is added linearly to the
scale uncertainty to define the total theoretical
uncertainty.

\section{The top cross-section: measurements, predictions and uncertainties}
\label{sec:results}

With the settings detailed in the previous section, 
we have computed the cross sections for the Tevatron and the LHC
 and their associated uncertainties. 
The most precise experimental
measurements of the top cross section are collected in Table~\ref{tab:sigma-tot-exp}, while the theoretical predictions
 are collected in Tables~\ref{tab:res-tev} to~\ref{tab:res-lhc14}. 
In each case,
we provide the NNLO+NNLL predictions for the total
top quark cross section, for
all the PDFs considered, and with the various
sources of theoretical uncertainties. 
The default
value of $\alpha_s(M_Z)$ from each PDF set
has been used in the computation of the central
predictions. Also included in these tables are the best available
experimental measurements at the various colliders.

\begin{table}[t]
\centering
\footnotesize
\begin{tabular}{c|c|c|c|c|c} \hline
Measurement & $\sigma_{t\bar t}$ (pb) & stat. (pb) & sys. (pb) & lumi. (pb) & total (pb) \\  [2ex] 
\hline \hline 
Tevatron CDF+D0 (Ref.~\cite{tevsigma}) & $7.65$ & $\pm~0.20$ & $\pm~ 0.29$ & $\pm~ 0.22$ & $7.65 \pm~ 0.42~(5.5\%)$ \\  [1ex] 
Atlas 7 TeV (Ref.~\cite{ATLAS:2012fja}) & $177$ & $\pm~ 3$ & $ ^{+8}_{-7} $ & $\pm~ 7$ &  $177^{+10}_{-11}~^{(+5.6\%)}_{(-6.2\%)}$ \\ [1ex]
CMS 7 TeV (Ref.~\cite{Chatrchyan:2012bra}) & $160.9$ & $\pm~ 2.5$ & $ ^{+5.1}_{-5.0} $ & $\pm~ 3.6$ & $160.9 \pm~ 6.6~(4.0\%)$ \\ [1ex] 
Atlas 8 TeV (Ref.~\cite{ATLAS-CONF-2012-149}) & $241$ & $\pm~ 2$ & $\pm~ 31$ & $\pm ~9$ & $241 \pm~ 32~(13.0\%)$ \\  [1ex] 
CMS 8 TeV (Refs.~\cite{CMS-PAS-TOP-12-007,CMS-PAS-TOP-12-006}) & $227$ & $\pm~ 3$ & $\pm~ 11$ & $\pm~ 10$ & $227 \pm~ 15~(6.7\%)$ \\  [1ex]
\hline
\end{tabular}
\caption{\small The experimental measurements from the Tevatron and LHC 7 and 8 TeV used in this paper. 
The statistical, systematic and
luminosity uncertainties are added in quadrature for the total
uncertainty in the last column.
See the text for more details.}
\label{tab:sigma-tot-exp}
\end{table}

For the Tevatron, we have used the
most up-to-date CDF and D0 combinations~\cite{tevsigma}
with 8.8 fb$^{-1}$ of data, see Table~\ref{tab:sigma-tot-exp}.
For LHC 7 TeV, we quote
separately the two most precise measurements from ATLAS and CMS.
From ATLAS we use the combination of all channels with luminosities
between 0.70 and 1.02 fb$^{-1}$~\cite{ATLAS:2012fja}, see Table~\ref{tab:sigma-tot-exp},
while for CMS we use the single most precise measurement\footnote{In this case, given that the dependence of the measured total cross section with 
$m_t$ (which enters through the acceptance correction) is provided, we have rescaled the central value
of the data to the same
value of the top quark mass used for the theory computations,
$m_t=173.3$ GeV. Other cross-section measurements do not
provide this information but rather include the $\delta_{m_t}$ uncertainty
in the computation of the acceptances as part of the total systematic
uncertainty. }, 
obtained in the dilepton channel with a luminosity of
2.03 fb$^{-1}$~\cite{Chatrchyan:2012bra}, also given in Table~\ref{tab:sigma-tot-exp}.
Note also that a preliminary combination of all
 channels between ATLAS and CMS for an integrated luminosity of
1.0 fb$^{-1}$  has been presented in~\cite{CMS-PAS-TOP-12-003}, yielding
$\sigma(t\bar{t}) =173.3 \pm 10.1$~pb.
For LHC 8 TeV, there are three available measurements,
two from CMS, in the dilepton channel, with 2.4 fb$^{-1}$~\cite{CMS-PAS-TOP-12-007} and in the lepton+jets channel with  2.8 fb$^{-1}$~\cite{CMS-PAS-TOP-12-006}, and one from ATLAS in the lepton+jets channel with, 5.8 fb$^{-1}$~\cite{ATLAS-CONF-2012-149}. We use the CMS dilepton and the
ATLAS lepton+jets results, collected in Table~\ref{tab:sigma-tot-exp}.

\begin{figure}
\centering
\includegraphics[width=0.49\textwidth]{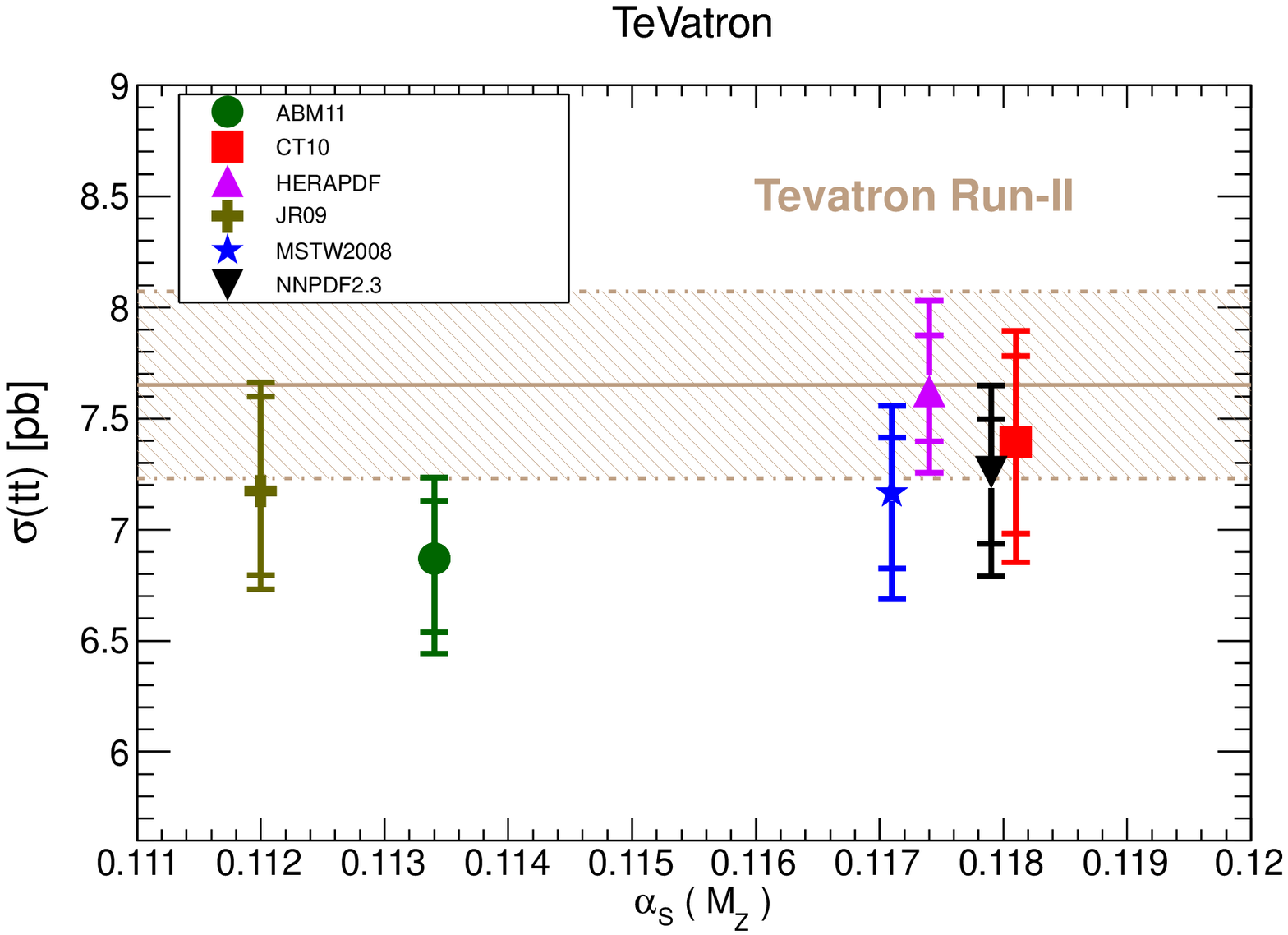}
\includegraphics[width=0.49\textwidth]{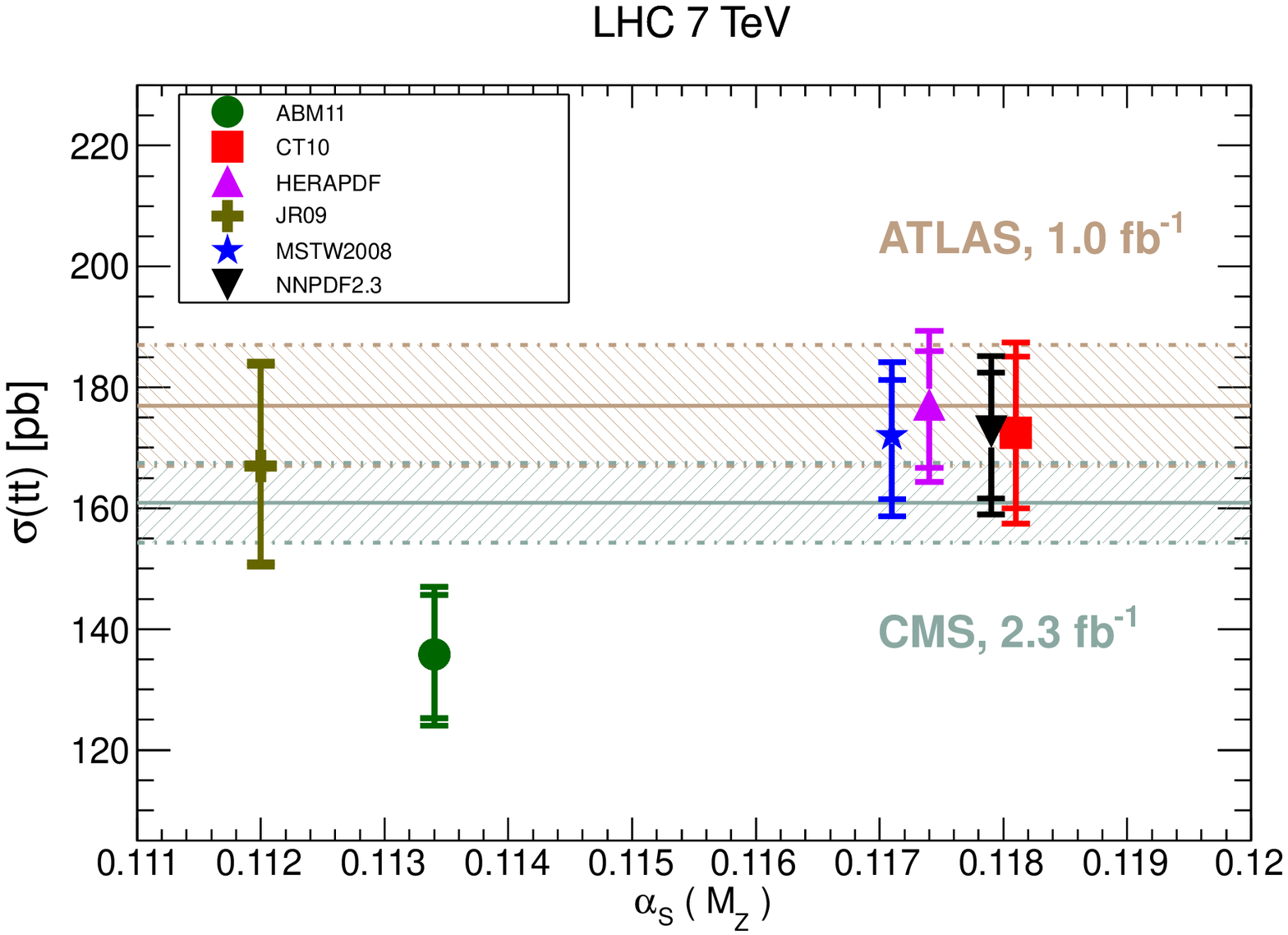}
\includegraphics[width=0.49\textwidth]{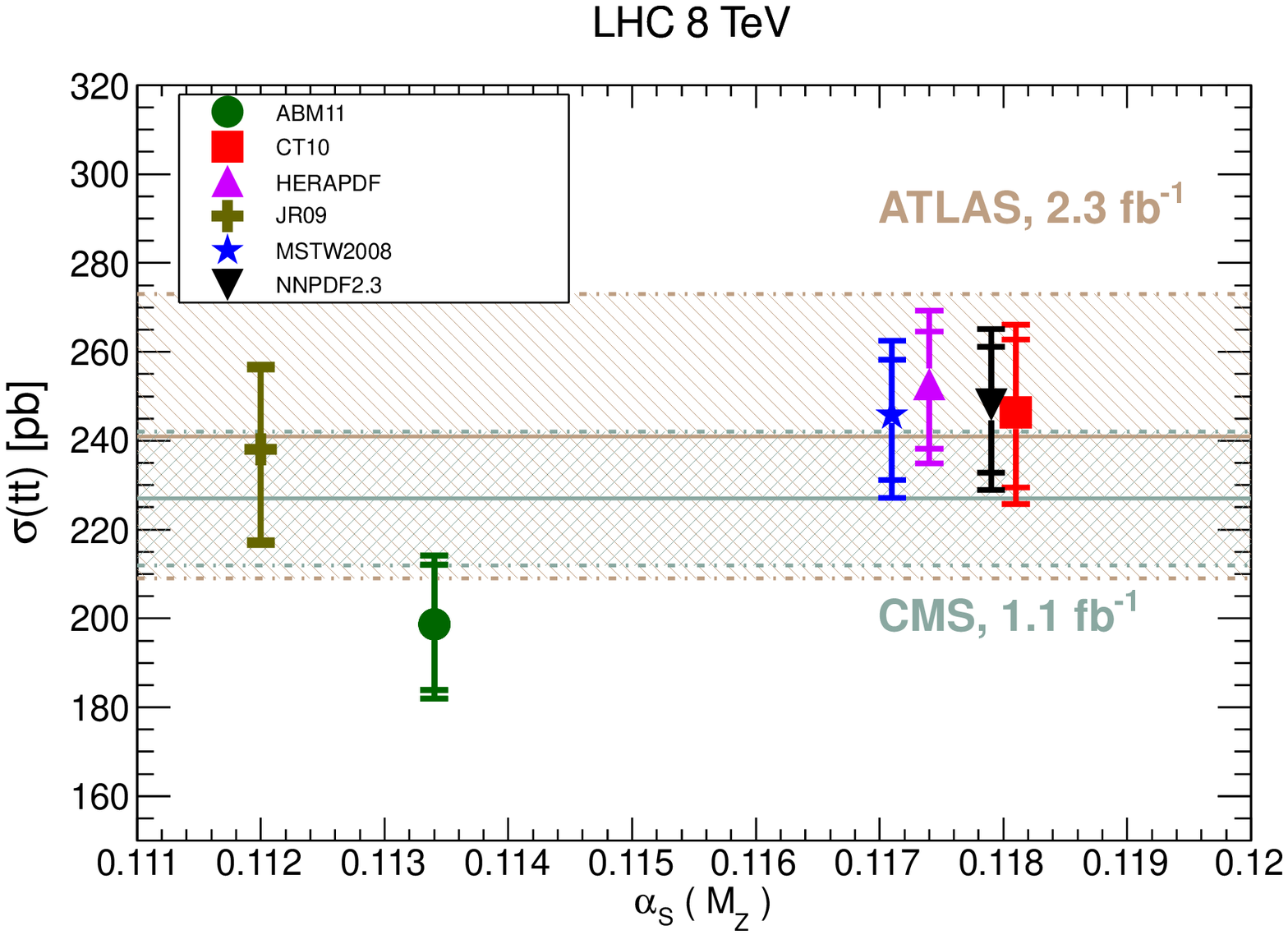}
\includegraphics[width=0.49\textwidth]{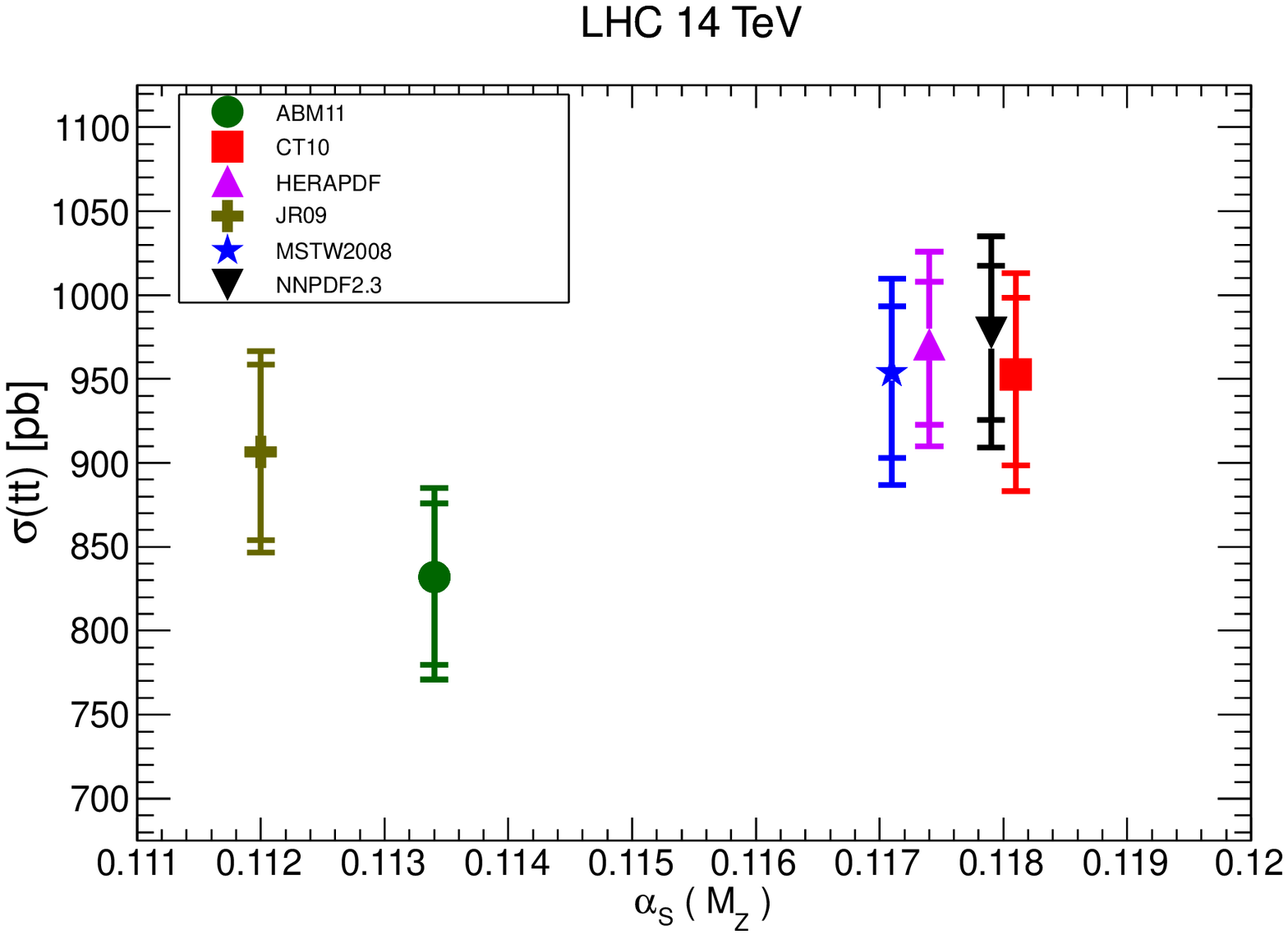}
\caption{\small The best predictions from each PDF set compared
to experimental data, as a function of the default
$\alpha_s(M_Z)$ value.
The inner 
error bar corresponds to the linear sum of
PDF and scale uncertainties, while the outer error
bar is the total theoretical uncertainty, computed as
described in the text.}
\label{fig:dataplots}
\end{figure}

In Fig.~\ref{fig:dataplots}
we show the predictions from each PDF set compared
to experimental data.
 The $x$-axis is the reference value
of $\alpha_s$ for each PDF set. The inner error bars correspond
to the linear sum of scale and PDF uncertainties, while the outer error
bar is the total theoretical uncertainty including also
the contributions from $\delta m_t$ and $\delta \alpha_s$.\footnote{The
inner error bar can thus be compared directly to previous
phenomenological studies, which consider only PDF and scale
uncertainties.}
Similar comparisons of the PDF and
$\alpha_s$ sensitivity of $\sigma(t\bar{t})$,
 based on previous approximate NNLO calculations,
have been reported in Refs.~\cite{Watt:2011kp,Watt:2012np,Forte:2013wc}.

In the remainder of this section we comment on various features of the above results.

\subsection{Comparison between the various PDF sets}

We notice that the results for NNPDF2.3, CT10,
MSTW08 and HERAPDF1.5 are all close
to each other, both in central values and 
theoretical uncertainties.
JR09 leads to similar central values, despite the smaller
value of $\alpha_s$, but is affected by larger
PDF+$\alpha_s$ uncertainties.
 On the other hand,
the ABM11 predictions are lower compared to other
sets and to experimental data.
 Fig.~\ref{fig:pdf-xsec-corr} suggests
that similarities or differences in the total cross section can be
understood in terms of the large-$x$ behavior of the gluon PDF, together
with the value of $\alpha_s$ used by each PDF set.

The agreement between CT, HERAPDF, MSTW and NNPDF can be
traced back to  (a) a similar default value of
$\alpha_s$ used and (b) a similar large-$x$ gluon PDF~\cite{Ball:2012wy}. 
On the other hand, regarding the differences
between ABM11 and the other sets, we note (a) the smaller value of
$\alpha_s$ used by ABM11 and (b) the softer large-$x$ gluon PDF in
the region relevant for top quark production~\cite{Ball:2012wy}. 
As shown below in
Fig.~\ref{fig:asdependence}, using a value for $\alpha_s$ closer to
the PDG average would improve the agreement of the ABM11 set both with
other PDF sets and with experimental data, though the ABM11 predictions at
the LHC are still lower
than that of other sets even for a common $\alpha_s$ value.

To understand the differences in the large-$x$ gluon in ABM11
as compared to other sets, 
Refs.~\cite{Thorne:2012az,Ball:2013gsa} have suggested that
the reason is that, keeping everything else
fixed, the use of a fixed-flavor
number scheme in a fit to deep-inelastic data (such as ABM11) leads to a softer
large-$x$ gluon as compared to the gluon obtained in PDF fits
based on variable-flavor number schemes~\cite{Forte:2010ta,thornehq,Guzzi:2011ew}.  
On the other hand, the  differences between the various implementations of
 variable-flavor number
schemes translate into much reduced differences in the PDFs~\cite{Thorne:2012az} and thus
into the top cross sections.

\begin{table}[t]
\centering
\footnotesize
 \begin{tabular}{c|c|c|c|c|c|c} \hline
\multicolumn{7}{c}{Tevatron Run II} \\ [1ex]
\hline
 PDF set & $\sigma_{tt}$ (pb) & $\delta_{\rm scale}$ (pb) & $\delta_{\rm PDF}$ (pb) & $\delta_{\alpha_s}$ (pb) & $\delta_{\rm m_t}$ (pb) & $\delta_{\rm tot}$ (pb) \\  [2ex] 
 \hline \hline 
ABM11                                    &    6.869 & $~^{+   0.104}_{-   0.174}$$~^{(+     1.5\%)}_{(-     2.5\%)}$  & $~^{+   0.157}_{-   0.157}$$~^{(+     2.3\%)}_{(-     2.3\%)}$  & $~^{+   0.000}_{-   0.000}$$~^{(+     0.0\%)}_{(-     0.0\%)}$  & $~^{+   0.207}_{-   0.201}$$~^{(+     3.0\%)}_{(-     2.9\%)}$  & 
$~^{+   0.364}_{-   0.429}$$~^{(+     5.3\%)}_{(-     6.2\%)}$  \\ [2ex] 
CT10                                     &    7.395 & $~^{+   0.116}_{-   0.210}$$~^{(+     1.6\%)}_{(-     2.8\%)}$  & $~^{+   0.270}_{-   0.203}$$~^{(+     3.6\%)}_{(-     2.7\%)}$  & $~^{+   0.136}_{-   0.136}$$~^{(+     1.8\%)}_{(-     1.8\%)}$  & $~^{+   0.235}_{-   0.227}$$~^{(+     3.2\%)}_{(-     3.1\%)}$  & 
$~^{+   0.499}_{-   0.544}$$~^{(+     6.7\%)}_{(-     7.4\%)}$  \\ [2ex] 
HERA1.5                                  &    7.624 & $~^{+   0.116}_{-   0.074}$$~^{(+     1.5\%)}_{(-     1.0\%)}$  & $~^{+   0.134}_{-   0.154}$$~^{(+     1.8\%)}_{(-     2.0\%)}$  & $~^{+   0.098}_{-   0.098}$$~^{(+     1.3\%)}_{(-     1.3\%)}$  & $~^{+   0.241}_{-   0.233}$$~^{(+     3.2\%)}_{(-     3.1\%)}$  & 
$~^{+   0.409}_{-   0.370}$$~^{(+     5.4\%)}_{(-     4.9\%)}$  \\ [2ex] 
JR09 &    7.174 & $~^{+   0.099}_{-   0.054}$$~^{(+     1.4\%)}_{(-     0.8\%)}$  & $~^{+   0.326}_{-   0.326}$$~^{(+     4.6\%)}_{(-     4.6\%)}$  & $~^{+   0.000}_{-   0.000}$$~^{(+     0.0\%)}_{(-     0.0\%)}$  & $~^{+   0.215}_{-   0.211}$$~^{(+     3.0\%)}_{(-     2.9\%)}$  & 
$~^{+   0.490}_{-   0.443}$$~^{(+     6.8\%)}_{(-     6.2\%)}$  \\ [2ex] 
MSTW08                                   &    7.164 & $~^{+   0.110}_{-   0.200}$$~^{(+     1.5\%)}_{(-     2.8\%)}$  & $~^{+   0.169}_{-   0.122}$$~^{(+     2.4\%)}_{(-     1.7\%)}$  & $~^{+   0.088}_{-   0.088}$$~^{(+     1.2\%)}_{(-     1.2\%)}$  & $~^{+   0.228}_{-   0.220}$$~^{(+     3.2\%)}_{(-     3.1\%)}$  & 
$~^{+   0.391}_{-   0.475}$$~^{(+     5.5\%)}_{(-     6.6\%)}$  \\ [2ex] 
NNPDF2.3                                 &    7.258 & $~^{+   0.117}_{-   0.202}$$~^{(+     1.6\%)}_{(-     2.8\%)}$  & $~^{+   0.121}_{-   0.121}$$~^{(+     1.7\%)}_{(-     1.7\%)}$  & $~^{+   0.090}_{-   0.090}$$~^{(+     1.2\%)}_{(-     1.2\%)}$  & $~^{+   0.229}_{-   0.221}$$~^{(+     3.1\%)}_{(-     3.0\%)}$  & 
$~^{+   0.390}_{-   0.469}$$~^{(+     5.4\%)}_{(-     6.5\%)}$  \\ [2ex] 
 \hline
CDF+D0 &    7.65 & & & & & $\pm$    0.42~(     5.5\%) \\ [1ex] 
 \hline
 \end{tabular}

\caption{\small The NNLO+NNLL predictions for the total
top quark pair cross-section at the Tevatron Run II, for
all the PDFs considered, and with the various
sources of theoretical uncertainties. 
The default
value of $\alpha_s(M_Z)$ from each collaboration
has been used in the computation of the central
predictions. 
The four theoretical uncertainties
are combined into a total theory error as discussed
in the text. The lower row shows the best
available experimental measurement. \label{tab:res-tev} }
\end{table}

\begin{table}[t]
\footnotesize
 \begin{tabular}{c|c|c|c|c|c|c} 
\hline
\multicolumn{7}{c}{LHC 7 TeV} \\ [1ex]
\hline
 PDF set & $\sigma_{tt}$ (pb) & $\delta_{\rm scale}$ (pb) & $\delta_{\rm PDF}$ (pb) & $\delta_{\alpha_s}$ (pb) & $\delta_{\rm m_t}$ (pb) & $\delta_{\rm tot}$ (pb) \\  [2ex] 
 \hline \hline 
ABM11                                    &    135.8 & $~^{+     3.5}_{-     4.2}$$~^{(+     2.6\%)}_{(-     3.1\%)}$  & $~^{+     6.4}_{-     6.4}$$~^{(+     4.7\%)}_{(-     4.7\%)}$  & $~^{+     0.0}_{-     0.0}$$~^{(+     0.0\%)}_{(-     0.0\%)}$  & $~^{+     4.3}_{-     4.2}$$~^{(+     3.2\%)}_{(-     3.1\%)}$  & 
$~^{+    11.2}_{-    11.8}$$~^{(+     8.2\%)}_{(-     8.7\%)}$  \\ [2ex] 
CT10                                     &    172.5 & $~^{+     4.6}_{-     6.0}$$~^{(+     2.7\%)}_{(-     3.5\%)}$  & $~^{+     8.0}_{-     6.5}$$~^{(+     4.6\%)}_{(-     3.8\%)}$  & $~^{+     3.7}_{-     3.7}$$~^{(+     2.2\%)}_{(-     2.2\%)}$  & $~^{+     5.3}_{-     5.1}$$~^{(+     3.1\%)}_{(-     3.0\%)}$  & 
$~^{+    14.9}_{-    15.0}$$~^{(+     8.6\%)}_{(-     8.7\%)}$  \\ [2ex] 
HERA1.5                                  &    177.2 & $~^{+     4.8}_{-     4.2}$$~^{(+     2.7\%)}_{(-     2.3\%)}$  & $~^{+     4.0}_{-     6.4}$$~^{(+     2.3\%)}_{(-     3.6\%)}$  & $~^{+     3.0}_{-     3.0}$$~^{(+     1.7\%)}_{(-     1.7\%)}$  & $~^{+     5.4}_{-     5.2}$$~^{(+     3.1\%)}_{(-     2.9\%)}$  & 
$~^{+    12.2}_{-    12.9}$$~^{(+     6.9\%)}_{(-     7.3\%)}$  \\ [2ex] 
JR09 &    167.0 & $~^{+     3.9}_{-     3.3}$$~^{(+     2.4\%)}_{(-     2.0\%)}$  & $~^{+    12.6}_{-    12.6}$$~^{(+     7.6\%)}_{(-     7.6\%)}$  & $~^{+     0.0}_{-     0.0}$$~^{(+     0.0\%)}_{(-     0.0\%)}$  & $~^{+     4.5}_{-     4.3}$$~^{(+     2.7\%)}_{(-     2.6\%)}$  & 
$~^{+    17.3}_{-    16.6}$$~^{(+    10.4\%)}_{(-     9.9\%)}$  \\ [2ex] 
MSTW08                                   &    172.0 & $~^{+     4.4}_{-     5.8}$$~^{(+     2.6\%)}_{(-     3.4\%)}$  & $~^{+     4.7}_{-     4.7}$$~^{(+     2.7\%)}_{(-     2.7\%)}$  & $~^{+     2.9}_{-     2.9}$$~^{(+     1.7\%)}_{(-     1.7\%)}$  & $~^{+     5.3}_{-     5.1}$$~^{(+     3.1\%)}_{(-     3.0\%)}$  & 
$~^{+    12.1}_{-    13.4}$$~^{(+     7.0\%)}_{(-     7.8\%)}$  \\ [2ex] 
NNPDF2.3                                 &    172.7 & $~^{+     4.6}_{-     6.0}$$~^{(+     2.7\%)}_{(-     3.5\%)}$  & $~^{+     5.2}_{-     5.2}$$~^{(+     3.0\%)}_{(-     3.0\%)}$  & $~^{+     2.7}_{-     2.7}$$~^{(+     1.6\%)}_{(-     1.6\%)}$  & $~^{+     5.3}_{-     5.2}$$~^{(+     3.1\%)}_{(-     3.0\%)}$  & 
$~^{+    12.5}_{-    13.7}$$~^{(+     7.2\%)}_{(-     8.0\%)}$  \\ [2ex] 
 \hline
ATLAS &    177 & & & & &      $~^{+10}_{-11}~^{(+5.6\%)}_{(-6.2\%)}$ \\ [1ex] 
CMS &    160.9&&&& &$\pm$       6.6~(     4.0\%) \\ [1ex] 
 \hline
 \end{tabular}

\caption{\small Same as Table~\ref{tab:res-tev}
for LHC 7 TeV.~\label{tab:res-lhc7}
}
\end{table}

\begin{table}[t]
\footnotesize
 \begin{tabular}{c|c|c|c|c|c|c} 
\hline
\multicolumn{7}{c}{LHC 8 TeV} \\ [1ex]
\hline
 PDF set & $\sigma_{tt}$ (pb) & $\delta_{\rm scale}$ (pb) & $\delta_{\rm PDF}$ (pb) & $\delta_{\alpha_s}$ (pb) & $\delta_{\rm m_t}$ (pb) & $\delta_{\rm tot}$ (pb) \\  [2ex] 
 \hline \hline 
ABM11                                    &    198.6 & $~^{+     5.0}_{-     6.2}$$~^{(+     2.5\%)}_{(-     3.1\%)}$  & $~^{+     8.5}_{-     8.5}$$~^{(+     4.3\%)}_{(-     4.3\%)}$  & $~^{+     0.0}_{-     0.0}$$~^{(+     0.0\%)}_{(-     0.0\%)}$  & $~^{+     6.1}_{-     5.9}$$~^{(+     3.1\%)}_{(-     3.0\%)}$  & 
$~^{+    15.5}_{-    16.6}$$~^{(+     7.8\%)}_{(-     8.4\%)}$  \\ [2ex] 
CT10                                     &    246.3 & $~^{+     6.4}_{-     8.6}$$~^{(+     2.6\%)}_{(-     3.5\%)}$  & $~^{+    10.1}_{-     8.2}$$~^{(+     4.1\%)}_{(-     3.3\%)}$  & $~^{+     4.9}_{-     4.9}$$~^{(+     2.0\%)}_{(-     2.0\%)}$  & $~^{+     7.4}_{-     7.1}$$~^{(+     3.0\%)}_{(-     2.9\%)}$  & 
$~^{+    19.8}_{-    20.5}$$~^{(+     8.1\%)}_{(-     8.3\%)}$  \\ [2ex] 
HERA1.5                                  &    252.7 & $~^{+     6.5}_{-     5.9}$$~^{(+     2.6\%)}_{(-     2.3\%)}$  & $~^{+     5.4}_{-     8.6}$$~^{(+     2.1\%)}_{(-     3.4\%)}$  & $~^{+     4.0}_{-     4.0}$$~^{(+     1.6\%)}_{(-     1.6\%)}$  & $~^{+     7.5}_{-     7.3}$$~^{(+     3.0\%)}_{(-     2.9\%)}$  & 
$~^{+    16.6}_{-    17.8}$$~^{(+     6.6\%)}_{(-     7.1\%)}$  \\ [2ex] 
JR09 &    238.0 & $~^{+     2.1}_{-     4.6}$$~^{(+     0.9\%)}_{(-     1.9\%)}$  & $~^{+    15.8}_{-    15.8}$$~^{(+     6.6\%)}_{(-     6.6\%)}$  & $~^{+     0.0}_{-     0.0}$$~^{(+     0.0\%)}_{(-     0.0\%)}$  & $~^{+     6.3}_{-     6.2}$$~^{(+     2.7\%)}_{(-     2.6\%)}$  & 
$~^{+    19.2}_{-    21.6}$$~^{(+     8.1\%)}_{(-     9.1\%)}$  \\ [2ex] 
MSTW08                                   &    245.8 & $~^{+     6.2}_{-     8.4}$$~^{(+     2.5\%)}_{(-     3.4\%)}$  & $~^{+     6.2}_{-     6.2}$$~^{(+     2.5\%)}_{(-     2.5\%)}$  & $~^{+     4.0}_{-     4.0}$$~^{(+     1.6\%)}_{(-     1.6\%)}$  & $~^{+     7.4}_{-     7.1}$$~^{(+     3.0\%)}_{(-     2.9\%)}$  & 
$~^{+    16.6}_{-    18.7}$$~^{(+     6.8\%)}_{(-     7.6\%)}$  \\ [2ex] 
NNPDF2.3                                 &    248.1 & $~^{+     6.4}_{-     8.7}$$~^{(+     2.6\%)}_{(-     3.5\%)}$  & $~^{+     6.6}_{-     6.6}$$~^{(+     2.7\%)}_{(-     2.7\%)}$  & $~^{+     3.7}_{-     3.7}$$~^{(+     1.5\%)}_{(-     1.5\%)}$  & $~^{+     7.5}_{-     7.2}$$~^{(+     3.0\%)}_{(-     2.9\%)}$  & 
$~^{+    17.1}_{-    19.1}$$~^{(+     6.9\%)}_{(-     7.7\%)}$  \\ [2ex] 
 \hline
ATLAS &    241.0 & & & & &$\pm$      32.0~(    13.3\%) \\ [1ex] 
CMS &    227.0&&&& &$\pm$      15.0~(     6.6\%) \\ [1ex] 
 \hline
 \end{tabular}

\caption{\small
 Same as Table~\ref{tab:res-tev}
for LHC 8 TeV.~\label{tab:res-lhc8}
}
\end{table}

\begin{table}[t]
\footnotesize
 \begin{tabular}{c|c|c|c|c|c|c} 
\hline
\multicolumn{7}{c}{LHC 14 TeV} \\ [1ex]
\hline
 PDF set & $\sigma_{tt}$ (pb) & $\delta_{\rm scale}$ (pb) & $\delta_{\rm PDF}$ (pb) & $\delta_{\alpha_s}$ (pb) & $\delta_{\rm m_t}$ (pb) & $\delta_{\rm tot}$ (pb) \\  [2ex] 
 \hline \hline 
ABM11                                    &    832.0 & $~^{+    18.7}_{-    27.4}$$~^{(+     2.2\%)}_{(-     3.3\%)}$  & $~^{+    25.1}_{-    25.1}$$~^{(+     3.0\%)}_{(-     3.0\%)}$  & $~^{+     0.0}_{-     0.0}$$~^{(+     0.0\%)}_{(-     0.0\%)}$  & $~^{+    23.3}_{-    22.5}$$~^{(+     2.8\%)}_{(-     2.7\%)}$  & 
$~^{+    52.9}_{-    61.1}$$~^{(+     6.4\%)}_{(-     7.3\%)}$  \\ [2ex] 
CT10                                     &    952.8 & $~^{+    23.3}_{-    34.5}$$~^{(+     2.4\%)}_{(-     3.6\%)}$  & $~^{+    22.4}_{-    19.9}$$~^{(+     2.3\%)}_{(-     2.1\%)}$  & $~^{+    14.0}_{-    14.0}$$~^{(+     1.5\%)}_{(-     1.5\%)}$  & $~^{+    26.1}_{-    25.2}$$~^{(+     2.7\%)}_{(-     2.6\%)}$  & 
$~^{+    60.3}_{-    69.5}$$~^{(+     6.3\%)}_{(-     7.3\%)}$  \\ [2ex] 
HERA1.5                                  &    970.5 & $~^{+    22.1}_{-    22.0}$$~^{(+     2.3\%)}_{(-     2.3\%)}$  & $~^{+    15.3}_{-    25.7}$$~^{(+     1.6\%)}_{(-     2.6\%)}$  & $~^{+    12.8}_{-    12.8}$$~^{(+     1.3\%)}_{(-     1.3\%)}$  & $~^{+    26.4}_{-    25.6}$$~^{(+     2.7\%)}_{(-     2.6\%)}$  & 
$~^{+    55.2}_{-    60.5}$$~^{(+     5.7\%)}_{(-     6.2\%)}$  \\ [2ex] 
JR09 &    906.5 & $~^{+    16.7}_{-    17.0}$$~^{(+     1.8\%)}_{(-     1.9\%)}$  & $~^{+    35.5}_{-    35.5}$$~^{(+     3.9\%)}_{(-     3.9\%)}$  & $~^{+     0.0}_{-     0.0}$$~^{(+     0.0\%)}_{(-     0.0\%)}$  & $~^{+    24.7}_{-    23.9}$$~^{(+     2.7\%)}_{(-     2.6\%)}$  & 
$~^{+    60.0}_{-    59.8}$$~^{(+     6.6\%)}_{(-     6.6\%)}$  \\ [2ex] 
MSTW08                                   &    953.6 & $~^{+    22.7}_{-    33.9}$$~^{(+     2.4\%)}_{(-     3.6\%)}$  & $~^{+    16.2}_{-    17.8}$$~^{(+     1.7\%)}_{(-     1.9\%)}$  & $~^{+    12.8}_{-    12.8}$$~^{(+     1.3\%)}_{(-     1.3\%)}$  & $~^{+    26.1}_{-    25.3}$$~^{(+     2.7\%)}_{(-     2.7\%)}$  & 
$~^{+    56.3}_{-    66.8}$$~^{(+     5.9\%)}_{(-     7.0\%)}$  \\ [2ex] 
NNPDF2.3                                 &    977.5 & $~^{+    23.6}_{-    35.4}$$~^{(+     2.4\%)}_{(-     3.6\%)}$  & $~^{+    16.4}_{-    16.4}$$~^{(+     1.7\%)}_{(-     1.7\%)}$  & $~^{+    12.2}_{-    12.2}$$~^{(+     1.3\%)}_{(-     1.3\%)}$  & $~^{+    26.9}_{-    26.1}$$~^{(+     2.8\%)}_{(-     2.7\%)}$  & 
$~^{+    57.4}_{-    68.5}$$~^{(+     5.9\%)}_{(-     7.0\%)}$  \\ [2ex] 
 \hline
 \hline
 \end{tabular}

\caption{\small
 Same as Table~\ref{tab:res-tev}
for LHC 14 TeV.~\label{tab:res-lhc14}
}
\end{table}

In order to quantify the agreement of data and theory, in the absence of theoretical uncertainties,
we compute the $\chi^2$ for the central values of the
five PDF sets under consideration and for the five available data points of Table~\ref{tab:sigma-tot-exp}: 
one for the Tevatron, two for 
LHC 7 TeV and two for LHC 8 TeV. For each of the five PDF sets studied, the $\chi^2$ is defined as
\begin{equation}
\label{eq:chi2}
\chi^2 = \sum_{i=1}^{N_{\rm dat}}\frac{\left( \sigma^{(\rm exp)}_{t\bar{t}} - \sigma^{(\rm th)}_{t\bar{t}} \right)^2}{\delta^{(\rm exp)2}_{\rm tot}} \, , 
\end{equation}
with $N_{\rm dat}=5$. We consider
all five data points as fully uncorrelated.\footnote{This is a necessary approximation, but as discussed in~\cite{CMS-PAS-TOP-12-003} there is
some degree of correlation between the ATLAS and CMS data, as there is between
LHC data at different c.m. energies.}
In addition, to quantify  the overall
consistency of theoretical predictions with experimental data,
we define a pull estimator as follows,
\begin{equation}
\label{eq:pull}
P = \frac{1}{N_{\rm dat}}\sum_{i=1}^{N_{\rm dat}}\frac{\left( \sigma^{(\rm exp)}_{t\bar{t}} - \sigma^{(\rm th)}_{t\bar{t}} \right)^2}{\delta^{(\rm exp)2}_{\rm tot}+\delta^{(\rm th)2}_{\rm tot}} \, , 
\end{equation}
where now in the denominator we add in quadrature the experimental and
theoretical total uncertainties.
For simplicity we have symmetrized the total theoretical systematic
error for each PDF set.

 The results for both the $\chi^2$ and the pull are summarized
in Table~\ref{tab:chi2}.
We provide both the total $\chi^2$ and the individual
contribution from the data at the Tevatron, LHC 7 and
8 TeV.
 As we can see, most PDF sets provide a good  description
of the total top quark cross section data, with $\chi^2/N_{\rm
  dat}\sim 1$.
The pull also shows the good consistency between experimental data
and theory prediction for most of the PDF sets considered.

\begin{table}[h]
\centering
\small
\begin{tabular}{c|c|c|c|c|c||c}
\hline
 &  $\chi^2_{\rm tev}$  &  $\chi^2_{\rm lhc7}$  &  $\chi^2_{\rm lhc8}$   & 
 $\chi^2_{\rm tot}$  & $\chi^2_{\rm tot}/N_{\rm dat}$  &  P\\
\hline
\hline
AMB11 & 3.5 & 31.4 & 5.3 & 40.2 &  8.0  &  3.2\\
CT10 & 0.4 & 3.3 & 1.7 & 5.3 & 1.1 & 0.3\\
HERAPDF15 & 0.0 & 6.1 & 3.1 & 9.2  & 1.8 & 0.5  \\
MSTW08 & 1.3 & 3.1 & 1.6 & 6.0  & 1.2 & 0.4\\
NNPDF2.3 & 0.9 & 3.4 & 2.0 &6.3 & 1.3 & 0.4\\
\hline
\end{tabular}
\caption{\small The $\chi^2$ between
data and theory, Eq.~(\ref{eq:chi2}), for
all the five PDF sets, both for the total
dataset and split into colliders.
Let us recall that the standard deviation of the $\chi^2$
distribution for $N_{\rm dat}=5$ data points is $\sqrt{2N_{\rm dat}}=3.1$.
The last column show the pull Eq.~(\ref{eq:pull}) between
experimental data and theory predictions.
\label{tab:chi2}}
\end{table}

\subsection{Uncertainty due to the value of $m_t$}

The uncertainty on the value of the top quark mass is now
a substantial fraction of the overall systematic uncertainty in the total
$t\bar{t}$ cross section.
 In Table~\ref{tab:nomt} we compare,
for the NNPDF2.3 set,  the total theory uncertainty with and without
including the uncertainty due to the top quark mass. 
We see that at the LHC, an
uncertainty of $\delta m_t =\pm 1$ GeV
translates into an increase between 1 and
1.5\% of the total theory uncertainty, and into
a somewhat larger increase at the Tevatron.
Given that PDFs,
scale and $m_t$ uncertainties are all now of similar size, the total
theory error would be 
only slightly reduced if one assumed that
the uncertainty due to $m_t$ could be completely neglected, for
example after more precise measurements of this parameter at the LHC.

\begin{table}[h]
\centering
\small
 \begin{tabular}{c|c|c|c} \hline
 Collider & $\sigma_{tt}$ (pb) & $\delta_{\rm PDF+scales+\alpha_s}$ (pb) & $\delta_{\rm tot}$ (pb)  \\ [2ex] 
 \hline \hline 
Tevatron             &    7.258 & $~^{+   0.267}_{-   0.352}$$~^{(+     3.7\%)}_{(-     4.9\%)}$  & $~^{+   0.390}_{-   0.469}$$~^{(+     5.4\%)}_{(-     6.5\%)}$  \\ [2ex] 
LHC 7 TeV            &    172.7 & $~^{+    10.4}_{-    11.8}$$~^{(+     6.0\%)}_{(-     6.8\%)}$  & $~^{+    12.5}_{-    13.7}$$~^{(+     7.2\%)}_{(-     8.0\%)}$  \\ [2ex] 
LHC 8 TeV            &    248.1 & $~^{+    14.0}_{-    16.2}$$~^{(+     5.6\%)}_{(-     6.5\%)}$  & $~^{+    17.1}_{-    19.1}$$~^{(+     6.9\%)}_{(-     7.7\%)}$  \\ [2ex] 
LHC 14 TeV           &    977.5 & $~^{+    44.1}_{-    55.8}$$~^{(+     4.5\%)}_{(-     5.7\%)}$  & $~^{+    57.4}_{-    68.5}$$~^{(+     5.9\%)}_{(-     7.0\%)}$  \\ [2ex] 
 \hline
 \end{tabular}

\caption{\small The NNLO+NNLL predictions for the total
top quark pair cross-section at the Tevatron 
and the LHC, with NNPDF2.3 as input.
The third column shows the theoretical
uncertainty when $\delta m_t$ is not taken into
account, while the last column is the total
theoretical uncertainty (the same as in 
Tables~\ref{tab:res-tev} to~\ref{tab:res-lhc14}).
 At the LHC, an
uncertainty of $\delta m_t =\pm 1$ GeV
translates into an increase between 1 and
1.5\% of the total theory uncertainty, and
a  bit more at the Tevatron.
 \label{tab:nomt} }
\end{table}

\subsection{Uncertainty due to the value of $\alpha_s$}
\label{sec:asdep}

\begin{figure}
\centering
\includegraphics[width=0.49\textwidth]{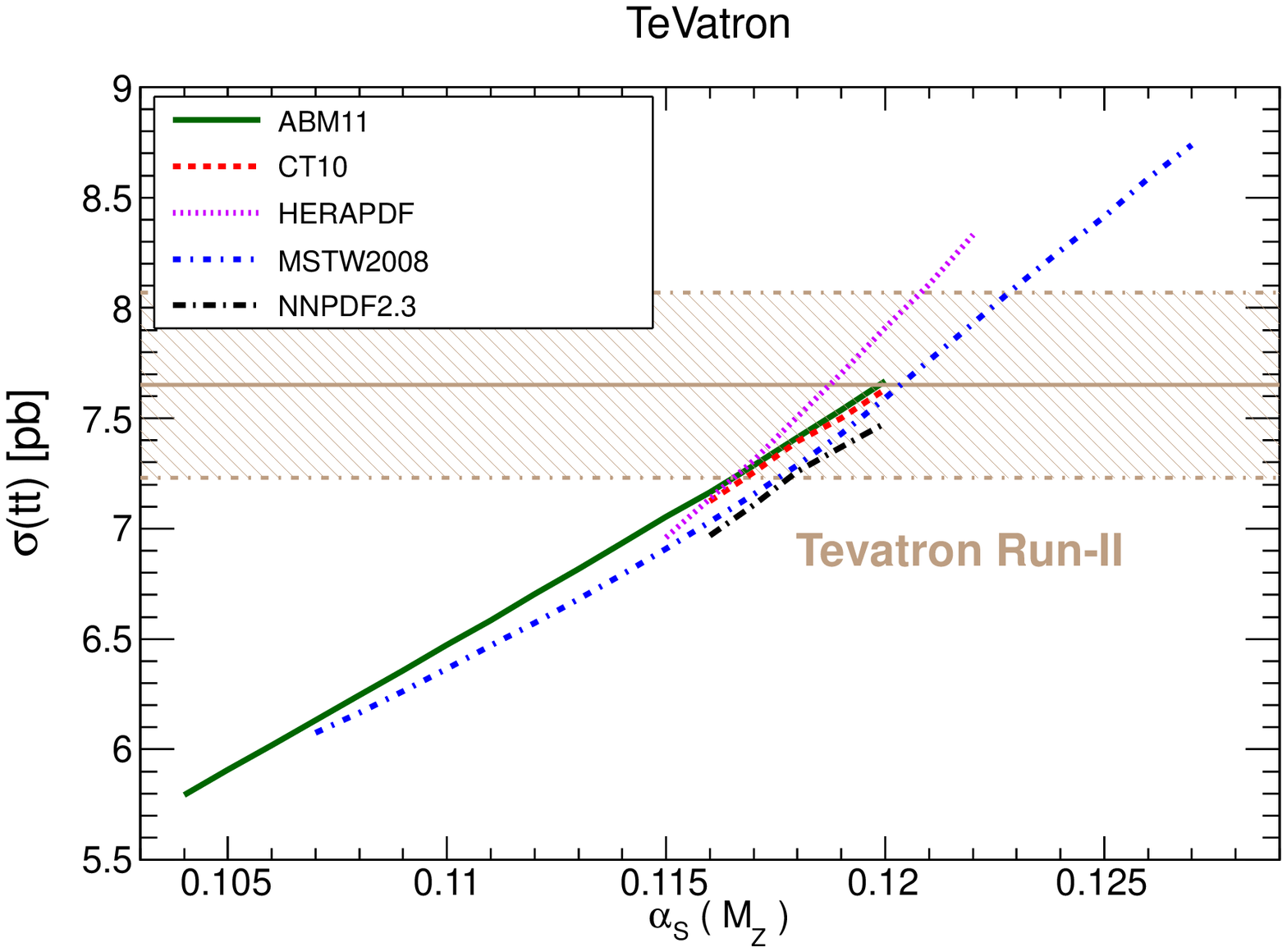}
\includegraphics[width=0.49\textwidth]{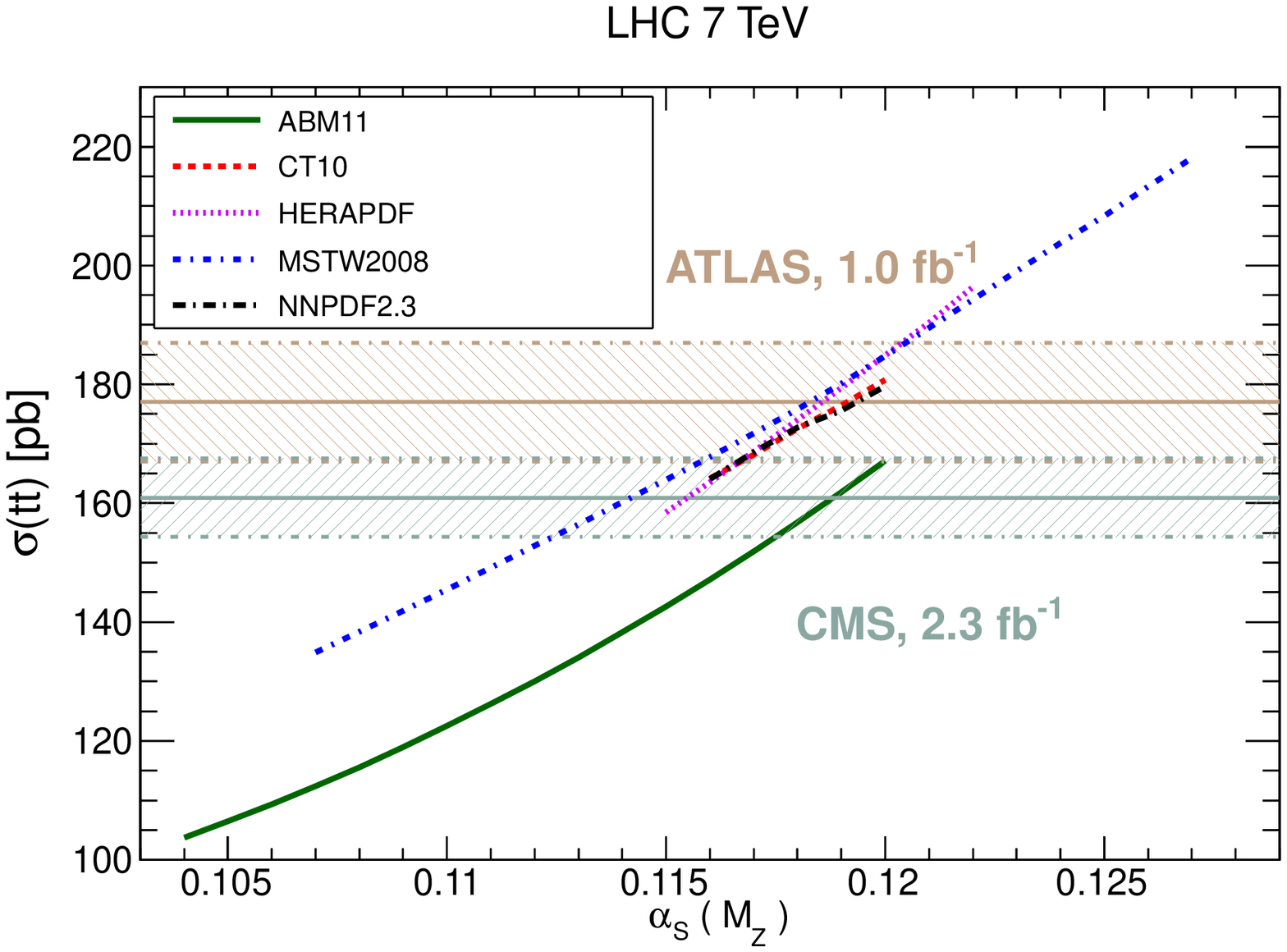}
\includegraphics[width=0.49\textwidth]{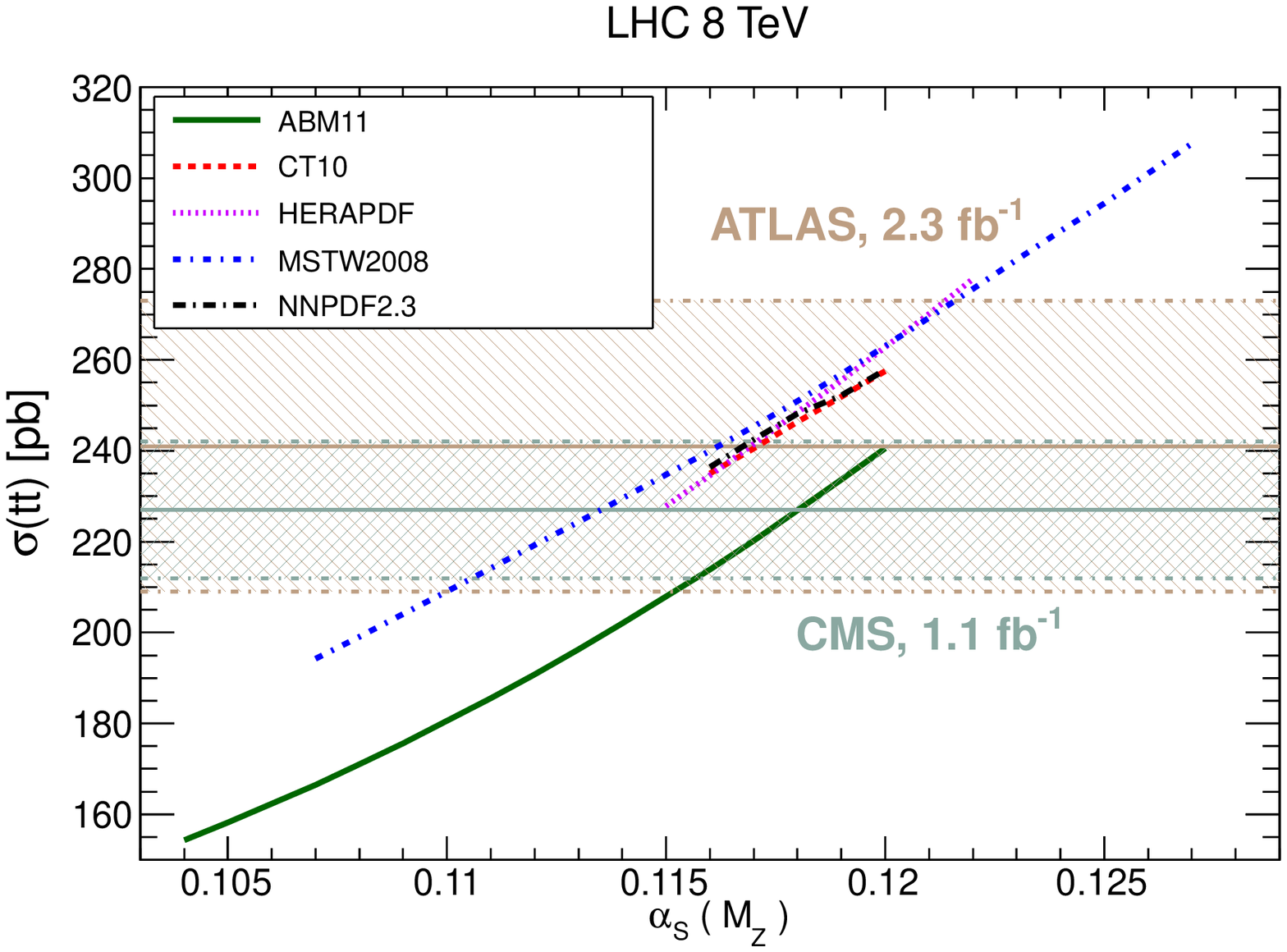}
\includegraphics[width=0.49\textwidth]{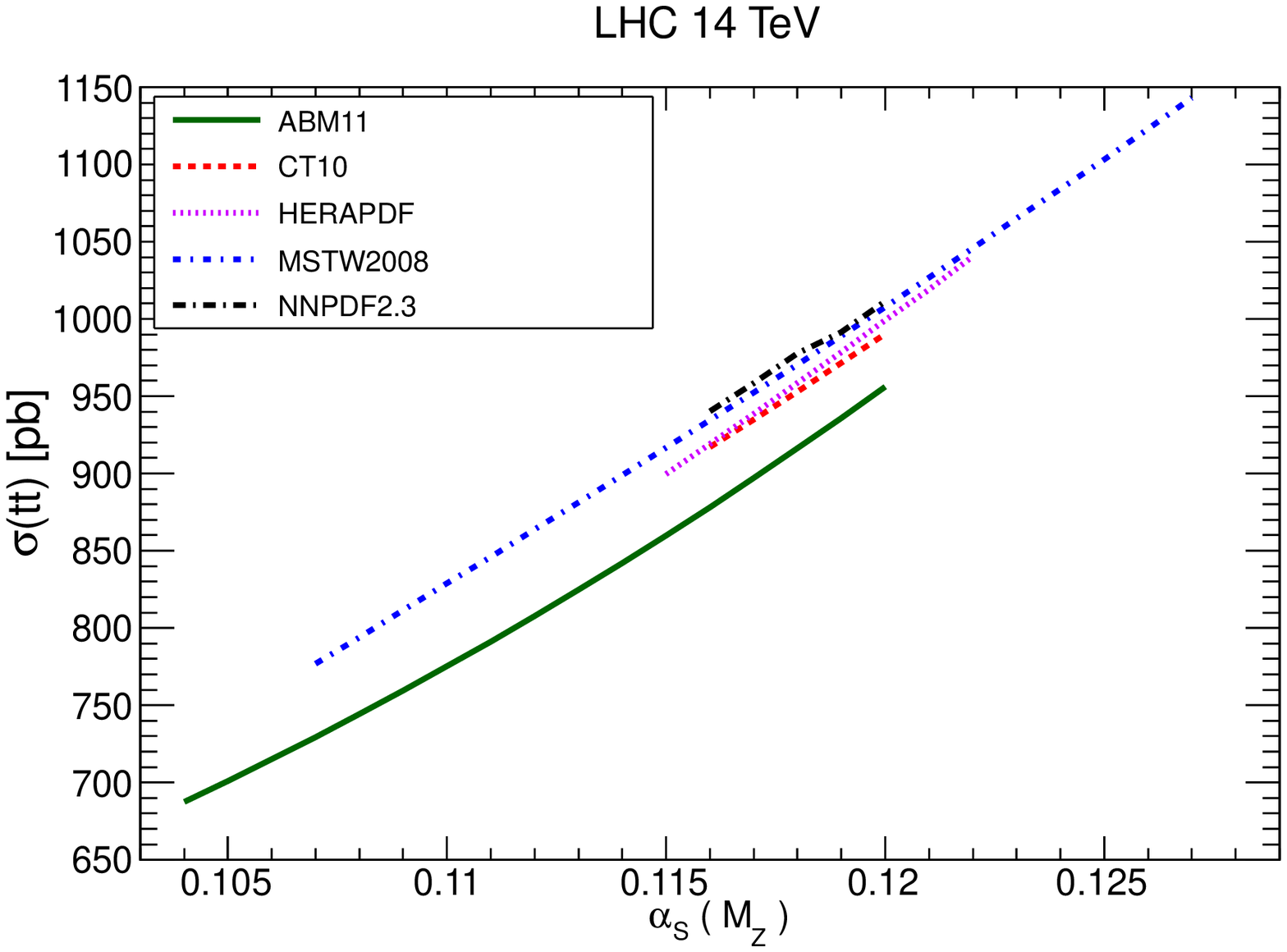}
\caption{\small The theoretical predictions for the various
PDF sets studied in this work as a function of the strong
coupling constant $\alpha_s(M_Z)$, compared to the best available
experimental data. From top to bottom we show
Tevatron, LHC 7 TeV, 8 TeV and 14 TeV.}
\label{fig:asdependence}
\end{figure}

In Fig.~\ref{fig:asdependence} we show the dependence
with $\alpha_s(M_Z)$ of the NNLO+NNLL cross sections for each
of the various PDF sets.\footnote{Except for the case of
JR09, where no varying-$\alpha_s$ PDF sets are provided.}
 We use consistently the same
value of $\alpha_s(M_Z)$ in the partonic cross sections
and in the PDFs, using all the range of varying $\alpha_s$ 
PDFs provided by each group. We also include for reference
the best experimental data. 
The slope with $\alpha_s$ is
similar for each of the PDF sets.
One can see that a variation of $\Delta \alpha_s$ by
0.001 increases the cross section by about 0.13 pb ($\sim 1.8\%$) at the
Tevatron, 4 pb ($\sim 2.3\%$) at LHC 7 TeV, 6 pb ($\sim 2.4\%$) at 8 TeV and
20 pb ($\sim 2.0\%$) at 14 TeV, with 
small differences between PDF sets.

\begin{figure}
\centering
\includegraphics[width=0.49\textwidth]{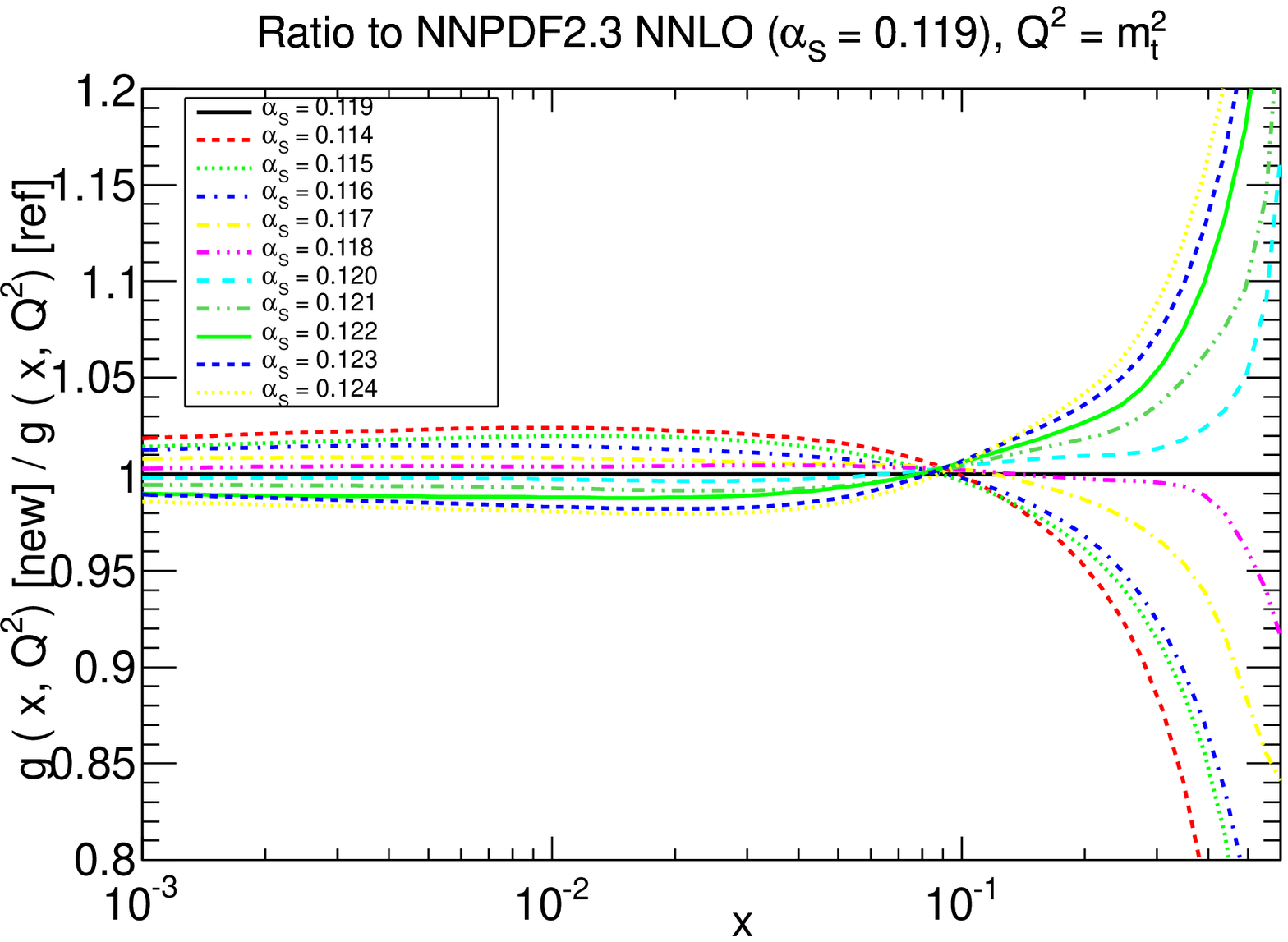}
\includegraphics[width=0.49\textwidth]{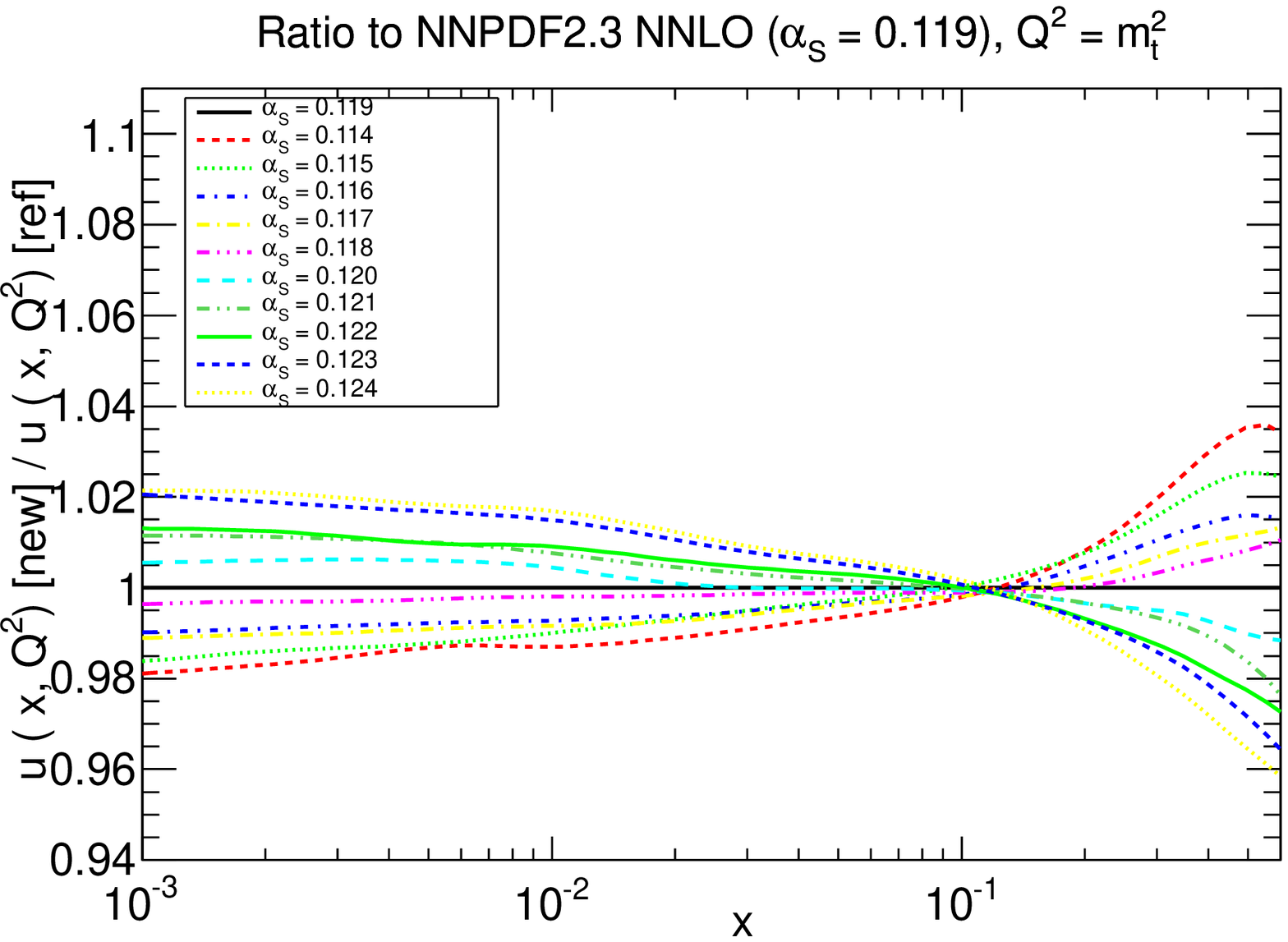}
\caption{\small Left plot: the gluon PDF
in the NNPDF2.3 fits obtained for
different input  values of $\alpha_s(M_Z)$, evaluated at
$Q^2=m_t^2$, shown as  the ratio with respect to the gluon PDF
obtained for $\alpha_s(M_Z)=0.119$.
Right plot: the same comparison, this time for the up quark PDF.
}
\label{fig:xg-nnpdf23-as-mtop}
\end{figure}

It is also worth noticing 
that the $\alpha_s$ dependence of the total cross section
reported in our tables is
slightly larger than what one would naively estimate based
on the power counting of the partonic cross section. 
The reason is
that, in the range of $x$  relevant for top quark production,
the value of the gluon density is larger for PDF fits with a larger
$\alpha_s$. 
This is shown in Fig.~\ref{fig:xg-nnpdf23-as-mtop}, for the
specific case of NNPDF2.3, but the behavior is
similar for other sets: larger $\alpha_s$ leads to smaller
$g(x)$ in the small/medium-$x$ region that is controlled by
deep-inelastic HERA data,
while DGLAP evolution and the momentum sum rule balance this
reduction with an increase at larger $x$.
In Fig.~\ref{fig:xg-nnpdf23-as-mtop} we also show the corresponding
plot for the case of the up quark density, relevant at the Tevatron.
In this case the correlation with $\alpha_s$ is rather less marked
and of opposite trend as in the case of the gluon.

The net effect of the positive
correlation between the
gluon PDF and $\alpha_s$ at large-$x$ 
is an enhanced sensitivity to $\alpha_s$ of the $t\bar{t}$
cross section, about 20\% larger than what one would have
obtained by keeping the PDFs fixed while changing only $\alpha_s$
in the partonic matrix element.\footnote{An alternative
way to conveying in a more quantitative way the information
contained in Fig.~\ref{fig:xg-nnpdf23-as-mtop} is to
compute the effective $\alpha_s$ exponent of the cross
section, $n_{\rm eff}\equiv d\ln \sigma(t\bar{t}) / d\ln \alpha_s$,
in the two cases, varying the PDF together with $\alpha_s$ and
keeping the PDFs fixed and varying only $\alpha_s$ in the matrix
element. At LHC 7 TeV, using NNPDF2.3 we find $n_{\rm eff}\sim 2.7$
in the former and $n_{\rm eff}\sim 2.3$ in the latter, confirming
the effect of the positive correlation between the gluon and
$\alpha_s$ in the top quark production region. } 
This
fact, together with the small, 2-3\%, scale uncertainty of the full
NNLO+NNLL result, suggest that the top production cross section could
provide a useful independent determination of $\alpha_s(M_Z)$. 
Such extraction would be
analogous to the determinations of $m_t$ from the same total cross
section~\cite{ATLAS:2011qga,CMS:2011lkd,Langenfeld:2009wd, Ahrens:2011px,Beneke:2012wb,Beneke:2011mq}.  
The
strong coupling can also be determined at the Tevatron and the LHC from jet
production~\cite{Abazov:2009nc,Malaescu:2012ts,CMS-PAS-QCD-11-003},
but results are affected by sizable scale uncertainties from the lack
of a complete NNLO calculation as well as by non-perturbative
corrections. 
 A first determination of $\alpha_s$ from the $t\bar{t}$
cross section, using previous approximate NNLO predictions, has been
performed by the CMS Collaboration~\cite{cmstopas}.

\section{Top quark data constraints on the gluon PDF}
\label{sec:rw}

The gluon PDF is one of the worse known partonic distributions.
Deep-inelastic data constrain the gluon only indirectly, via
scaling violations, and direct information comes only from
collider inclusive jet production
 data~\cite{D0:2008hua,Aaltonen:2008eq,Chatrchyan:2012bja,Aad:2011fc}.
Recently, the use of isolated photon data has also been advocated
to pin down the gluon~\cite{d'Enterria:2012yj}. 
However, direct constrains at hadron collider
from photon and jet data are affected
by substantial scale uncertainties due to the missing
full NNLO result, and are complicated by various
non-perturbative uncertainties. 
The availability of
a full NNLO calculation makes the
total top pair cross section the only collider
observable which, at present, is both directly sensitive to the gluon
PDF and can be consistently included in a NNLO
QCD analysis. 
The fact  that non-perturbative corrections
are much reduced in the total top quark production cross-section as compared to
photons and jets is another good motivation to use tops
as probes of the gluon PDF.
The possibility of using top cross section data in PDF analysis was
already
discussed at the qualitative level in Ref.~\cite{Nadolsky:2008zw}.
More recently, Ref.~\cite{Beneke:2012wb} provided
 a first estimate of the impact of top quark
data on the gluon PDF 
 based on approximate NNLO results.

In Sect.~\ref{sec:results} we have shown at the
qualitative level that available data already discriminates
between different PDF sets, that is, between different large-$x$ gluon
PDFs. 
Now we will be more quantitative and determine
if  available
data can help in reducing the gluon uncertainties within a single PDF set. 
In order to quantify this impact, we will use the Bayesian
PDF reweighting method of Refs.~\cite{Ball:2011gg,Ball:2010gb}
on the NNPDF2.3 set.\footnote{An alternative possibility would
have been a direct inclusion of the top quark data in the
NNPDF framework using {\tt MCFM} code together
with  the fast interface for $t\bar{t}$ production provided
by {\tt APPLgrid}~\cite{Carli:2010rw}. This fast interface will be used in
future NNPDF releases which will include  relevant
top quark production data.}
The same study could be carried out with Hessian
PDF sets as discussed in~\cite{Watt:2012tq}.
In principle,
one could also use the top quark differential distributions
data from ATLAS and CMS~\cite{top:2012hg,:2012qka}. 
However, these data are
less precise than the total cross section measurement and the corresponding
theoretical predictions are currently only available at NLO.\footnote{For some
specific differential distributions, results at NLO supplemented by threshold
resummation are available, see~\cite{Kidonakis:2011ca,Ahrens:2011mw} and references therein.}

Therefore, we have included the
$N_{\rm dat}=5$ experimental data points available
from the Tevatron and the LHC into the NNPDF2.3 NNLO fit.
The definition of $\chi^2$ that we use is Eq.~(\ref{eq:chi2}).
The effective number of replicas after reweighting
(the exponential of the Shannon entropy) is $N_{\rm eff}=86$,
out of the starting sample of 100 replicas, indicating
the moderate constraining power of the data.

\begin{figure}[t]
\centering
\includegraphics[width=0.49\textwidth]{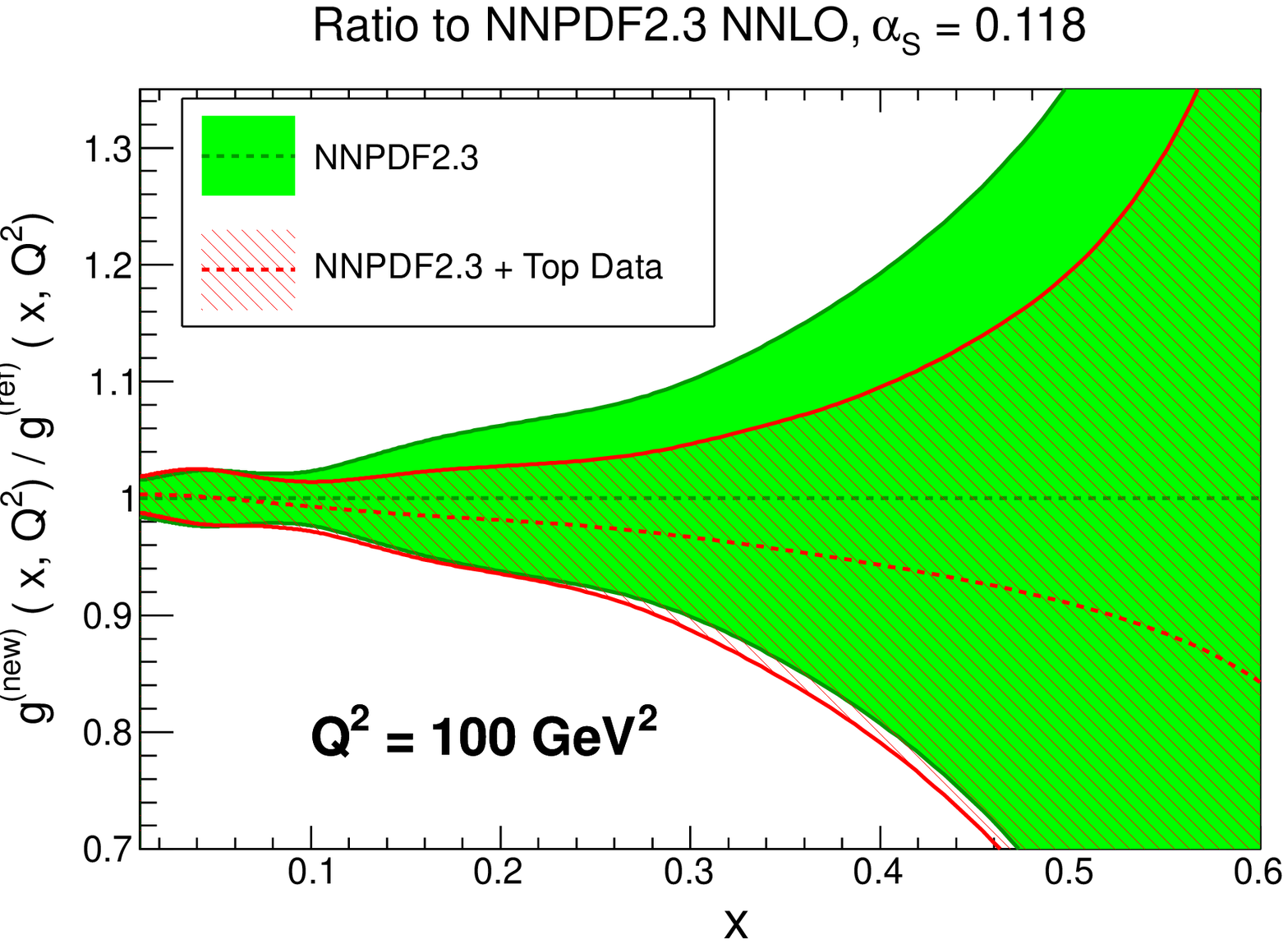}
\includegraphics[width=0.49\textwidth]{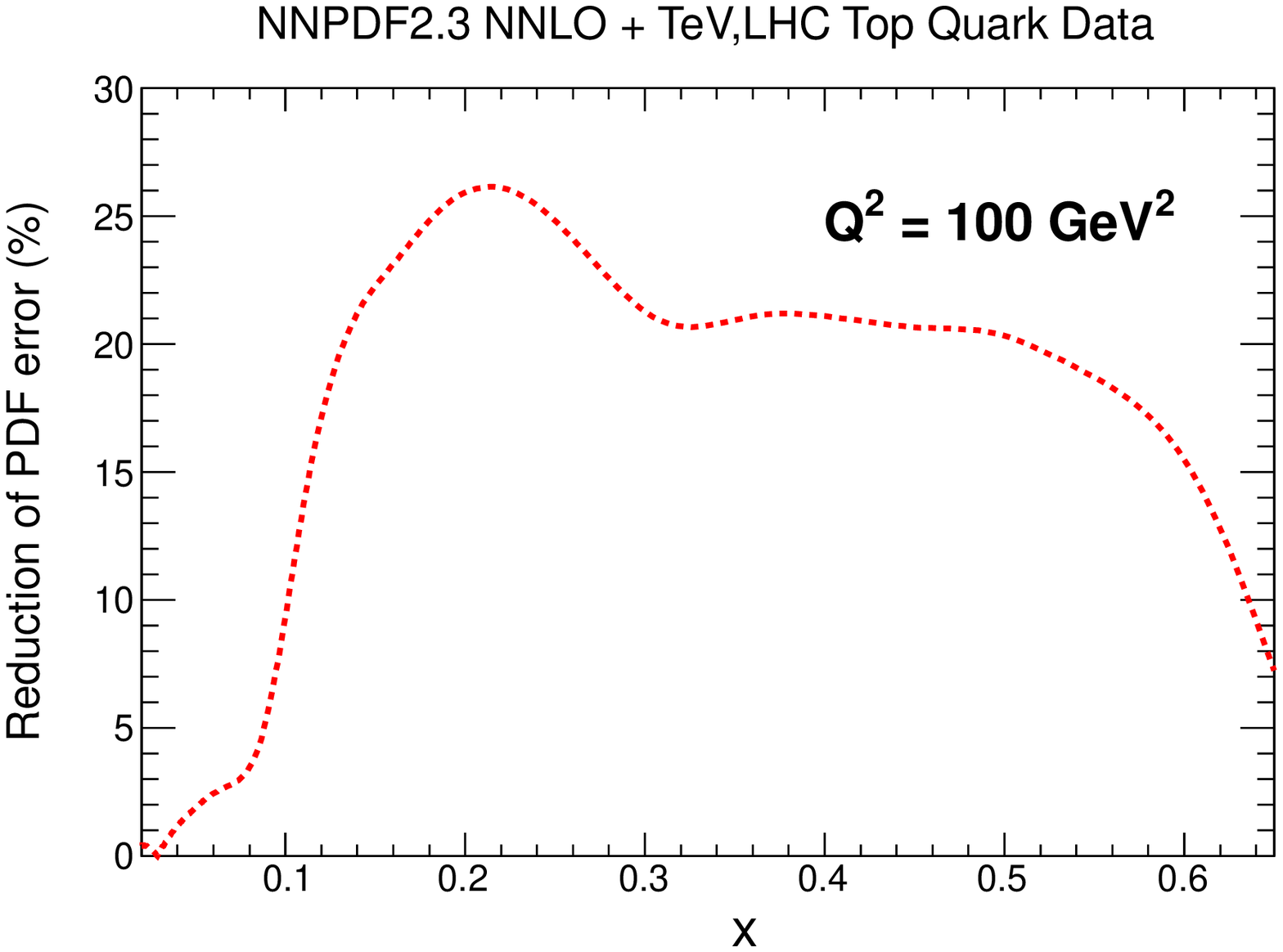}
\caption{\small Left plot: the ratio of the NNPDF2.3 NNLO gluon PDF at $Q^2=100$ GeV$^2$
 between the default fit and after including the Tevatron and LHC top quark
 cross section data. Right plot: the relative reduction of
PDF uncertainties thanks to the inclusion of top data
in the PDF fit. }
\label{fig:rwgluon}
\end{figure}

The results of adding the top quark data into the gluon PDF are shown
in Fig.~\ref{fig:rwgluon}. 
We show the NNPDF2.3 NNLO gluon at
$Q^2=100$ GeV$^2$, in the default fit and after including the Tevatron
and LHC top quark production cross section data.
  We observe that the large-$x$ gluon PDF uncertainties decrease. This is
expected since in that region the correlation between the gluon and
the top-quark cross section is maximal (see Fig.~\ref{fig:pdf-xsec-corr}).
In Fig.~\ref{fig:rwgluon} we also show the relative reduction in PDF
uncertainties from the addition of top data. 
This reduction coincides,
within the finite statistics of the original NNPDF Monte Carlo sample,
with the correlation profile of Fig.~\ref{fig:pdf-xsec-corr}, and in
particular confirms that the top quark data have a  small impact
below $x\sim 0.1$ or so, as well as for very large values of $x$.
 Therefore, we conclude that the available top
data can already help reduce the uncertainties on the gluon PDF by a
factor of up to 20\%, and are thus an important ingredient to future
global PDF analyses.
On the other hand, the impact of top data on the quark PDFs is
essentially negligible.

It is  interesting to study  the modifications of the theory predictions
 after the top quark data have been added into
the NNPDF2.3 fit. 
In Table~\ref{tab:chi2table}
we show the $t\bar{t}$ cross section
for NNPDF2.3, comparing the default
prediction with the  predictions after adding different
subsets of the top quark data. 
We show only the entries
which correspond to pure predictions.
By including top data from lower energy colliders,
we can provide arguably the most accurate theoretical prediction
for the total $t\bar{t}$ cross section at higher energies,
given that  PDF uncertainties will be reduced in the same
kinematical range from lower energy data.\footnote{Note that, as shown by Fig.~\ref{fig:pdf-xsec-corr}, the typical $x$ ranges covered by the theory
predictions at LHC 7, 8 and 14 TeV
are quite similar, justifying the extrapolation of lower LHC energy data to improve the
theory predictions at higher LHC center of mass energies. }

These predictions are collected in 
Table~\ref{tab:chi2table}. 
As an illustration,
the NNPDF2.3 prediction including Tevatron and LHC 7 
top data would be the best available theory prediction for LHC 8 TeV.
Note that not only
PDF uncertainties are reduced, but that also the central value
is shifted to improve the agreement with the experimental data.
As can be seen, the precise 7 TeV data carry most of the constraining power,
though of course improved power of the 8 TeV data will be provided
with the analysis of the full 2012 dataset.

\begin{table}[h]
\centering
\small
\begin{tabular}{c|c|c|c|c}
\hline
Collider  &  Ref  &
 Ref+TeV & Ref +TeV+LHC7 & Ref+TeV+LHC7+8  \\
\hline
\hline
Tevatron & 7.26 $\pm$ 0.12  & ({\it 7.29  $\pm$ 0.12 }) &   ({\it 7.27  $\pm$ 0.12 }) &   ({\it 7.27  $\pm$ 0.12 }) \\
\hline
LHC 7 TeV  &  172.5    $\pm$ 5.2      &   172.7 $\pm$ 5.1 &   ({\it 170.5  $\pm$ 3.6 }) &   ({\it  170.5 $\pm$ 3.5 }) \\
LHC 8 TeV  &   247.8   $\pm$  6.6     &   248.0 $\pm$ 6.5 &  245.0 $\pm$ 4.6 &   ({\it 245.2 $\pm$ 4.4 }) \\
LHC 14 TeV  &   976.5   $\pm$  16.4    &   976.2 $\pm$ 16.3 &  969.8 $\pm$ 12.0 &  969.6 $\pm$ 11.6 \\
\hline
\end{tabular}
\caption{\small The $t\bar{t}$ cross section $\sigma_{t\bar{t}}$, in
picobarns,
for the NNPDF2.3 NNLO set, together with the associated 
PDF uncertainties.
We show both the reference predictions with NNPDF2.3, and the predictions once
Tevatron, LHC7 and LHC8 data are added sequentially to the
fitted dataset. 
We show both the predictions corresponding
to beam energies whose data have not been included
in the fit, and the post-dictions (in parenthesis and italics)
for the beam energies whose data has been used in the fit.
The fourth column corresponds
to the best theory prediction for LHC 8 TeV, while the last
column is the best theory prediction for LHC 14 TeV.
 \label{tab:chi2table} }
\end{table}


Then in Table~\ref{tab:chi2table2} we provide NNPDF2.3 $\chi^2$ compared
to the  top quark  data, before
adding any data, after adding all Tevatron
and LHC data and adding only the
Tevatron and LHC 7 TeV data points. 
The slight improvement of an
already good quantitative description can be seen. As expected,
the agreement of the prediction with LHC8 data, when only Tevatron
and LHC7 data are used, is a non-trivial consistency check
of the whole procedure.\footnote{The small change of the $\chi^2$ between
TEV+LHC data and TEV+LHC7 data is due to statistical fluctuations, reflecting the fact that the 8 TeV data are still not
precise enough to provide constraints on the gluon PDF.}

\begin{table}[h]
\centering
\small
\begin{tabular}{c|c|c|c}
\hline
Collider  &  NNPDF2.3 &  
NNPDF2.3  &
NNPDF2.3  \\
  &  &  
+ TeV, LHC data &
+ TeV, LHC 7 TeV data \\
\hline
\hline
$\chi^{2}$ (Total, $N_{\rm dat}=5$)   & 6.28 &  4.88 & 4.87 \\
$\chi^{2}$ (LHC 8 TeV, $N_{\rm dat}=2$)  & 1.64  & 1.24  & 1.24  \\
\hline
\end{tabular}
\caption{\small The NNPDF2.3 $\chi^2$ compared
to the hadron collider top quark production data, before
adding any data (first column), after adding all Tevatron
and LHC data (second column) and adding only the
Tevatron and LHC 7 TeV data points (third column).
Note that the $\chi^2$ is not normalized by the
number of data points. \label{tab:chi2table2} }
\end{table}

Given that the constraints from top quark data in a global
PDF fit such as NNPDF2.3 are already substantial, we expect
even larger constrains in PDF sets based on reduced
datasets.
To quantify the impact of
top data into a DIS--only PDF fit,
we have performed a similar analysis as the one
with NNPDF2.3, but now
starting from a PDF set based on
a  reduced dataset, the NNPDF2.1 
HERA-only set~\cite{Ball:2011mu,Ball:2011uy}.
As the name indicates, this PDF set includes only HERA data,
and thus is affected by larger PDF uncertainties
than the NNPDF global fits, in particular
for the large-$x$ gluon.

\begin{figure}[h]
\centering
\includegraphics[width=0.49\textwidth]{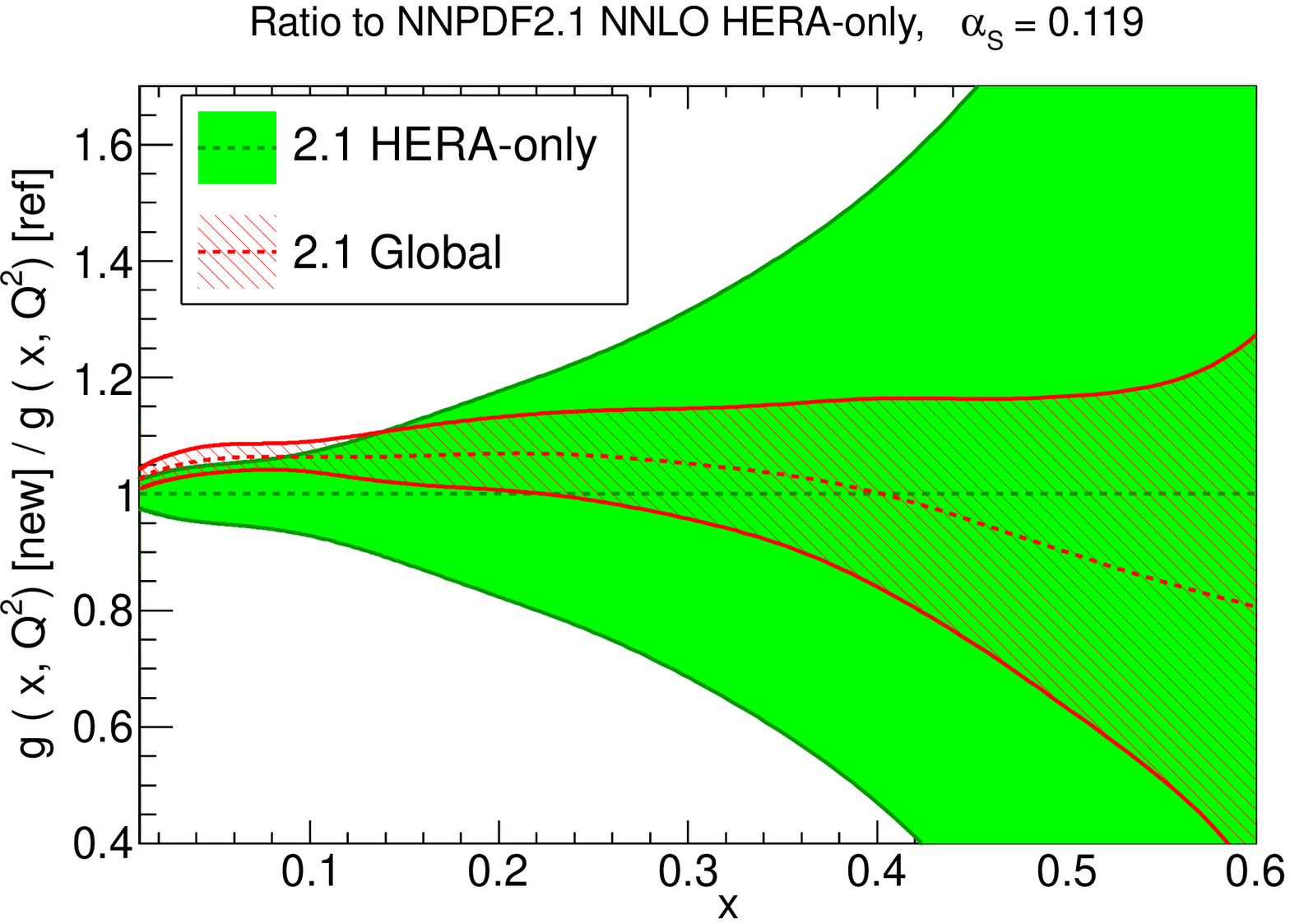}
\includegraphics[width=0.49\textwidth]{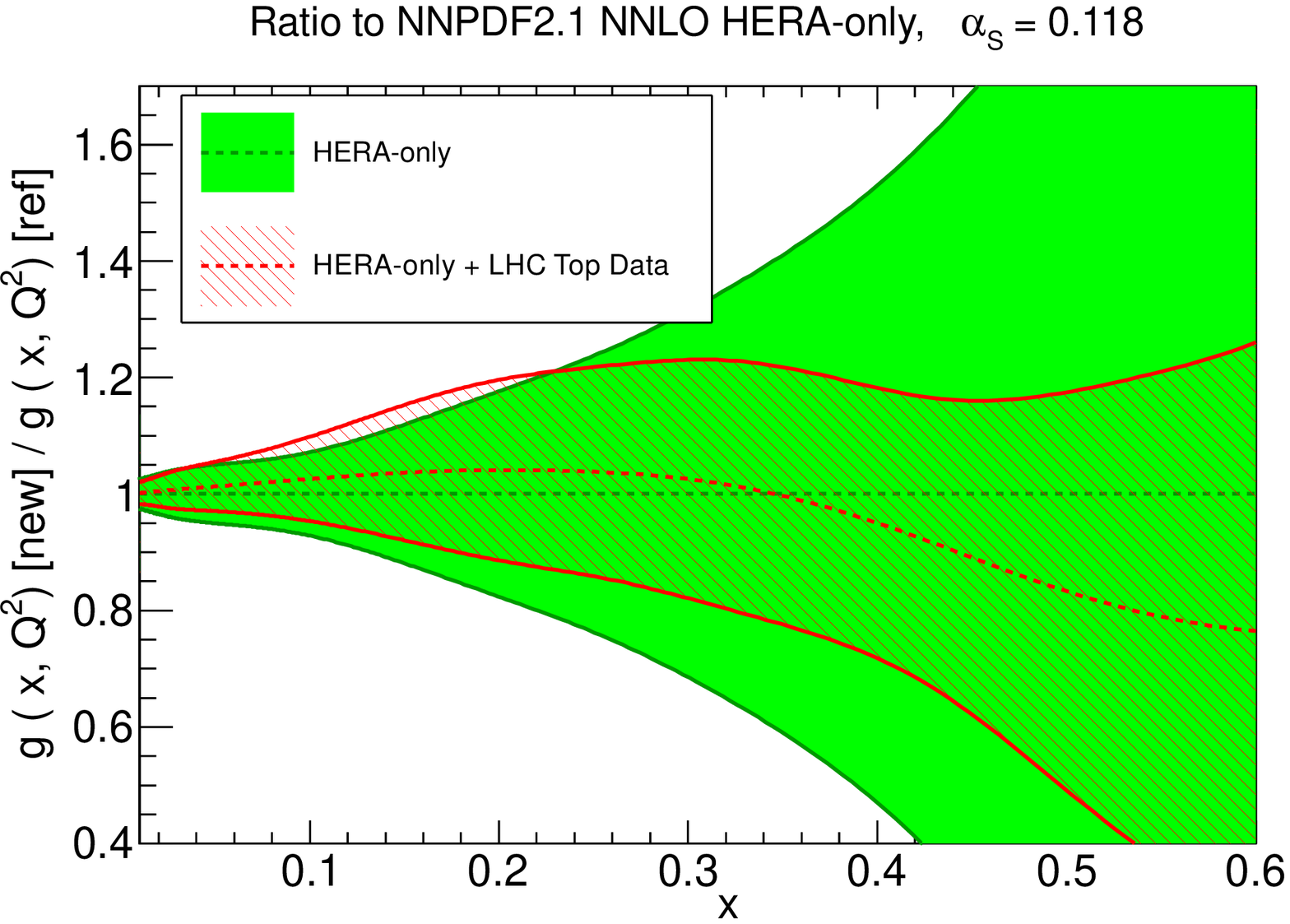}
\caption{\small Left plot: the NNPDF2.1 HERA-only gluon PDF, at $Q^2=100$ GeV$^2$, compared to the reference global NNPDF2.1 gluon.
Right plot: the NNPDF2.1 HERA-only gluon 
before and after including the LHC top quark
production cross sections. }
\label{fig:rwgluon_heraonly}
\end{figure}

In Fig.~\ref{fig:rwgluon_heraonly} we compare first of all the
gluon from the NNPDF2.1 HERA-only fit with that of
 the NNPDF2.1 global fit, to show the large differences
 in PDF uncertainties due 
to the reduced dataset in the former case. 
Then we show the improvements
in the gluon PDF in the HERA-only fit after the addition of
the top quark data. 
It is clear that the impact is substantial.
Remarkably, the top quark data bring the gluon from the HERA-only fit closer
to the gluon from the global fit.

\subsection{Impact on predictions for BSM particle production}
\label{sec:bsm}

Many scenarios of BSM physics predict the production of massive
final states in gluon-initiated processes.
The improvement in the large-$x$ gluon PDF uncertainties
seen in Fig.~\ref{fig:rwgluon} implies a similar
improvement in any cross section that is dominated by initial-state gluons
in a similar kinematical region.
Production kinematics determines that any high-mass final state
that is gluon initiated will benefit from the improvement
in PDF uncertainties.
Therefore, now we explore in two cases the phenomenological implications of
the improved large-$x$ gluon: the production of heavy Kaluza-Klein
resonances in warped extra dimensions scenarios, and the PDF
uncertainties in the high invariant mass distribution of
top quark pairs.

\begin{figure}[h]
\centering
\includegraphics[width=0.55\textwidth]{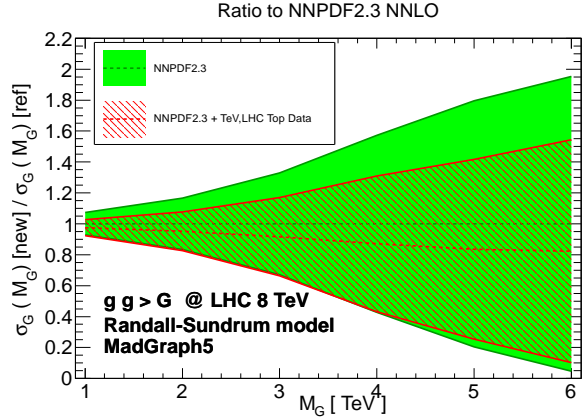}
\caption{\small The PDF uncertainties for the production of Randall-Sundrum
Kaluza-Klein gravitons at the LHC 8 TeV, with NNPDF2.3 before and
after including the top quark data, as a function of the graviton mass $M_G$.
 We have assumed that the graviton
couples only to gluons.
The cross section has been computed at leading order
with {\tt MadGraph5}. }
\label{fig:rwgraviton}
\end{figure}

To begin with, we have considered Kaluza-Klein massive graviton production
in warped extra dimension scenarios~\cite{Randall:1999vf},
in particular in the so-called {\it bulk} models in which
the coupling of the graviton 
to fermions is suppressed, and thus production is
driven by gluon-gluon annihilation~\cite{Giudice:2000av}. 
We have computed the
cross sections  for Randall-Sundrum graviton production at the
LHC 8 TeV at leading
order using the  {\tt MadGraph5} program~\cite{Alwall:2007st},
for a range of values of the graviton mass $M_G$.
The improvement in theoretical (PDF) uncertainties in the
graviton production cross section thanks to top quark
data is shown in Fig.~\ref{fig:rwgraviton}.
We see that, to begin with, PDF uncertainties are  large,
almost 40\% for  $M_{G}=3$ TeV, and growing as we approach
the kinematic threshold, reflecting our lack of knowledge of the
large-$x$ gluon PDF. 
The top quark data reduce moderately the
production uncertainties. 
Of course, future, more precise
top data will render these constraints more stringent, which
in turn translate into the possibility of better characterizing
eventual BSM high mass particles that the LHC could find.
 Similar
improvements are found for 14 TeV.

\begin{figure}[h]
\centering
\includegraphics[width=0.49\textwidth]{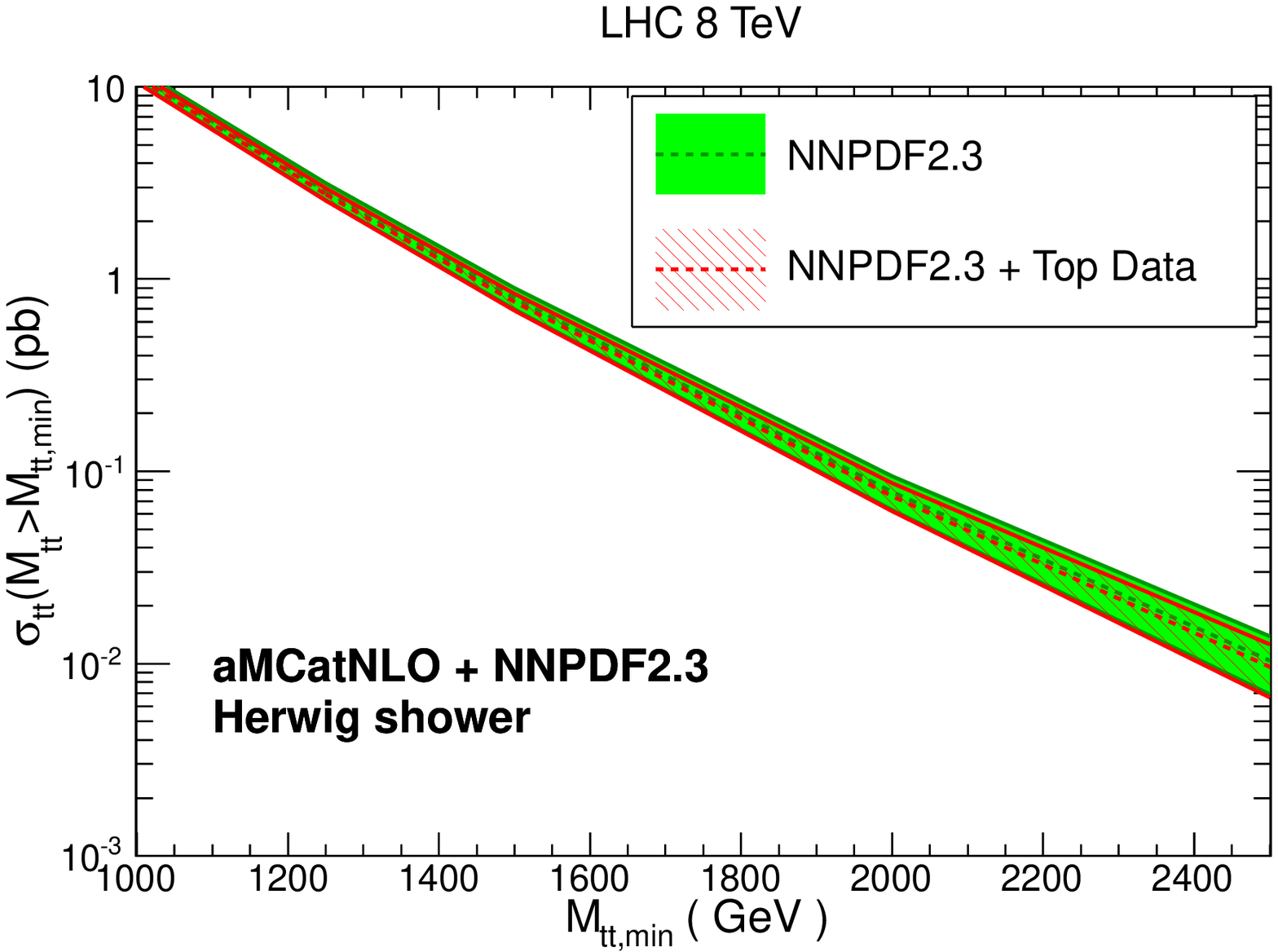}
\includegraphics[width=0.49\textwidth]{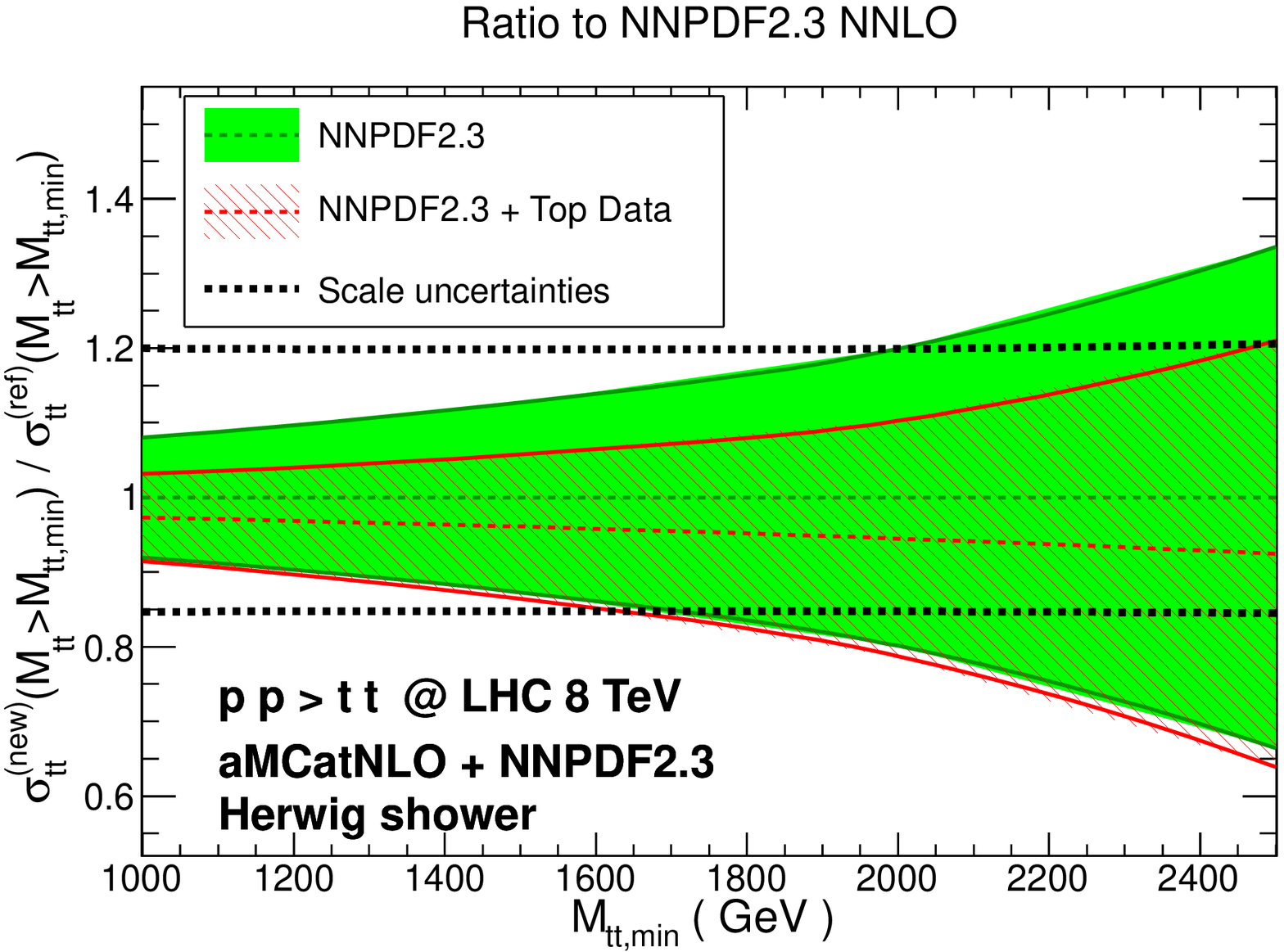}
\caption{\small Left plot: the $t\bar{t}$ cross section
above a minimum value of the $t\bar{t}$ invariant mass
$M_{tt}$, computed with  {\tt aMCatNLO} with the
{\tt Herwig} parton shower, and the
NNPDF2.3 as input set, at the LHC 8 TeV. 
The error band
is the PDF uncertainty band.
Right plot: the relative scale  and PDF uncertainties for the
cross section of top quark pairs with
invariant mass above $M_{tt,\rm min}$,
at LHC 8 TeV. 
In both cases we show the predictions with NNPDF2.3 before and
after including the top quark data into the PDF fit. }
\label{fig:rwamcatnlo}
\end{figure}

The second example is the high mass tail of the
 invariant mass distribution in top-quark
production, relevant for many BSM 
searches~\cite{Barger:2006hm,Frederix:2007gi}, which is also
related with the top quark forward-backward
asymmetry measured at the Tevatron. 
One example
is the searches for heavy resonances that decay
into a top-antitop pair~\cite{Chatrchyan:2012cx,Aad:2012dpa, Aad:2012wm, Aad:2012raa}.
To study the impact of the improvement of gluon PDF uncertainties there,
we have computed top quark pair production with the {\tt aMC@NLO} 
program~\cite{Frederix:2011ss,Frixione:2002ik,Frederix:2009yq},
at NLO matched to the {\tt Herwig6} shower~\cite{Corcella:2000bw}. 
The renormalization and factorization scales are set equal
to the sum  of transverse masses of all final-state particles.
Using NNPDF2.3, we have evaluated the cross section with 
a cut in the minimum invariant mass
of the $t\bar{t}$ pair, $M_{tt}$,  before and after including the top quark data into the PDF fit.
We show the results in Fig.~\ref{fig:rwamcatnlo}, for the
absolute cross section (left) and for the relative scale and PDF
uncertainties (right) as a function of the minimum value of $M_{tt}$ allowed.

As we can see, the addition of the total $t\bar{t}$ cross
section data reduces the PDF uncertainty in the high-end tail of the
$t\bar{t}$ mass distribution. 
It must be remarked that the data in
this high-mass tail are only a negligible fraction of the total
$t\bar{t}$ cross section, and therefore do not play any role in the
PDF fit itself.
 The PDF fit including $\sigma(t\bar{t})$ 
reduces the gluon uncertainty at large $x$ only because of the
overall constraints on the PDF evolution, which correlate the $x$
behavior in the $x\sim 0.1$ region (which dominates the total
production cross section) and the large-$x$ region, which is relevant
to the high-mass behavior. 

As in the case of dijet cross sections, we
expect that rate measurements in kinematical
regions where, for example, the $t\bar{t}$ system has a large
rapidity, can be used to further improve the knowledge of large-$x$
gluons, and improve even more the precision of predictions for
the production of large-mass objects in $gg$-initiated channels.

\section{Cross-section ratios between different LHC beam energies}
\label{sec:ratios}

The measurement of cross-section ratios between different center of mass energies at the
LHC has
 two main motivations~\cite{Mangano:2012mh}: first, they
 are interesting  for precision SM studies and second, they
have the potential
to enhance the possible BSM sensitivity of absolute cross sections.
In Ref.~\cite{Mangano:2012mh}, results for
top quark cross sections based on NLO+NNLL theory were provided; here
we update them to NNLO+NNLL and compare them
with experimental data.

\begin{figure}[h]
\centering
\includegraphics[width=0.49\textwidth]{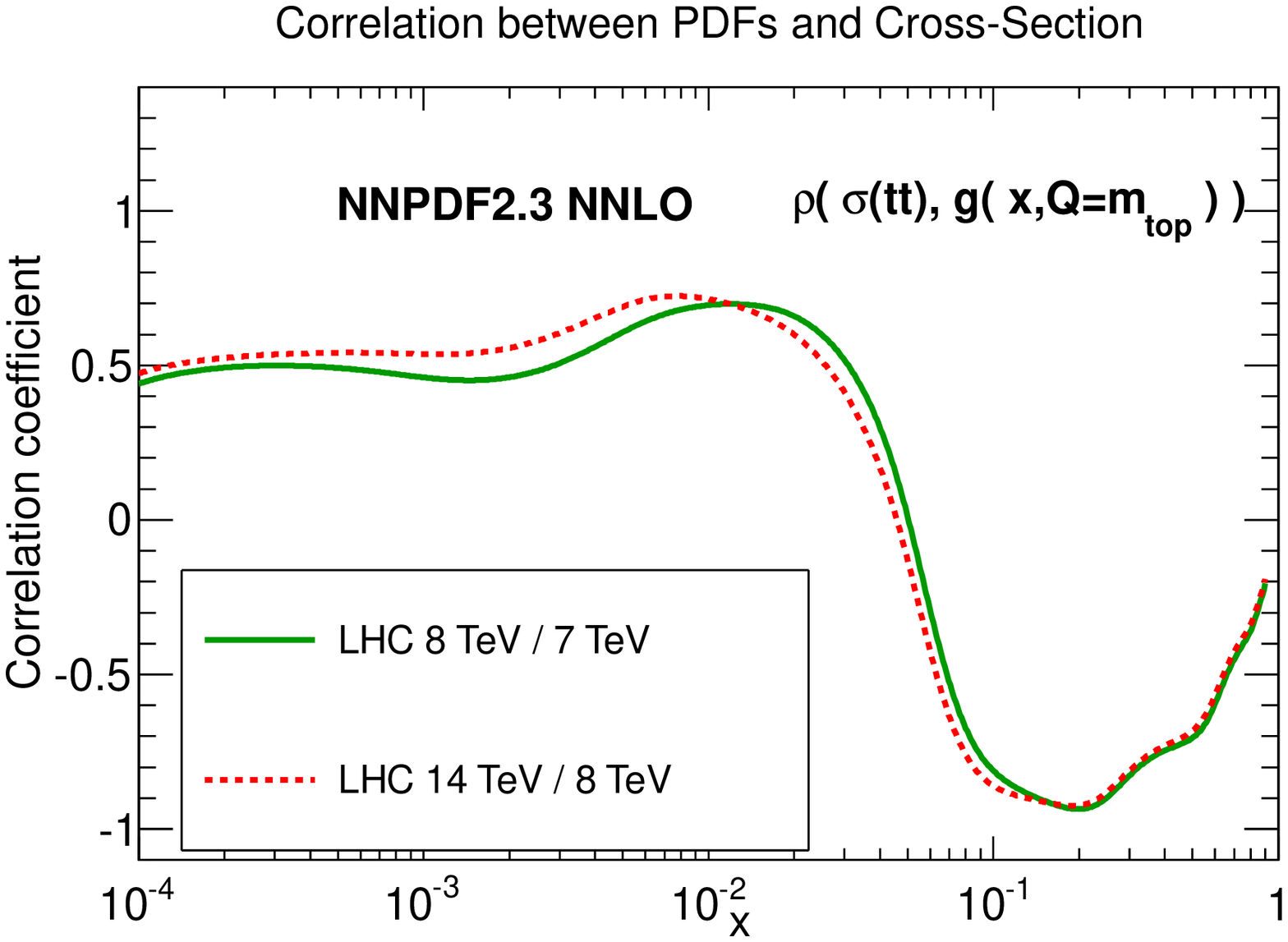}
\includegraphics[width=0.49\textwidth]{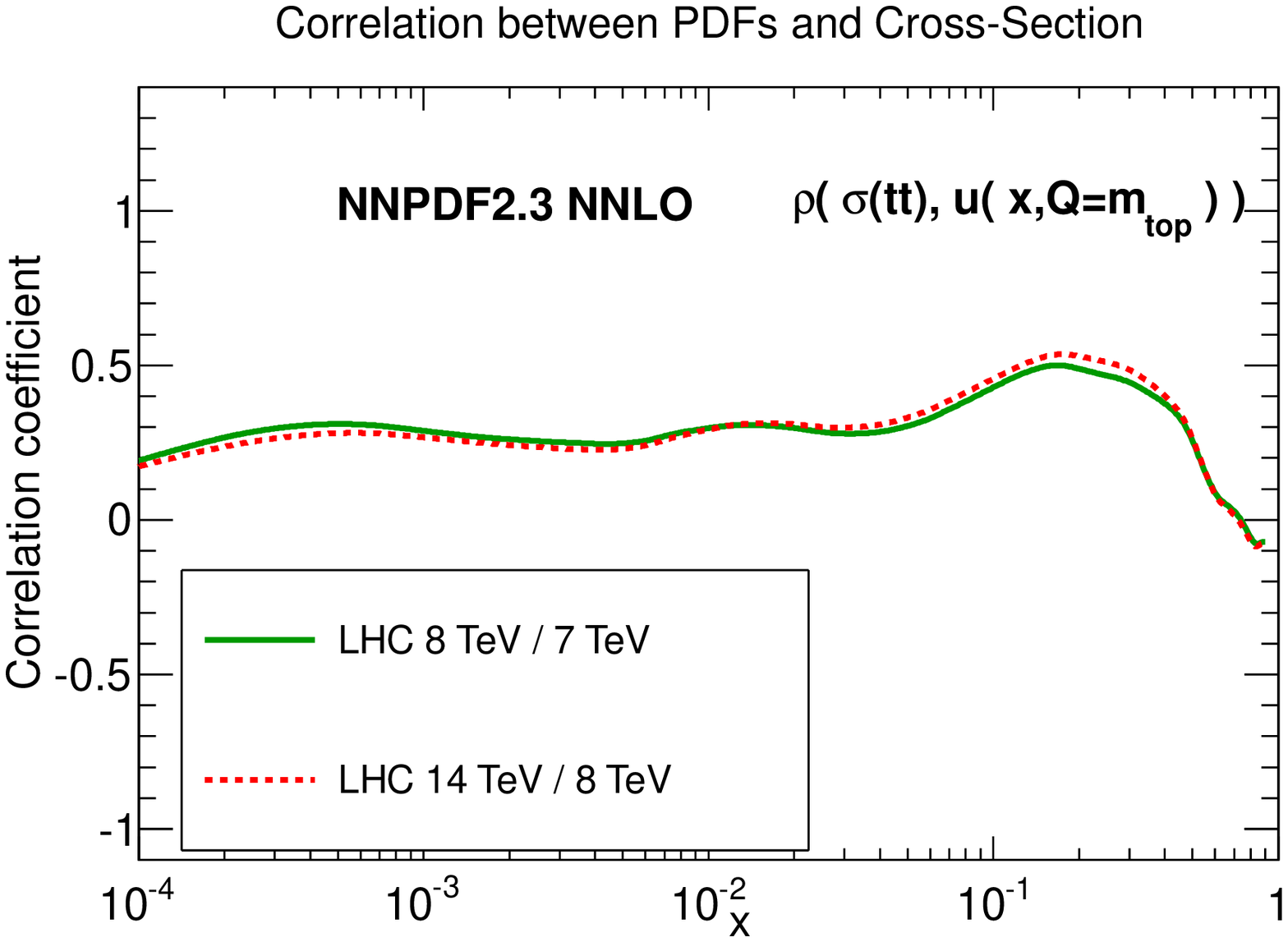}
\caption{\small The correlation between the gluon PDF (left plot)
and the up quark PDF (right plot) with the
 top quark production cross-section ratios
between different LHC center of mass energies.
  The correlations are computed
for $Q=m_t=173.3$ GeV. }
\label{fig:pdf-xsec-corr-rat}
\end{figure}

First of all we
 show the correlations between PDFs and the cross-section ratios in Fig.~\ref{fig:pdf-xsec-corr-rat}.
We see that the PDF correlation is approximately
the inverse of that of the absolute cross sections, shown in 
Fig.~\ref{fig:pdf-xsec-corr}. 
The reason for this  anti-correlation at large
$x$ is that when going to higher energies, the average probed values of $x$
are smaller.
%

Using the same settings
as before, we now provide NNLO+NNLL 
predictions for the cross-section ratios 
in Tables~\ref{tab:res-lhcrat-8-7} and~\ref{tab:res-lhcrat-14-8}.
We note that all systematic theoretical uncertainties are  small: in the
ratio between 8 and 7 TeV the total theory uncertainty is
at the permille level, while in the ratio between 14 and 8 TeV,
the total error on the ratio is at most around 2\%. 
This is dominated by PDF uncertainty, which is four times larger
than the combined scale, $\alpha_s$ and $m_t$ uncertainties. 
Notice the great improvement due to the NNLO result:
in the NLO+NNLL analysis of top cross-section ratios of Ref.~\cite{Mangano:2012mh}, scale and PDF uncertainties were of similar size.
We note that, as can be seen from
Table~\ref{tab:res-lhcrat-14-8}, in the case of the
14 over 8 TeV ratios the differences between PDF sets
are at the level of 10\%. Therefore, such measurement would provide 
powerful PDF constraints even within the foreseeable 
experimental accuracies.

\begin{table}[h]
\centering
\footnotesize
 \begin{tabular}{c|c|c|c|c|c|c} 
\hline
\multicolumn{7}{c}{LHC 8 TeV / 7 TeV ratio} \\ [1ex]
\hline
 PDF set & $\sigma_{tt}$ & $\delta_{\rm scale}$ & $\delta_{\rm PDF}$ & $\delta_{\alpha_s}$ (pb) & $\delta_{\rm m_t}$  & $\delta_{\rm tot}$  \\  [2ex] 
 \hline \hline 
ABM11                                    &    1.463 & $~^{+   0.001}_{-   0.002}$$~^{(+     0.1\%)}_{(-     0.1\%)}$  & $~^{+   0.006}_{-   0.006}$$~^{(+     0.4\%)}_{(-     0.4\%)}$  & $~^{+   0.000}_{-   0.000}$$~^{(+     0.0\%)}_{(-     0.0\%)}$  & $~^{+   0.001}_{-   0.001}$$~^{(+     0.1\%)}_{(-     0.1\%)}$  & 
$~^{+   0.007}_{-   0.008}$$~^{(+     0.5\%)}_{(-     0.5\%)}$  \\ [2ex] 
CT10                                     &    1.428 & $~^{+   0.001}_{-   0.001}$$~^{(+     0.1\%)}_{(-     0.1\%)}$  & $~^{+   0.008}_{-   0.010}$$~^{(+     0.5\%)}_{(-     0.7\%)}$  & $~^{+   0.002}_{-   0.002}$$~^{(+     0.2\%)}_{(-     0.2\%)}$  & $~^{+   0.001}_{-   0.001}$$~^{(+     0.1\%)}_{(-     0.1\%)}$  & 
$~^{+   0.009}_{-   0.011}$$~^{(+     0.6\%)}_{(-     0.8\%)}$  \\ [2ex] 
HERA1.5                                  &    1.426 & $~^{+   0.001}_{-   0.002}$$~^{(+     0.0\%)}_{(-     0.1\%)}$  & $~^{+   0.003}_{-   0.003}$$~^{(+     0.2\%)}_{(-     0.2\%)}$  & $~^{+   0.001}_{-   0.001}$$~^{(+     0.1\%)}_{(-     0.1\%)}$  & $~^{+   0.001}_{-   0.001}$$~^{(+     0.1\%)}_{(-     0.1\%)}$  & 
$~^{+   0.004}_{-   0.005}$$~^{(+     0.3\%)}_{(-     0.4\%)}$  \\ [2ex] 
JR09  &    1.426 & $~^{+   0.001}_{-   0.001}$$~^{(+     0.0\%)}_{(-     0.0\%)}$  & $~^{+   0.014}_{-   0.014}$$~^{(+     1.0\%)}_{(-     1.0\%)}$  & $~^{+   0.000}_{-   0.000}$$~^{(+     0.0\%)}_{(-     0.0\%)}$  &  $~^{+   0.001}_{-   0.001}$$~^{(+     0.1\%)}_{(-     0.1\%)}$   & 
$~^{+   0.014}_{-   0.014}$$~^{(+     1.0\%)}_{(-     1.0\%)}$  \\ [2ex] 
MSTW08                                   &    1.429 & $~^{+   0.001}_{-   0.001}$$~^{(+     0.1\%)}_{(-     0.1\%)}$  & $~^{+   0.004}_{-   0.004}$$~^{(+     0.2\%)}_{(-     0.2\%)}$  & $~^{+   0.001}_{-   0.001}$$~^{(+     0.1\%)}_{(-     0.1\%)}$  & $~^{+   0.001}_{-   0.001}$$~^{(+     0.1\%)}_{(-     0.1\%)}$  & 
$~^{+   0.005}_{-   0.005}$$~^{(+     0.3\%)}_{(-     0.3\%)}$  \\ [2ex] 
NNPDF2.3                                 &    1.437 & $~^{+   0.001}_{-   0.001}$$~^{(+     0.1\%)}_{(-     0.1\%)}$  & $~^{+   0.006}_{-   0.006}$$~^{(+     0.4\%)}_{(-     0.4\%)}$  & $~^{+   0.001}_{-   0.001}$$~^{(+     0.1\%)}_{(-     0.1\%)}$  & $~^{+   0.001}_{-   0.001}$$~^{(+     0.1\%)}_{(-     0.1\%)}$  & 
$~^{+   0.007}_{-   0.007}$$~^{(+     0.5\%)}_{(-     0.5\%)}$  \\ [2ex] 
 \hline
ATLAS &    1.36 && && &$\pm$     0.11~(     8\%) \\ [1ex] 
CMS &    1.40 & & & && $\pm$    0.08~(     6\%) \\ [1ex] 
 \hline
 \end{tabular}

\caption{\small
The NNLO+NNLL predictions for the ratio
of  top quark cross section at the LHC between
8 and 7 TeV, with
all the PDFs considered, and with the various
sources of theoretical uncertainties. 
The default
value of $\alpha_s(M_Z)$ from each collaboration
has been used in the computation of the central
predictions. 
The lower row shows our estimate of the best
available experimental measurement. 
The total theoretical uncertainty is the linear sum
of scale and parametric uncertainties, as
discussed in the text.
 \label{tab:res-lhcrat-8-7}
}
\end{table}

\begin{table}[h]
\centering
\footnotesize
 \begin{tabular}{c|c|c|c|c|c|c}
\hline
\multicolumn{7}{c}{LHC 14 TeV / 8 TeV ratio} \\ [1ex]
\hline
 PDF set & $\sigma_{tt}$ & $\delta_{\rm scale}$ & $\delta_{\rm PDF}$ & $\delta_{\alpha_s}$ (pb) & $\delta_{\rm m_t}$  & $\delta_{\rm tot}$  \\  [2ex] 
 \hline \hline 
ABM11                                    &    4.189 & $~^{+   0.008}_{-   0.016}$$~^{(+     0.2\%)}_{(-     0.4\%)}$  & $~^{+   0.057}_{-   0.057}$$~^{(+     1.4\%)}_{(-     1.4\%)}$  & $~^{+   0.000}_{-   0.000}$$~^{(+     0.0\%)}_{(-     0.0\%)}$  & $~^{+   0.012}_{-   0.012}$$~^{(+     0.3\%)}_{(-     0.3\%)}$  & 
$~^{+   0.067}_{-   0.074}$$~^{(+     1.6\%)}_{(-     1.8\%)}$  \\ [2ex] 
CT10                                     &    3.869 & $~^{+   0.006}_{-   0.009}$$~^{(+     0.2\%)}_{(-     0.2\%)}$  & $~^{+   0.068}_{-   0.088}$$~^{(+     1.8\%)}_{(-     2.3\%)}$  & $~^{+   0.020}_{-   0.020}$$~^{(+     0.5\%)}_{(-     0.5\%)}$  & $~^{+   0.010}_{-   0.010}$$~^{(+     0.2\%)}_{(-     0.2\%)}$  & 
$~^{+   0.077}_{-   0.100}$$~^{(+     2.0\%)}_{(-     2.6\%)}$  \\ [2ex] 
HERA1.5                                  &    3.841 & $~^{+   0.005}_{-   0.012}$$~^{(+     0.1\%)}_{(-     0.3\%)}$  & $~^{+   0.033}_{-   0.025}$$~^{(+     0.9\%)}_{(-     0.7\%)}$  & $~^{+   0.010}_{-   0.010}$$~^{(+     0.3\%)}_{(-     0.3\%)}$  & $~^{+   0.009}_{-   0.010}$$~^{(+     0.2\%)}_{(-     0.2\%)}$  & 
$~^{+   0.041}_{-   0.041}$$~^{(+     1.1\%)}_{(-     1.1\%)}$  \\ [2ex] 
JR09  &    3.808 & $~^{+   0.005}_{-   0.005}$$~^{(+     0.1\%)}_{(-     0.1\%)}$  & $~^{+   0.117}_{-   0.117}$$~^{(+     3.1\%)}_{(-     3.1\%)}$  & $~^{+   0.000}_{-   0.000}$$~^{(+     0.0\%)}_{(-     0.0\%)}$  &  $~^{+   0.012}_{-   0.012}$$~^{(+     0.3\%)}_{(-     0.3\%)}$     & 
$~^{+   0.122}_{-   0.122}$$~^{(+     3.2\%)}_{(-     3.2\%)}$  \\ [2ex] 
MSTW08                                   &    3.880 & $~^{+   0.006}_{-   0.009}$$~^{(+     0.2\%)}_{(-     0.2\%)}$  & $~^{+   0.036}_{-   0.036}$$~^{(+     0.9\%)}_{(-     0.9\%)}$  & $~^{+   0.011}_{-   0.011}$$~^{(+     0.3\%)}_{(-     0.3\%)}$  & $~^{+   0.010}_{-   0.010}$$~^{(+     0.2\%)}_{(-     0.2\%)}$  & 
$~^{+   0.045}_{-   0.048}$$~^{(+     1.2\%)}_{(-     1.2\%)}$  \\ [2ex] 
NNPDF2.3                                 &    3.940 & $~^{+   0.006}_{-   0.010}$$~^{(+     0.2\%)}_{(-     0.3\%)}$  & $~^{+   0.048}_{-   0.048}$$~^{(+     1.2\%)}_{(-     1.2\%)}$  & $~^{+   0.009}_{-   0.009}$$~^{(+     0.2\%)}_{(-     0.2\%)}$  & $~^{+   0.010}_{-   0.010}$$~^{(+     0.3\%)}_{(-     0.3\%)}$  & 
$~^{+   0.056}_{-   0.060}$$~^{(+     1.4\%)}_{(-     1.5\%)}$  \\ [2ex] 
 \hline
 \end{tabular}

\caption{\small
 Same as Table~\ref{tab:res-lhcrat-8-7}
for the ratio of 14 over 8 TeV cross sections
at the LHC. ~\label{tab:res-lhcrat-14-8}
}
\end{table}

We have checked that the results computed at NNLO+NNLL
with NNPDF2.1 are fully consistent with the 
computation of Ref.~\cite{Mangano:2012mh}, performed
at NLO+NNLL accuracy. 
This proves the reliability of the theoretical uncertainty
on the cross section ratios for $t\bar{t}$ production
presented in~\cite{Mangano:2012mh}, and underscores the
stability of this ratio under higher-order corrections.
Indeed, central values and PDF uncertainties are unchanged, 
while scale uncertainties are further decreased.
For instance, at NLO+NNLL the scale uncertainty in the
14/8 ratio was about 1\%, while with the NNLO+NNLL
computation it decreases down to 0.3\%.

While no measurement of the ratio between 8 and 7 TeV
is available, we can estimate the expected precision from
the absolute cross-section measurements of
Table~\ref{tab:sigma-tot-exp}
by assuming that all systematic uncertainties (but not
the luminosity) are fully correlated between 8 and 7 TeV.
This leads to
\beq
\sigma^{\rm (Atlas)}_{\rm LHC8/7}(t\bar{t}) &=& 1.36 \pm  0.11~{\rm pb}~({8\%}) \, , \nonumber\\
\sigma^{\rm (CMS)}_{\rm LHC8/7}(t\bar{t}) &=& 1.40 \pm 0.08~{\rm pb}~({6\%}) \, .
\eeq
Note that while ATLAS is higher than CMS at 8 and 7 TeV, this trend is inverted in the cross section ratio.

If no such correlation between the systematic uncertainties is assumed we get
\beq
\sigma^{\rm (Atlas)}_{\rm LHC8/7}(t\bar{t}) &=& 1.36 \pm  0.20~{\rm pb}~({15\%}) \, ,  \nonumber\\
\sigma^{\rm (CMS)}_{\rm LHC8/7}(t\bar{t}) &=& 1.40 \pm 0.11~{\rm pb}~({8\%}) \, ,
\eeq 
which illustrates the importance of maximizing the cancellation
of systematics. 
This requires a dedicated strategy for the ratio
measurement, rather than simply combining data at two different energies.

\begin{figure}
\centering
\includegraphics[width=0.49\textwidth]{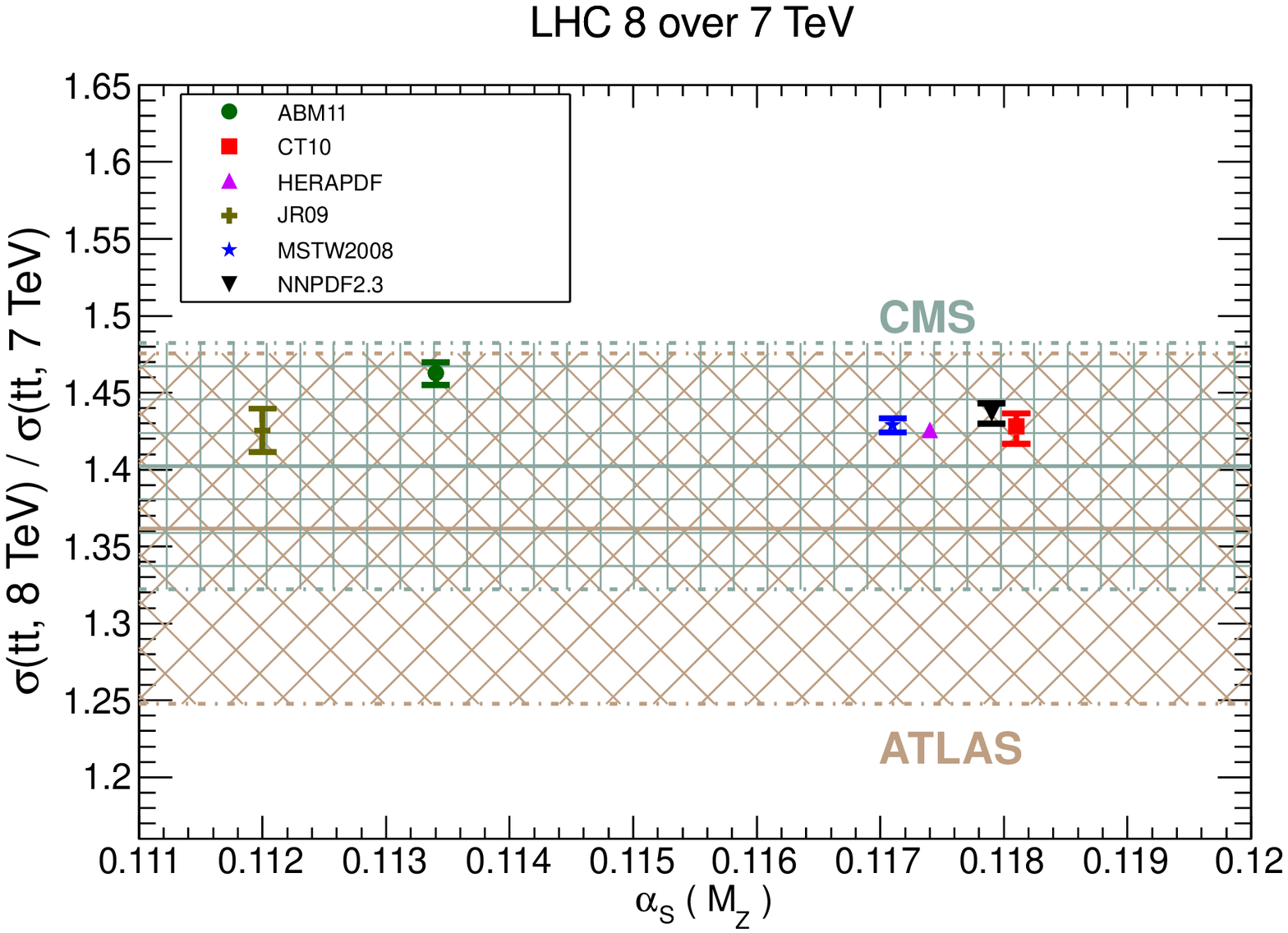}
\includegraphics[width=0.49\textwidth]{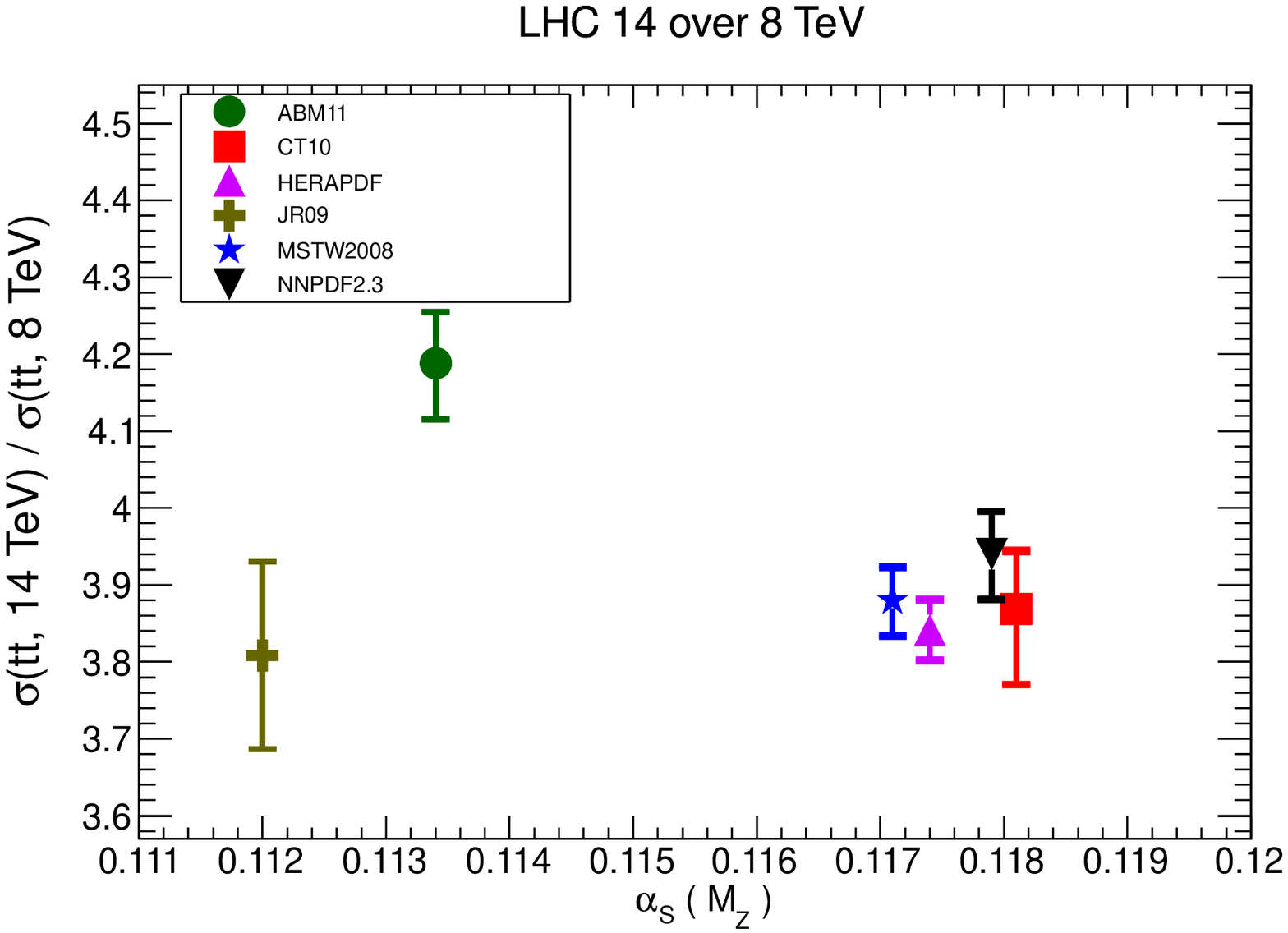}
\caption{\small The best predictions from each PDF set 
for the LHC ratios between 8 and 7 TeV (left plot)
and 14 and 8 TeV (right plot). 
The error bars correspond
to the total theoretical uncertainty.
For the 8/7 ratio,
the ATLAS and CMS experimental results have been obtained
taking the ratio of the central data and assuming that
experimental systematics (but not luminosity) is fully
correlated between different beam energies.
}
\label{fig:dataplots-rat}
\end{figure}

\begin{figure}
\centering
\includegraphics[width=0.49\textwidth]{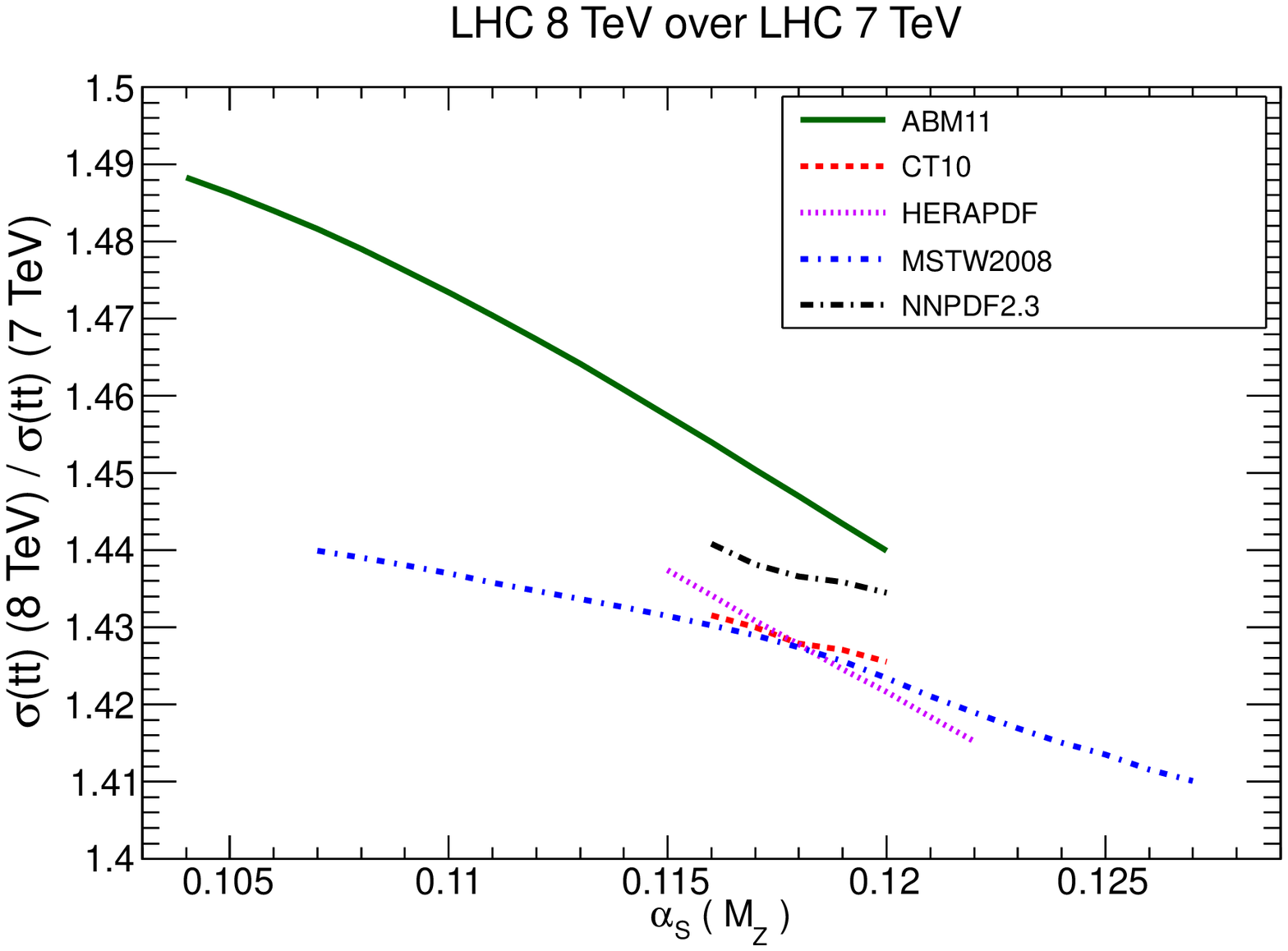}
\includegraphics[width=0.49\textwidth]{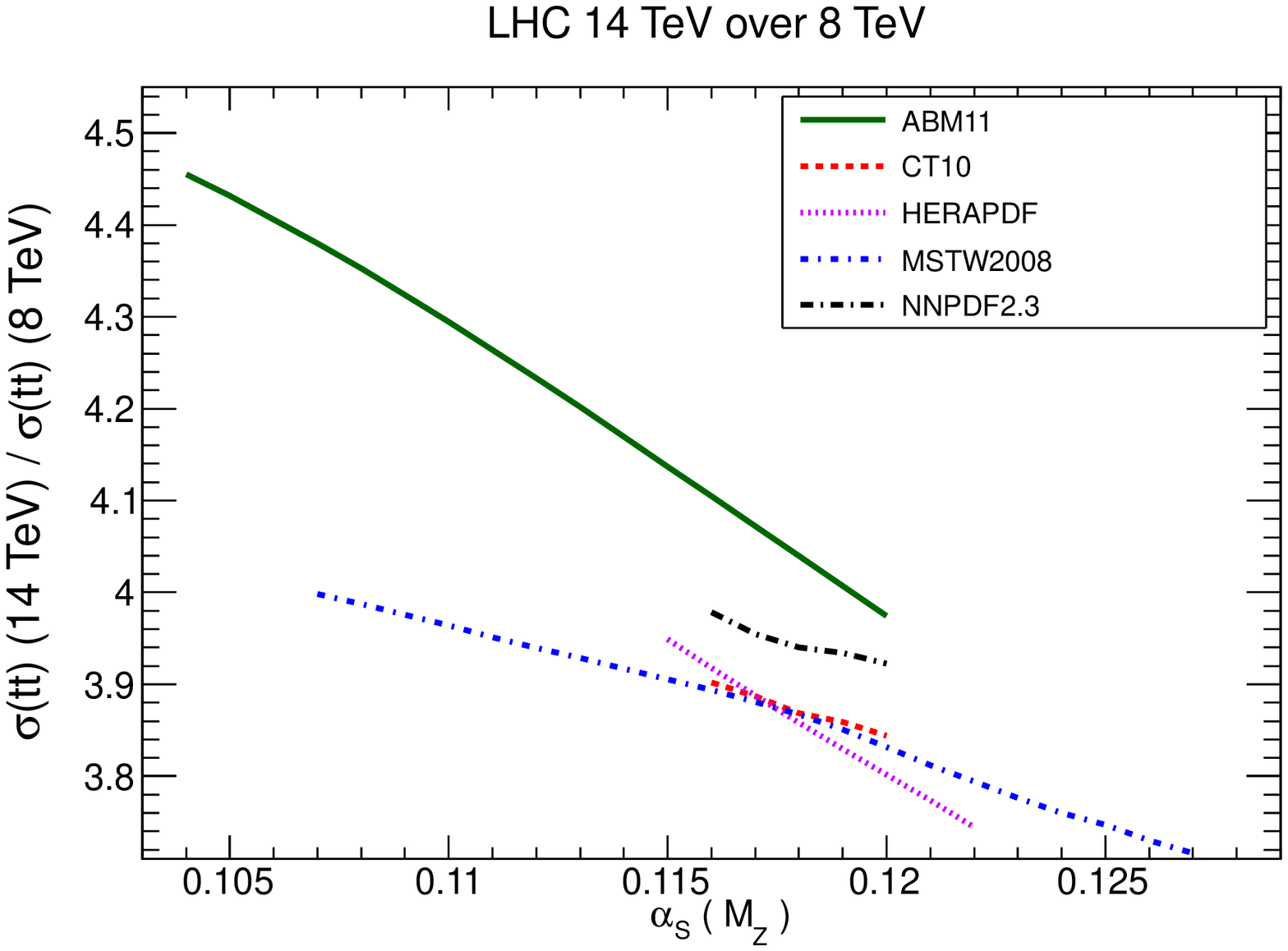}
\caption{\small The theoretical predictions for
the cross section ratios between different LHC beam energies for the
 the various
PDF sets, as a function of the strong
coupling constant $\alpha_s(M_Z)$.}
\label{fig:asdependencerat}
\end{figure}

We compare graphically the theoretical predictions for the
ratios between 8 and 7 TeV and 14 and 8 TeV in
Fig.~\ref{fig:dataplots-rat}.
 In the first case, we
also show the estimate of the ATLAS and CMS results,
assuming full correlation of the systematic uncertainties.
Then we show in Fig.~\ref{fig:asdependencerat} the dependence
on $\alpha_s(M_Z)$ of the NNLO+NNLL cross-section ratios for each
of the various PDF sets.
The slope with $\alpha_s$ for the prediction
of  each of the PDF sets varies more than for the
absolute cross sections, Fig.~\ref{fig:asdependence}.
However, with the assumed uncertainty on $\alpha_s$, the contribution
$\delta_{\alpha_s}$ to the total theory uncertainty of the cross-section ratios is much smaller than the PDF uncertainty.

Finally, it is useful to provide predictions for the
cross section ratios in the case of NNPDF2.3 supplemented
by the Tevatron and LHC top data, discussed in the previous
section.
The results are shown in Table~\ref{tab:ratiorw}.
As can be seen, while the differences are non negligible, they
are probably well beyond foreseeable experimental accuracies.
This is expected since cross-section ratios are 
mostly sensitive to more dramatic differences between
PDF sets, such as those between AMB11 and the other sets
in the case of the 14 over 8 TeV cross-section ratio, 
as shown in Fig~\ref{fig:dataplots-rat}.

\begin{table}[h]
\centering
\small
\begin{tabular}{c|c|c}
\hline
Collider  &  NNPDF2.3  & NNPDF2.3 + TeV,LHC top data   \\
\hline
\hline
$\sigma_{\rm LHC8/7}(t\bar{t})$ & 1.437 $\pm$ 0.006 ~       (0.4 \%)   &
1.439 $\pm$ 0.005 ~       (0.3 \%) \\
$\sigma_{\rm LHC14/8}(t\bar{t})$ & 3.94  $\pm$ 0.05 ~      (1.2 \%)   & 
3.96 $\pm$  0.04 ~      (1.0 \%)\\
\hline
\end{tabular}
\caption{\small The $t\bar{t}$ cross section ratios
for the NNPDF2.3 NNLO set, together with the associated 
PDF uncertainties, both in the reference fit and when
NNPDF2.3 is supplemented by Tevatron and LHC top quark
production data.
 \label{tab:ratiorw} }
\end{table}


\section{Hypothetical fourth-generation heavy quark production at the LHC}
\label{sec:highmass}

Following Ref.~\cite{Cacciari:2008zb}, we provide also 
the total cross section at NNLO+NNLL accuracy 
for a pair of hypothetical heavy fourth-generation quarks, belonging to the fundamental representation
of $SU(3)$.
Such new massive fermions
arise naturally in BSM
theories with strongly-coupled dynamics.
We denote these hypothetical heavy fourth-generation quarks  by $T$.
 Our aim
is to assess
 the scale and PDF uncertainties affecting the QCD contribution to the 
production of such heavy fermions using the most up-to-date
theoretical inputs.
 We have used both MSTW2008 and NNPDF2.3 NNLO PDFs as input
in the computation, and provide
predictions for masses $M_T$ in the range between 200 GeV and 1.3 TeV.
The PDF and scale uncertainties are defined as in Sect.~\ref{sec:results},
with the difference than now the central renormalization
and factorization scales are set to the heavy quark mass,
$\mu_F=\mu_R=M_T$.

\begin{figure}
\centering
\includegraphics[width=0.49\textwidth]{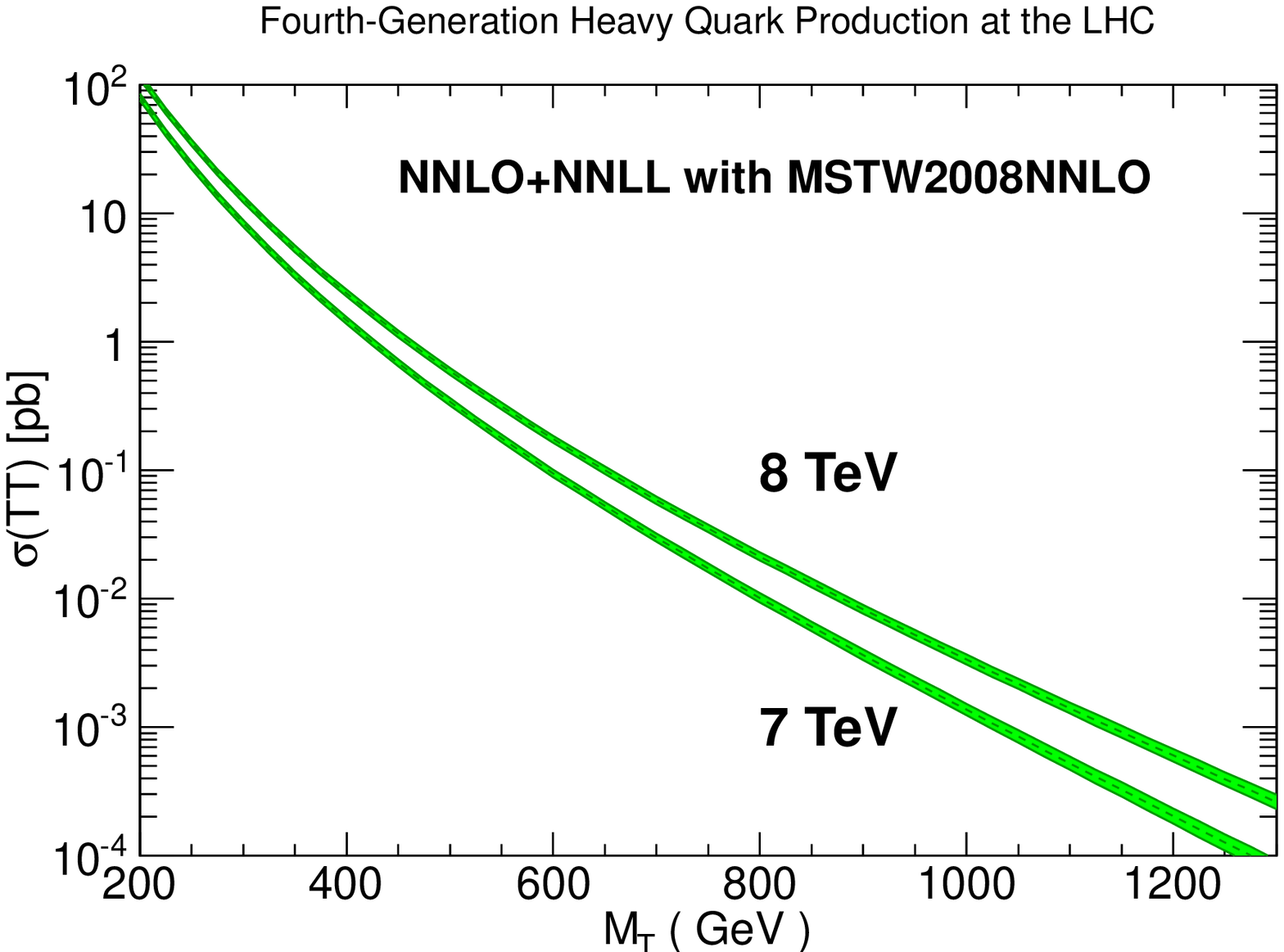}
\includegraphics[width=0.49\textwidth]{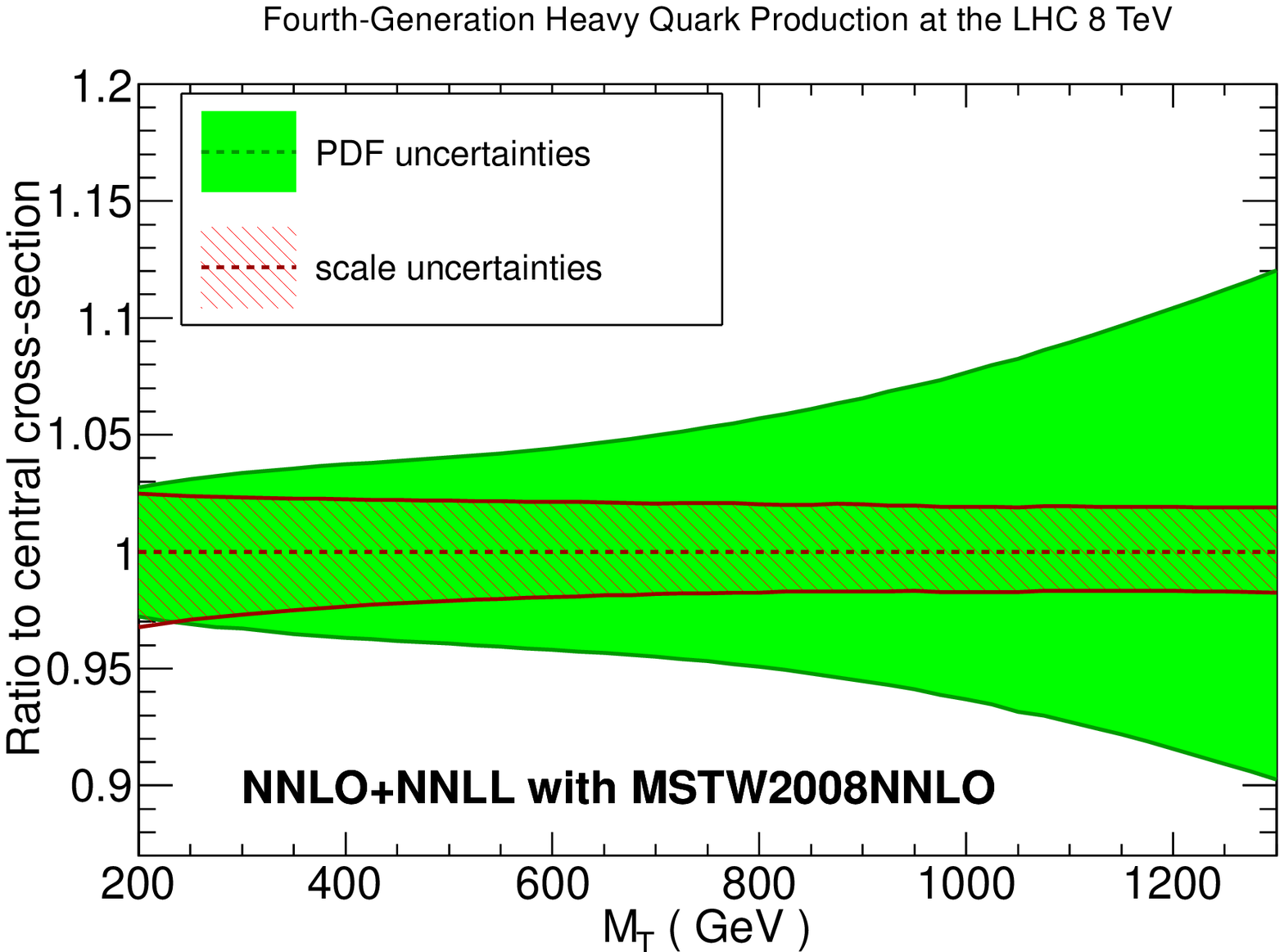}
\caption{\small Left plot: the cross sections for the
production of a $T\overline{T}$ pair of hypothetical heavy fourth-generation quarks at the
LHC 7 and 8 TeV, computed at NNLO+NNLL with  MSTW2008NNLO, as a function
of the heavy quark mass $M_T$.
The uncertainty band is the linear sum of
 PDF and scale
uncertainties. 
Right plot: the relative PDF and scale
uncertainties
as a function of $M_T$ for LHC 8 TeV. }
\label{fig:heavyt}
\end{figure}

The numerical results for the total cross
sections and associated scale and PDF uncertainties have been tabulated 
and they are available in the source of the arXiv submission
of this paper.\footnote{The data files are {\tt heavyfermion\_7tev\_mstw08.data}, {\tt heavyfermion\_8tev\_mstw08.data}, {\tt heavyfermion\_7tev\_nnpdf23.data}
and {\tt heavyfermion\_7tev\_nnpdf23.data}.}
In Fig.~\ref{fig:heavyt} (left) we show the production cross sections
at LHC 7 and 8 TeV as a function of $M_T$, where the
uncertainty band is the linear sum of scale and PDF uncertainties.
The MSTW08 PDF set was used as input.
 In
Fig.~\ref{fig:heavyt} (right) we also show
the relative
PDF and scale uncertainties,  at the LHC 8 TeV. 
We notice that for large heavy fermion masses $M_T\gg m_t$,
PDF uncertainties become the dominant source
of theoretical error.

\begin{figure}
\centering
\includegraphics[width=0.49\textwidth]{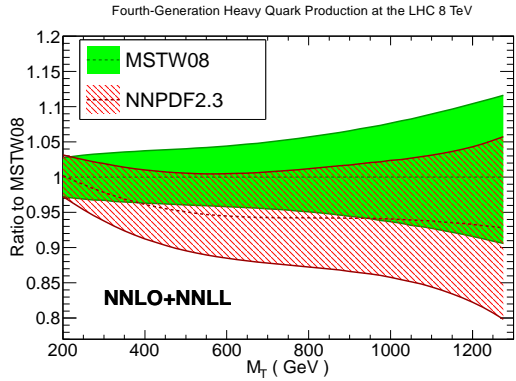}
\caption{\small Same as Fig.~\ref{fig:heavyt} (right), now
comparing the predictions of MSTW08 and NNPDF2.3.
Only PDF uncertainties are shown.
 }
\label{fig:heavyt_pdfdep}
\end{figure}

Another useful comparison is provided by the PDF dependence of the
hypothetical heavy fourth-generation quark production cross
section.
In Fig.~\ref{fig:heavyt_pdfdep} we compare, as a function of $M_T$, the
predictions for the NNLO+NNLL cross-sections in MSTW08 and
NNPDF2.3.
Only PDF uncertainties are shown.
While the two uncertainty bands overlap, the envelope of the two
sets is substantially larger than the bands of the individual sets.
The NNPDF2.3 predictions are smaller than the MSTW08 ones, by about
1-sigma, for $M_T\ge 400$ GeV.

\section{Summary and outlook}

In this work we have presented a detailed assessment of
the theoretical systematic uncertainties that
affect the total top quark pair production cross section at hadron
colliders.
 We have also compared the theoretical predictions with the most 
recent experimental data
from the Tevatron and the LHC 7 and 8 TeV, and provided
predictions for LHC 14 TeV.  

For our analysis we have used the most precise perturbative
and non-perturbative information available, in particular
the recently computed NNLO+NNLL cross sections and the most
up-to-date NNLO PDF sets. 
Thanks to the significantly reduced scale dependence
of the full NNLO calculation, now the three main
sources of theoretical uncertainty, scales, PDFs
and the top quark mass, are roughly of the same
size, around 2-3\%.
We find that most PDF sets agree
both in their central values and in the size of their
PDF uncertainties.

Given the high accuracy of the perturbative $t\bar t$ cross section,
we have demonstrated that the available data provide a strong
constraint on existing NNLO PDF sets, in particular on the large-$x$
gluon PDF. 
 The inclusion of Tevatron and LHC top quark data in the
NNPDF2.3 set results in a reduction on the large-$x$ gluon PDF
uncertainty, which has further important phenomenological
consequences. 
We have explicitly studied two such cases, both relevant
for BSM searches: high mass graviton production in warped extra
dimensions scenarios, and the high mass tail of the top pair
invariant mass distribution.  
We have also given updated predictions
for the pair production of a hypothetical fourth-generation heavy quark $T$,
which should be helpful in interpreting the results of 
ongoing experimental searches
\cite{Chatrchyan:2012vu,ATLAS:2012qe,Chatrchyan:2012fp}.

For completeness, we have also provided up-to-date predictions
for the cross-section ratios of $\sigma_{t\bar t}$ at different LHC 
center-of-mass energies.

To conclude, we emphasize that the availability of the NNLO calculation for
the total $t\bar{t}$ cross section makes top pair production the first
hadron collider process which is both directly sensitive to the gluon
PDF and can be included consistently into a NNLO global PDF analysis
without any approximation.  
This leads us to believe that top quark
data will be an important ingredient of future PDF fits, especially as
more precise data on the total cross section and on various differential
distributions becomes available.

\begin{acknowledgments}
We acknowledge Tancredi Carli, Maria Jos\'e Costa, Roberto Tenchini
and Roberto Chierici for information about the LHC top quark
cross-section data, and Stefano Forte, Stefano Frixione, Katerina Lipka and Sebastian Naumann-Emme
for useful discussions.
The work of M.~C. was supported by the Heisenberg and by the Gottfried Wilhelm Leibniz programmes of the Deutsche Forschungsgemeinschaft, and by the DFG Sonderforschungsbereich/Transregio 9 ``Computergest\"utzte Theoretische Teilchenphysik".
 J.~R. is supported by a Marie Curie Intra--European Fellowship of the European Community's 7th Framework 
Programme under contract number PIEF-GA-2010-272515.
The work of M.~L.~M. and A.~M. is 
supported by the ERC grant 291377 ``LHCtheory: Theoretical predictions and analyses of LHC physics: advancing the precision frontier".
\end{acknowledgments}

\providecommand{\href}[2]{#2}\begingroup\raggedright\endgroup


\begin{thebibliography}{100}

\bibitem{Czakon:2013goa}
M.~Czakon, P.~Fiedler and A.~Mitov,
 {\it The total top quark pair production cross-section at hadron colliders through ${\cal O}(\alpha_S^4)$},
 CERN-PH-TH/2013-056,  TTK-13-08, \href{http://xxx.lanl.gov/abs/1303.6254}{{\tt
  arXiv:1303.6254}}.

\bibitem{Baernreuther:2012ws}
P.~Baernreuther, M.~Czakon, and A.~Mitov, {\it {Percent Level Precision Physics
  at the Tevatron: First Genuine NNLO QCD Corrections to $q \bar{q} \to t
  \bar{t} + X$}},  {\em Phys.Rev.Lett.} {\bf 109} (2012) 132001,
  [\href{http://xxx.lanl.gov/abs/1204.5201}{{\tt arXiv:1204.5201}}].

\bibitem{Czakon:2012zr}
M.~Czakon and A.~Mitov, {\it {NNLO corrections to top-pair production at hadron
  colliders: the all-fermionic scattering channels}},
  \href{http://xxx.lanl.gov/abs/1207.0236}{{\tt arXiv:1207.0236}}.

\bibitem{Czakon:2012pz}
M.~Czakon and A.~Mitov, {\it {NNLO corrections to top pair production at hadron
  colliders: the quark-gluon reaction}},
  \href{http://xxx.lanl.gov/abs/1210.6832}{{\tt arXiv:1210.6832}}.

\bibitem{Beneke:2009rj}
M.~Beneke, P.~Falgari, and C.~Schwinn, {\it {Soft radiation in heavy-particle
  pair production: All-order colour structure and two-loop anomalous
  dimension}},  {\em Nucl.Phys.} {\bf B828} (2010) 69--101,
  [\href{http://xxx.lanl.gov/abs/0907.1443}{{\tt arXiv:0907.1443}}].

\bibitem{Czakon:2009zw}
M.~Czakon, A.~Mitov, and G.~F. Sterman, {\it {Threshold Resummation for
  Top-Pair Hadroproduction to Next-to-Next-to-Leading Log}},  {\em Phys.Rev.}
  {\bf D80} (2009) 074017, [\href{http://xxx.lanl.gov/abs/0907.1790}{{\tt
  arXiv:0907.1790}}].

\bibitem{Cacciari:2011hy}
M.~Cacciari, M.~Czakon, M.~L. Mangano, A.~Mitov, and P.~Nason, {\it {Top-pair
  production at hadron colliders with next-to-next-to-leading logarithmic
  soft-gluon resummation}},  {\em Phys.Lett.} {\bf B710} (2012) 612--622,
  [\href{http://xxx.lanl.gov/abs/1111.5869}{{\tt arXiv:1111.5869}}].

\bibitem{Cacciari:2008zb}
M.~Cacciari, S.~Frixione, M.~L. Mangano, P.~Nason, and G.~Ridolfi, {\it
  {Updated predictions for the total production cross sections of top and of
  heavier quark pairs at the Tevatron and at the LHC}},  {\em JHEP} {\bf 0809}
  (2008) 127, [\href{http://xxx.lanl.gov/abs/0804.2800}{{\tt
  arXiv:0804.2800}}].

\bibitem{Kidonakis:2011ca} 
  N.~Kidonakis, B.~D.~Pecjak and ,
  {\it Top-quark production and QCD},
  {\rm Eur.\ Phys.\ J.\ C} {\bf 72}, 2084 (2012),
  [\href{http://xxx.lanl.gov/abs/1108.6063}{{\tt
  arXiv:1108.6063}}].
  
  
\bibitem{Forte:2013wc}
 S.~Forte and G.~Watt, {\it {Progress in the Determination of the Partonic Structure of the Proton}}, \href{http://xxx.lanl.gov/abs/1301.6754}{{\tt arXiv:1301.6754}}.

\bibitem{Watt:2011kp}
G.~Watt, {\it {Parton distribution function dependence of benchmark Standard Model total cross sections at the 7 TeV LHC}},  {\em JHEP} {\bf 1109} (2011) 069,
  [\href{http://xxx.lanl.gov/abs/1106.5788}{{\tt arXiv:1106.5788}}].

\bibitem{Watt:2012np}
G.~Watt, {\it {MSTW PDFs and impact of PDFs on cross sections at Tevatron and LHC}},  {\em  Nucl.\ Phys.\ Proc.\ Suppl.} {\bf 222} (2012) 061,
  [\href{http://xxx.lanl.gov/abs/1201.1295 }{{\tt arXiv:1201.1295 }}].
 

\bibitem{Moch:2012mk}
S.~Moch, P.~Uwer, and A.~Vogt, {\it {On top-pair hadro-production at
  next-to-next-to-leading order}},  {\em Phys.Lett.} {\bf B714} (2012) 48--54,
  [\href{http://xxx.lanl.gov/abs/1203.6282}{{\tt arXiv:1203.6282}}].

\bibitem{Aliev:2010zk}
M.~Aliev et~al., {\it {-- HATHOR -- HAdronic Top and Heavy quarks crOss section
  calculatoR}},  {\em Comput. Phys. Commun.} {\bf 182} (2011) 1034--1046,
  [\href{http://xxx.lanl.gov/abs/1007.1327}{{\tt arXiv:1007.1327}}].
  
  \bibitem{Beneke:2011mq} 
  M.~Beneke, P.~Falgari, S.~Klein, C.~Schwinn and ,
  {\it Hadronic top-quark pair production with NNLL threshold resummation}.
  {\em Nucl.\ Phys.\ B} {\bf 855}, 695 (2012),
  [\href{http://xxx.lanl.gov/abs/1109.1536}{{\tt arXiv:1109.1536}}].

\bibitem{Beneke:2012wb}
M.~Beneke, P.~Falgari, S.~Klein, J.~Piclum, C.~Schwinn, et~al., {\it {Inclusive
  Top-Pair Production Phenomenology with TOPIXS}},  {\em JHEP} {\bf 1207}
  (2012) 194, [\href{http://xxx.lanl.gov/abs/1206.2454}{{\tt
  arXiv:1206.2454}}].
  
\bibitem{Ahrens:2011px}
V.~Ahrens, A.~Ferroglia, M.~Neubert, B.~D. Pecjak, and L.~L. Yang, {\it
  {Precision predictions for the t+t(bar) production cross section at hadron
  colliders}},  {\em Phys.Lett.} {\bf B703} (2011) 135--141,
  [\href{http://xxx.lanl.gov/abs/1105.5824}{{\tt arXiv:1105.5824}}].

\bibitem{Kramer:2012bx}
M.~Kramer, A.~Kulesza, R.~van~der Leeuw, M.~Mangano, S.~Padhi, et~al., {\it
  {Supersymmetry production cross sections in $pp$ collisions at $\sqrt{s}=7$
  TeV}},  \href{http://xxx.lanl.gov/abs/1206.2892}{{\tt arXiv:1206.2892}}.

\bibitem{Agashe:2007zd}
K.~Agashe, H.~Davoudiasl, G.~Perez, and A.~Soni, {\it {Warped Gravitons at the
  LHC and Beyond}},  {\em Phys.Rev.} {\bf D76} (2007) 036006,
  [\href{http://xxx.lanl.gov/abs/hep-ph/0701186}{{\tt hep-ph/0701186}}].

\bibitem{Randall:1999vf}
L.~Randall and R.~Sundrum, {\it {An Alternative to compactification}},  {\em
  Phys.Rev.Lett.} {\bf 83} (1999) 4690--4693,
  [\href{http://xxx.lanl.gov/abs/hep-th/9906064}{{\tt hep-th/9906064}}].

\bibitem{Giudice:2000av}
G.~F. Giudice, R.~Rattazzi, and J.~D. Wells, {\it {Graviscalars from higher
  dimensional metrics and curvature Higgs mixing}},  {\em Nucl.Phys.} {\bf
  B595} (2001) 250--276, [\href{http://xxx.lanl.gov/abs/hep-ph/0002178}{{\tt
  hep-ph/0002178}}].

\bibitem{Frederix:2007gi}
R.~Frederix and F.~Maltoni, {\it {Top pair invariant mass distribution: A
  Window on new physics}},  {\em JHEP} {\bf 0901} (2009) 047,
  [\href{http://xxx.lanl.gov/abs/0712.2355}{{\tt arXiv:0712.2355}}].

\bibitem{Barger:2006hm}
V.~Barger, T.~Han, and D.~G. Walker, {\it {Top Quark Pairs at High Invariant
  Mass: A Model-Independent Discriminator of New Physics at the LHC}},  {\em
  Phys.Rev.Lett.} {\bf 100} (2008) 031801,
  [\href{http://xxx.lanl.gov/abs/hep-ph/0612016}{{\tt hep-ph/0612016}}].

\bibitem{Chiappetta:1990jd}
P.~Chiappetta and M.~Perrottet, {\it {Possible bounds on compositeness from
  inclusive one jet production in large hadron colliders}},  {\em Phys.Lett.}
  {\bf B253} (1991) 489--493.

\bibitem{Chatrchyan:2013muj}
{\bf CMS Collaboration}, S.~Chatrchyan et~al., {\it {Search for
  contact interactions using the inclusive jet pT spectrum in pp collisions at
  sqrt(s)=7 TeV}},  \href{http://xxx.lanl.gov/abs/1301.5023}{{\tt
  arXiv:1301.5023}}.

\bibitem{Chatrchyan:2012bf}
{\bf CMS Collaboration}, S.~Chatrchyan et~al., {\it {Search for
  quark compositeness in dijet angular distributions from $pp$ collisions at
  $\sqrt{s}=7$ TeV}},  {\em JHEP} {\bf 1205} (2012) 055,
  [\href{http://xxx.lanl.gov/abs/1202.5535}{{\tt arXiv:1202.5535}}].

\bibitem{ATLAS:2012pu}
{\bf ATLAS Collaboration}, G.~Aad et~al., {\it {ATLAS search for
  new phenomena in dijet mass and angular distributions using $pp$ collisions
  at $\sqrt{s}=7$ TeV}},  {\em JHEP} {\bf 1301} (2013) 029,
  [\href{http://xxx.lanl.gov/abs/1210.1718}{{\tt arXiv:1210.1718}}].

\bibitem{D0:2008hua}
{\bf D0} Collaboration, V.~M. Abazov et~al., {\it {Measurement of the inclusive
  jet cross-section in $p \bar{p}$ collisions at $\sqrt{s}$ =1.96~TeV}},  {\em
  Phys. Rev. Lett.} {\bf 101} (2008) 062001,
  [\href{http://xxx.lanl.gov/abs/0802.2400}{{\tt arXiv:0802.2400}}].

\bibitem{Aaltonen:2008eq}
{\bf CDF} Collaboration, T.~Aaltonen et~al., {\it {Measurement of the Inclusive
  Jet Cross Section at the Fermilab Tevatron p-pbar Collider Using a Cone-Based
  Jet Algorithm}},  {\em Phys. Rev.} {\bf D78} (2008) 052006,
  [\href{http://xxx.lanl.gov/abs/0807.2204}{{\tt arXiv:0807.2204}}].

\bibitem{Chatrchyan:2012bja}
{\bf CMS} Collaboration, S.~Chatrchyan et~al., {\it {Measurements of
  differential jet cross sections in proton-proton collisions at $\sqrt{s}=7$
  TeV with the CMS detector}},  \href{http://xxx.lanl.gov/abs/1212.6660}{{\tt
  arXiv:1212.6660}}.

\bibitem{Aad:2011fc}
{\bf ATLAS} Collaboration, G.~Aad et~al., {\it {Measurement of inclusive jet
  and dijet production in pp collisions at sqrt(s) = 7 TeV using the ATLAS
  detector}},  {\em Phys. Rev.} {\bf D86} (2012) 014022,
  [\href{http://xxx.lanl.gov/abs/1112.6297}{{\tt arXiv:1112.6297}}].

\bibitem{d'Enterria:2012yj}
D.~d'Enterria and J.~Rojo, {\it {Quantitative constraints on the gluon
  distribution function in the proton from collider isolated-photon data}},
  {\em Nucl.Phys.} {\bf B860} (2012) 311--338,
  [\href{http://xxx.lanl.gov/abs/1202.1762}{{\tt arXiv:1202.1762}}].

\bibitem{Carminati:2012mm}
L.~Carminati, C.~Costa, D.~d'Enterria, I.~Koletsou, G.~Marchiori, et~al., {\it
  {Sensitivity of the LHC isolated-gamma+jet data to the parton distribution
  functions of the proton}},  \href{http://xxx.lanl.gov/abs/1212.5511}{{\tt
  arXiv:1212.5511}}.

\bibitem{Ridder:2013mf}
A.~G.-D. Ridder, T.~Gehrmann, E.~Glover, and J.~Pires, {\it {Second order QCD
  corrections to jet production at hadron colliders: the all-gluon
  contribution}},  \href{http://xxx.lanl.gov/abs/1301.7310}{{\tt
  arXiv:1301.7310}}.

\bibitem{Ball:2010de}
{\bf NNPDF} Collaboration, R.~D. Ball et~al., {\it {A first
  unbiased global NLO determination of parton distributions and their
  uncertainties}},  {\em Nucl. Phys.} {\bf B838} (2010) 136--206,
  [\href{http://xxx.lanl.gov/abs/1002.4407}{{\tt arXiv:1002.4407}}].

\bibitem{Czakon:2011xx}
M.~Czakon and A.~Mitov, {\it {Top++: a program for the calculation of the
  top-pair cross-section at hadron colliders}},
  \href{http://xxx.lanl.gov/abs/1112.5675}{{\tt arXiv:1112.5675}}.

\bibitem{Alekhin:2012ig}
S.~Alekhin, J.~Blumlein, and S.~Moch, {\it {Parton distribution functions and
  benchmark cross sections at NNLO}},
  {\em  Phys.\ Rev.\ D} {\bf 86}, 054009 (2012),
  [\href{http://xxx.lanl.gov/abs/1202.2281}{{\tt arXiv:1202.2281}}].

\bibitem{Nadolsky:2012ia}
J.~Gao, M.~Guzzi, J.~Huston, H.~-L.~Lai, Z.~Li, P.~Nadolsky, J.~Pumplin and D.~Stump et al.,
 {\it { The CT10 NNLO Global Analysis of QCD}},  \href{http://xxx.lanl.gov/abs/1302.6246}{{\tt
  arXiv:1302.6246}}.

\bibitem{CooperSarkar:2011aa}
{\bf ZEUS , H1} Collaborations, A.~Cooper-Sarkar, {\it {PDF Fits at HERA}},
  {\em PoS} {\bf EPS-HEP2011} (2011) 320,
  [\href{http://xxx.lanl.gov/abs/1112.2107}{{\tt arXiv:1112.2107}}].

\bibitem{JimenezDelgado:2008hf} 
  P.~Jimenez-Delgado and E.~Reya,
  {\it Dynamical NNLO parton distributions,}
  {\em Phys.\ Rev.\ D} {\bf 79}, 074023 (2009),
 [\href{http://xxx.lanl.gov/abs/0810.4274}{{\tt arXiv:0810.4274}}].


\bibitem{Martin:2009iq}
A.~D. Martin, W.~J. Stirling, R.~S. Thorne, and G.~Watt, {\it {Parton
  distributions for the LHC}},  {\em Eur. Phys. J.} {\bf C63} (2009) 189--285,
  [\href{http://xxx.lanl.gov/abs/0901.0002}{{\tt arXiv:0901.0002}}].

\bibitem{Ball:2012cx}
{\bf NNPDF} Collaboration, R.~D. Ball, V.~Bertone, S.~Carrazza, C.~S. Deans, L.~Del~Debbio, et~al., {\it
  {Parton distributions with LHC data}},  {\em Nucl.Phys.} {\bf B867} (2013)
  244--289, [\href{http://xxx.lanl.gov/abs/1207.1303}{{\tt arXiv:1207.1303}}].

\bibitem{Ball:2012wy}
R.~D. Ball, S.~Carrazza, L.~Del~Debbio, S.~Forte, J.~Gao, et~al., {\it {Parton
  Distribution Benchmarking with LHC Data}},
  \href{http://xxx.lanl.gov/abs/1211.5142}{{\tt arXiv:1211.5142}}.

\bibitem{Nadolsky:2001yg}
P.~M. Nadolsky and Z.~Sullivan, {\it {PDF uncertainties in W H production at
  Tevatron}},  \href{http://xxx.lanl.gov/abs/hep-ph/0110378}{{\tt
  hep-ph/0110378}}.

\bibitem{Beringer:1900zz}
{\bf Particle Data Group} Collaboration, J.~Beringer et~al., {\it {Review of
  Particle Physics (RPP)}},  {\em Phys.Rev.} {\bf D86} (2012) 010001.

\bibitem{Lai:2010nw}
H.-L. Lai et~al., {\it {Uncertainty induced by QCD coupling in the CTEQ global
  analysis of parton distributions}},  {\em Phys. Rev.} {\bf D82} (2010)
  054021, [\href{http://xxx.lanl.gov/abs/1004.4624}{{\tt arXiv:1004.4624}}].

\bibitem{Demartin:2010er}
F.~Demartin, S.~Forte, E.~Mariani, J.~Rojo, and A.~Vicini, {\it {The impact of
  PDF and alphas uncertainties on Higgs Production in gluon fusion at hadron
  colliders}},  {\em Phys. Rev.} {\bf D82} (2010) 014002,
  [\href{http://xxx.lanl.gov/abs/1004.0962}{{\tt arXiv:1004.0962}}].



\bibitem{lhapdf} 
  M.~R.~Whalley, D.~Bourilkov, R.~C.~Group and ,
  {\it The Les Houches accord PDFs (LHAPDF) and LHAGLUE},  hep-ph/0508110.

\bibitem{Aaltonen:2012ra}
{\bf CDF and D0} Collaborations, T.~Aaltonen et~al., {\it {Combination of the
  top-quark mass measurements from the Tevatron collider}},  {\em Phys.Rev.}
  {\bf D86} (2012) 092003, [\href{http://xxx.lanl.gov/abs/1207.1069}{{\tt
  arXiv:1207.1069}}].

\bibitem{Dittmaier:2011ti}
{\bf LHC Higgs Cross Section Working Group}, S.~Dittmaier et~al.,
  {\it {Handbook of LHC Higgs Cross Sections: 1. Inclusive Observables}},
  \href{http://xxx.lanl.gov/abs/1101.0593}{{\tt arXiv:1101.0593}}.

\bibitem{tevsigma}
{\bf The Tevatron electroweak working group}, T.~Aaltonen et~al.,
  {\it {Combination of the tt production cross section measurements from the
  Tevatron Collider}},  {\tt D0 Note 6363}.

\bibitem{ATLAS:2012fja}
{\bf ATLAS} Collaboration, {\it {Statistical combination of top quark pair
  production cross-section measurements using dilepton, single-lepton, and
  all-hadronic final states at √s = 7 TeV with the ATLAS detector}}, {\tt 
ATLAS-CONF-2012-024}.

\bibitem{Chatrchyan:2012bra}
{\bf CMS} Collaboration, S.~Chatrchyan et~al., {\it {Measurement of the
  $t\bar{t}$ production cross section in the dilepton channel in $pp$
  collisions at $\sqrt{s}=7$ TeV}},  {\em JHEP} {\bf 1211} (2012) 067,
  [\href{http://xxx.lanl.gov/abs/1208.2671}{{\tt arXiv:1208.2671}}].

\bibitem{ATLAS-CONF-2012-149}
{\bf ATLAS} Collaboration, {\it Measurement of the top quark pair production
  cross section in the single-lepton channel with atlas in proton-proton
  collisions at 8 tev using kinematic fits with b-tagging},
 {\tt  ATLAS-CONF-2012-149}.

\bibitem{CMS-PAS-TOP-12-007}
{\bf CMS} Collaboration, {\it Top pair cross section in dileptons},
  {\tt CMS-PAS-TOP-12-003}.

\bibitem{CMS-PAS-TOP-12-006}
{\bf CMS} Collaboration, {\it Top pair cross section in e/mu+jets at 8 tev},
  {\tt CMS-PAS-TOP-12-006}.

\bibitem{CMS-PAS-TOP-12-003}
{\bf ATLAS and CMS} Collaborations, {\it Combination of Atlas and CMS top-quark
  pair cross section measurements using proton-proton collisions at sqrt(s) = 7
  TeV},  {\tt CMS-PAS-TOP-12-003}.



\bibitem{Forte:2010ta}
S.~Forte, E.~Laenen, P.~Nason, and J.~Rojo, {\it {Heavy quarks in
  deep-inelastic scattering}},  {\em Nucl. Phys.} {\bf B834} (2010) 116--162,
  [\href{http://xxx.lanl.gov/abs/1001.2312}{{\tt arXiv:1001.2312}}].

\bibitem{thornehq}
R.~S. Thorne, {\it A variable-flavour number scheme for nnlo},  {\em Phys.
  Rev.} {\bf D73} (2006) 054019,
  [\href{http://xxx.lanl.gov/abs/hep-ph/0601245}{{\tt hep-ph/0601245}}].

\bibitem{Guzzi:2011ew}
M.~Guzzi, P.~M. Nadolsky, H.-L. Lai, and C.-P. Yuan, {\it {General-Mass
  Treatment for Deep Inelastic Scattering at Two-Loop Accuracy}},  {\em
  Phys.Rev.} {\bf D86} (2012) 053005,
  [\href{http://xxx.lanl.gov/abs/1108.5112}{{\tt arXiv:1108.5112}}].

\bibitem{Thorne:2012az}
R.~Thorne, {\it {The Effect of Changes of Variable Flavour Number Scheme on
  PDFs and Predicted Cross Sections}},  {\em Phys. Rev.} {\bf D86} (2012)
  074017, [\href{http://xxx.lanl.gov/abs/1201.6180}{{\tt arXiv:1201.6180}}].

\bibitem{Ball:2013gsa}
{\bf NNPDF} Collaboration, R.~D. Ball et~al., {\it
  {Theoretical issues in PDF determination and associated uncertainties}},
  {\em Phys.Lett.}
  {\bf B723} (2013) 330, [\href{http://xxx.lanl.gov/abs/1303.1189}{{\tt arXiv:1303.1189}}].

\bibitem{ATLAS:2011qga}
{\bf ATLAS} Collaboration, {\it {Determination of the Top-Quark Mass from the
  ttbar Cross Section Measurement in pp Collisions at sqrt(s)=7 TeV with the
  ATLAS detector}}, {\tt  ATLAS-CONF-2011-054}.

\bibitem{CMS:2011lkd}
{\bf CMS} Collaboration, {\it {Determination of the Top Quark Mass from the
  ttbar Cross Section at sqrt(s) = 7 TeV}}, {\tt  CMS-PAS-TOP-11-008}.

\bibitem{Langenfeld:2009wd}
U.~Langenfeld, S.~Moch, and P.~Uwer, {\it {Measuring the running top-quark
  mass}},  {\em Phys.Rev.} {\bf D80} (2009) 054009,
  [\href{http://xxx.lanl.gov/abs/0906.5273}{{\tt arXiv:0906.5273}}].



\bibitem{Abazov:2009nc}
{\bf D0} Collaboration, V.~Abazov et~al., {\it {Determination of the strong
  coupling constant from the inclusive jet cross section in $p\bar{p}$
  collisions at sqrt(s)=1.96 TeV}},  {\em Phys.Rev.} {\bf D80} (2009) 111107,
  [\href{http://xxx.lanl.gov/abs/0911.2710}{{\tt arXiv:0911.2710}}].

\bibitem{Malaescu:2012ts}
B.~Malaescu and P.~Starovoitov, {\it {Evaluation of the Strong Coupling
  Constant $\alpha_s$ Using the ATLAS Inclusive Jet Cross-Section Data}},  {\em
  Eur.Phys.J.} {\bf C72} (2012) 2041,
  [\href{http://xxx.lanl.gov/abs/1203.5416}{{\tt arXiv:1203.5416}}].

\bibitem{CMS-PAS-QCD-11-003}
{\bf CMS} Collaboration, {\it Measurement of the ratio of the inclusive 3-jet
  to 2-jet cross-sections in pp collisions at 7 tev and first determination of
  the strong coupling at transverse momenta in the tev range},
  {{\tt CMS-PAS-QCD-11-003}}.

\bibitem{cmstopas}
{\bf CMS} Collaboration, {\it First determination of the strong coupling
  constant from the ttbar cross section}, {{\tt CMS-PAS-TOP-12-022}}.

\bibitem{Nadolsky:2008zw} 
  P.~M.~Nadolsky, H.~-L.~Lai, Q.~-H.~Cao, J.~Huston et~al., {\it {
  Implications of CTEQ global analysis for collider observables}},
 {\em Phys.\ Rev.\ }  {\bf D78}, 013004 (2008),
[\href{http://xxx.lanl.gov/abs/0802.0007}{{\tt
  arXiv:0802.0007}}].

\bibitem{Ball:2011gg}
{\bf NNPDF} Collaboration, R.~D. Ball, V.~Bertone, F.~Cerutti, L.~Del~Debbio,
  S.~Forte, et~al., {\it {Reweighting and Unweighting of Parton Distributions
  and the LHC W lepton asymmetry data}},  {\em Nucl.Phys.} {\bf B855} (2012)
  608--638, [\href{http://xxx.lanl.gov/abs/1108.1758}{{\tt arXiv:1108.1758}}].

\bibitem{Ball:2010gb}
{\bf NNPDF} Collaboration, R.~D. Ball et~al., {\it {Reweighting NNPDFs: the W
  lepton asymmetry}},  {\em Nucl. Phys.} {\bf B849} (2011) 112--143,
  [\href{http://xxx.lanl.gov/abs/1012.0836}{{\tt arXiv:1012.0836}}].

\bibitem{Carli:2010rw} 
  T.~Carli, D.~Clements, A.~Cooper-Sarkar, C.~Gwenlan, G.~P.~Salam, F.~Siegert, P.~Starovoitov and M.~Sutton,
  {\it A posteriori inclusion of parton density functions in NLO QCD final-state calculations at hadron colliders: The APPLGRID Project,}
  {\em Eur.\ Phys.\ J.\ C }{\bf 66}, 503 (2010),
 [\href{http://xxx.lanl.gov/abs/0911.2985}{{\tt arXiv:0911.2985}}].

\bibitem{Watt:2012tq}
G.~Watt and R.~Thorne, {\it {Study of Monte Carlo approach to experimental
  uncertainty propagation with MSTW 2008 PDFs}},  {\em JHEP} {\bf 1208} (2012)
  052, [\href{http://xxx.lanl.gov/abs/1205.4024}{{\tt arXiv:1205.4024}}].

\bibitem{top:2012hg}
{\bf ATLAS} Collaboration, G.~Aad et~al., {\it {Measurements of top quark pair
  relative differential cross-sections with ATLAS in pp collisions at sqrt(s) =
  7 TeV}},  \href{http://xxx.lanl.gov/abs/1207.5644}{{\tt arXiv:1207.5644}}.

\bibitem{:2012qka}
{\bf CMS} Collaboration, S.~Chatrchyan et~al., {\it {Measurement of
  differential top-quark pair production cross sections in $pp$ colisions at
  $\sqrt{s}=7$ TeV}},  \href{http://xxx.lanl.gov/abs/1211.2220}{{\tt
  arXiv:1211.2220}}.
  
  \bibitem{Ahrens:2011mw} 
  V.~Ahrens, A.~Ferroglia, M.~Neubert, B.~D.~Pecjak, L.~-L.~Yang and ,
  {\it RG-improved single-particle inclusive cross sections and forward-backward asymmetry in $t\bar t$ production at hadron colliders,},
  {\em JHEP} {\bf 1109}, 070 (2011),
  \href{http://xxx.lanl.gov/abs/1103.0550}{{\tt
  arXiv:1103.0550}}.



\bibitem{Ball:2011mu}
{\bf NNPDF} Collaboration, R.~D. Ball et~al., {\it {Impact of Heavy Quark
  Masses on Parton Distributions and LHC Phenomenology}},  {\em Nucl. Phys.}
  {\bf B849} (2011) 296--363, [\href{http://xxx.lanl.gov/abs/1101.1300}{{\tt
  arXiv:1101.1300}}].

\bibitem{Ball:2011uy}
{\bf NNPDF} Collaboration, R.~D. Ball et~al., {\it {Unbiased global
  determination of parton distributions and their uncertainties at NNLO and at
  LO}},  {\em Nucl.Phys.} {\bf B855} (2012) 153--221,
  [\href{http://xxx.lanl.gov/abs/1107.2652}{{\tt arXiv:1107.2652}}].

\bibitem{Alwall:2007st}
J.~Alwall et~al., {\it {MadGraph/MadEvent v4: The New Web Generation}},  {\em
  JHEP} {\bf 09} (2007) 028, [\href{http://xxx.lanl.gov/abs/0706.2334}{{\tt
  arXiv:0706.2334}}].

\bibitem{Chatrchyan:2012cx}
{\bf CMS} Collaboration, S.~Chatrchyan et~al., {\it {Search for resonant
  $t\bar{t}$ production in lepton+jets events in $pp$ collisions at
  $\sqrt{s}=7$ TeV}},  {\em JHEP} {\bf 1212} (2012) 015,
  [\href{http://xxx.lanl.gov/abs/1209.4397}{{\tt arXiv:1209.4397}}].


\bibitem{Aad:2012dpa} 
{\bf ATLAS} Collaboration,  G.~Aad { et al.}, {\it
 A search for $t\bar{t}$ resonances in lepton+jets events with highly boosted top quarks collected in $pp$ collisions at $\sqrt{s} = 7$ TeV with the ATLAS detector},
  {\em JHEP} {\bf 1209}, 041 (2012),
 [\href{http://xxx.lanl.gov/abs/1207.2409}{{\tt arXiv:1207.2409}}].

\bibitem{Aad:2012wm} 
{\bf ATLAS} Collaboration,  G.~Aad { et al.}, {\it
 A search for $t\bar{t}$ resonances with the ATLAS detector in 2.05 
fb$^{-1}$ of proton-proton collisions at $\sqrt{s}=7$ TeV},
{\em  Eur.\ Phys.\ J.}\  {\bf C72}, 2083 (2012),
 [\href{http://xxx.lanl.gov/abs/1205.5371}{{\tt arXiv:1205.5371}}].
 
\bibitem{Aad:2012raa} 
{\bf ATLAS} Collaboration,  G.~Aad { et al.}, {\it
 Search for resonances decaying into top-quark pairs using fully hadronic decays in $pp$ collisions with ATLAS at $\sqrt{s}=7$ TeV,}
  {\em JHEP }{\bf 1301}, 116 (2013),
[\href{http://xxx.lanl.gov/abs/1211.2202}{{\tt arXiv:1211.2202}}].

\bibitem{Frederix:2011ss}
R.~Frederix, S.~Frixione, V.~Hirschi, F.~Maltoni, R.~Pittau, et~al., {\it
  {Four-lepton production at hadron colliders: aMC@NLO predictions with
  theoretical uncertainties}},  {\em JHEP} {\bf 1202} (2012) 099,
  [\href{http://xxx.lanl.gov/abs/1110.4738}{{\tt arXiv:1110.4738}}].

\bibitem{Frixione:2002ik}
S.~Frixione and B.~R. Webber, {\it {Matching NLO QCD computations and parton
  shower simulations}},  {\em JHEP} {\bf 0206} (2002) 029,
  [\href{http://xxx.lanl.gov/abs/hep-ph/0204244}{{\tt hep-ph/0204244}}].

\bibitem{Frederix:2009yq}
R.~Frederix, S.~Frixione, F.~Maltoni, and T.~Stelzer, {\it {Automation of
  next-to-leading order computations in QCD: The FKS subtraction}},  {\em JHEP}
  {\bf 0910} (2009) 003, [\href{http://xxx.lanl.gov/abs/0908.4272}{{\tt
  arXiv:0908.4272}}].

\bibitem{Corcella:2000bw}
G.~Corcella, I.~Knowles, G.~Marchesini, S.~Moretti, K.~Odagiri, et~al., {\it
  {HERWIG 6: An Event generator for hadron emission reactions with interfering
  gluons (including supersymmetric processes)}},  {\em JHEP} {\bf 0101} (2001)
  010, [\href{http://xxx.lanl.gov/abs/hep-ph/0011363}{{\tt hep-ph/0011363}}].

\bibitem{Mangano:2012mh}
M.~L. Mangano and J.~Rojo, {\it {Cross Section Ratios between different CM
  energies at the LHC: opportunities for precision measurements and BSM
  sensitivity}},  {\em JHEP} {\bf 1208} (2012) 010,
  [\href{http://xxx.lanl.gov/abs/1206.3557}{{\tt arXiv:1206.3557}}].

\bibitem{Chatrchyan:2012vu}
{\bf CMS } Collaboration, S.~Chatrchyan et~al., {\it {Search for
  pair produced fourth-generation up-type quarks in $pp$ collisions at
  $\sqrt{s}=7$ TeV with a lepton in the final state}},  {\em Phys.Lett.} {\bf
  B718} (2012) 307--328, [\href{http://xxx.lanl.gov/abs/1209.0471}{{\tt
  arXiv:1209.0471}}].

\bibitem{ATLAS:2012qe}
{\bf ATLAS } Collaboration, G.~Aad et~al., {\it {Search for pair
  production of heavy top-like quarks decaying to a high-pT $W$ boson and a $b$
  quark in the lepton plus jets final state at $\sqrt{s}=7$ TeV with the ATLAS
  detector}},  {\em Phys.Lett.} {\bf B718} (2013) 1284--1302,
  [\href{http://xxx.lanl.gov/abs/1210.5468}{{\tt arXiv:1210.5468}}].

\bibitem{Chatrchyan:2012fp}
{\bf CMS } Collaboration, S.~Chatrchyan et~al., {\it {Combined
  search for the quarks of a sequential fourth generation}},  {\em Phys.Rev.}
  {\bf D86} (2012) 112003, [\href{http://xxx.lanl.gov/abs/1209.1062}{{\tt
  arXiv:1209.1062}}].













\end{thebibliography}
\end{document}